\documentclass[iop,numberedappendix,revtex4-1]{emulateapj}

\usepackage[]{amsmath}

\shorttitle{MACS1149 Forecasts}
\shortauthors{Treu et al. (2015)}

\usepackage{color}			      
\definecolor{midgray}{gray}{0.4}		
\definecolor{orange}{rgb}{1,0.5,0}    

\usepackage[colorlinks=true,citecolor=midgray,linkcolor=midgray,urlcolor=midgray]{hyperref}        
\usepackage{color}
\usepackage{url}

\newcommand{\Nzhst}{389}
\newcommand{\Nzmem}{170}
\newcommand{\Nzboth}{70}
\newcommand{\Nzmuse}{111}
\newcommand{\Npmem}{136}

\newcommand{\Nztot}{429}

\newcommand{\simgt}{\,\rlap{\lower 3.5 pt \hbox{$\mathchar \sim$}} \raise
1pt \hbox {$>$}\,}
\newcommand{\simlt}{\,\rlap{\lower 3.5 pt \hbox{$\mathchar \sim$}} \raise
1pt \hbox {$<$}\,}

\newcommand{\HST}{\textit{HST}}

\newcommand{\M}{MACSJ1149.5+2223}

\begin{document}

\title{``Refsdal'' Meets Popper: Comparing Predictions of the Reappearance of the Multiply Imaged Supernova Behind \M}


\newcommand{\HubbleFellow}{Hubble Fellow}
\newcommand{\Packard}{Packard Fellow}
\newcommand{\CalTech}{California Institute of Technology, 1200 East California Boulevard, Pasadena, CA 91125, USA}
\newcommand{\Cantabria}{IFCA, Instituto de F\'isica de Cantabria (UC-CSIC), Av. de Los Castros s/n, 39005 Santander, Spain}
\newcommand{\IFCA}{\Cantabria}
\newcommand{\JHU}{Department of Physics and Astronomy, The Johns Hopkins University, 3400 N. Charles St., Baltimore, MD 21218, USA}
\newcommand{\Michigan}{Department of Astronomy, University of Michigan, 1085 S. University Avenue, Ann Arbor, MI 48109, USA}
\newcommand{\UCDavis}{University of California Davis, 1 Shields Avenue, Davis, CA 95616, USA}
\newcommand{\UCLA}{Department of Physics and Astronomy, University of California, Los Angeles, CA 90095, USA; tt@astro.ucla.edu}
\newcommand{\USC}{Department of Physics and Astronomy, University of South Carolina, 712 Main St., Columbia, SC 29208, USA}
\newcommand{\RCEU}{Research Center for the Early Universe, University of Tokyo, 7-3-1 Hongo, Bunkyo-ku, Tokyo 113-0033, Japan}
\newcommand{\TokyoPhys}{Department of Physics, University of Tokyo, 7-3-1 Hongo, Bunkyo-ku, Tokyo 113-0033, Japan}
\newcommand{\IPMU}{Kavli Institute for the Physics and Mathematics of the Universe (Kavli IPMU, WPI), University of Tokyo, 5-1-5 Kashiwanoha, Kashiwa, Chiba 277-8583, Japan}
\newcommand{\TokyoAstro}{Department of Astronomy, Graduate School of Science, The University of Tokyo, 7-3-1 Hongo, Bunkyo-ku, Tokyo 113-0033, Japan}
\newcommand{\DARK}{Dark Cosmology Centre, Niels Bohr Institute, University of Copenhagen, Juliane Maries Vej 30, DK-2100 Copenhagen, Denmark} 
\newcommand{\INFN}{INFN, Sezione di Bologna, Viale Berti Pichat 6/2, I-40127 Bologna, Italy}
\newcommand{\EHU}{Fisika Teorikoa, Zientzia eta Teknologia Fakultatea, Euskal Herriko Unibertsitatea UPV/EHU}
\newcommand{\Basque}{IKERBASQUE, Basque Foundation for Science, Alameda Urquijo, 36-5 48008 Bilbao, Spain}
\newcommand{\Berkeley}{Department of Astronomy, University of California, Berkeley, CA 94720-3411, USA}
\newcommand{\STScI}{Space Telescope Science Institute, 3700 San Martin Dr., Baltimore, MD 21218, USA}
\newcommand{\Ferrara}{Dipartimento di Fisica e Scienze della Terra, Universit\`{a} degli Studi di Ferrara, via Saragat 1, I-44122, Ferrara, Italy}
\newcommand{\INAF}{INAF, Osservatorio Astronomico di Bologna, via Ranzani 1, I-40127 Bologna, Italy}
\newcommand{\UCSB}{Department of Physics, University of California, Santa Barbara, CA 93106-9530, USA}
\newcommand{\SantaBarbara}{\UCSB}
\newcommand{\Kapteyn}{Kapteyn Astronomical Institute, University of Groningen, Postbus 800, 9700 AV Groningen, the Netherlands}
\newcommand{\WKU}{Department of Physics, Western Kentucky University, Bowling Green, KY 42101, USA}
\newcommand{\IAP}{Institut d’Astrophysique de Paris, UMR7095 CNRS-Universit\'{e} Pierre et Marie Curie, 98bis bd Arago, F-75014 Paris, France}
\newcommand{\ASIAA}{Institute of Astronomy and Astrophysics, Academia Sinica, P.O. Box 23-141, Taipei 10617, Taiwan}
\newcommand{\TokyoKashiwa}{Institute for Cosmic Ray Research, The University of Tokyo, Kashiwa, Chiba 277-8582, Japan}
\newcommand{\Munich}{University Observatory Munich, Scheinerstrasse 1, D-81679 Munich, Germany} 

\newcommand{\Rutgers}{Department of Physics and Astronomy, Rutgers, The State University of New Jersey, Piscataway, NJ 08854, USA}
\newcommand{\IllinoisAstro}{ Astronomy Department, University of Illinois at Urbana-Champaign, 1002 W.\ Green Street, Urbana, IL 61801, USA }
\newcommand{\IllinoisPhysics}{ Department of Physics, University of Illinois at Urbana-Champaign, 1110 W.\ Green Street, Urbana, IL 61801, USA }


\newcounter{affilct}
\setcounter{affilct}{0}

\makeatletter
\newcommand{\affilref}[1]{%
  \@ifundefined{c@#1}%
    {\newcounter{#1}%
     \setcounter{#1}{\theaffilct}%
     \refstepcounter{affilct}%
     \label{#1}%
     }{}%
  \ref{#1}%
 }
\makeatother

\makeatletter
\newcommand*\affilreftxt[2]{%
  \@ifundefined{c@#1txt}
    {\newcounter{#1txt}%
     \setcounter{#1txt}{1}
     \altaffiltext{\ref{#1}}{#2}
     }{
     }
  }
\makeatother

\author{T.~Treu\altaffilmark{\affilref{UCLA},\affilref{Packard}}}
\affilreftxt{UCLA}{\UCLA}
\affilreftxt{Packard}{\Packard}
\email{tt@astro.ucla.edu}

\author{G.~Brammer\altaffilmark{\affilref{STScI}}}
\affilreftxt{STScI}{\STScI}

\author{J.~M.~Diego\altaffilmark{\affilref{Cantabria}}}
\affilreftxt{Cantabria}{\Cantabria}

\author{C.~Grillo\altaffilmark{\affilref{DARK}}}
\affilreftxt{DARK}{\DARK}

\author{P.~L.~Kelly\altaffilmark{\affilref{Berkeley}}}
\affilreftxt{Berkeley}{\Berkeley}

\author{M.~Oguri\altaffilmark{\affilref{IPMU},\affilref{TokyoPhys},\affilref{RCEU}}}
\affilreftxt{IPMU}{\IPMU}
\affilreftxt{TokyoPhys}{\TokyoPhys}
\affilreftxt{RCEU}{\RCEU}

\author{S.~A.~Rodney\altaffilmark{\affilref{USC},\affilref{JHU},\affilref{HubbleFellow}}}
\affilreftxt{USC}{\USC}
\affilreftxt{JHU}{\JHU}
\affilreftxt{HubbleFellow}{\HubbleFellow}

\author{P.~Rosati\altaffilmark{\affilref{Ferrara}}}
\affilreftxt{Ferrara}{\Ferrara}

\author{K.~Sharon\altaffilmark{\affilref{Michigan}}}
\affilreftxt{Michigan}{\Michigan}

\author{A.~Zitrin\altaffilmark{\affilref{CalTech},\affilref{HubbleFellow}}}
\affilreftxt{CalTech}{\CalTech}

\author{I.~Balestra\altaffilmark{\affilref{Munich}}}
\affilreftxt{Munich}{\Munich}

\author{M. Brada\v{c}\altaffilmark{\affilref{UCDavis}}}
\affilreftxt{UCDavis}{\UCDavis}

\author{T.~Broadhurst\altaffilmark{\affilref{EHU},\affilref{Basque}}}
\affilreftxt{EHU}{\EHU}
\affilreftxt{Basque}{\Basque}

\author{G.~B.~Caminha\altaffilmark{\affilref{Ferrara}}}
\affilreftxt{Ferrara}{\Ferrara}


\author{A.~Halkola}

\author{A.~Hoag\altaffilmark{\affilref{UCDavis}}}
\affilreftxt{UCDavis}{\UCDavis}

\author{M.~Ishigaki\altaffilmark{\affilref{TokyoKashiwa},\affilref{TokyoPhys}}}
\affilreftxt{TokyoKashiwa}{\TokyoKashiwa}
\affilreftxt{TokyoPhys}{\TokyoPhys}

\author{T.~L.~Johnson\altaffilmark{\affilref{Michigan}}}
\affilreftxt{Michigan}{\Michigan}

\author{W.~Karman\altaffilmark{\affilref{Kapteyn}}}
\affilreftxt{Kapteyn}{\Kapteyn}

\author{R.~Kawamata\altaffilmark{\affilref{TokyoAstro}}}
\affilreftxt{TokyoAstro}{\TokyoAstro}

\author{A.~Mercurio\altaffilmark{\affilref{INAF}}}
\affilreftxt{INAF}{\INAF}


\author{K.~B.~Schmidt\altaffilmark{\affilref{UCSB}}}
\affilreftxt{UCSB}{\UCSB}

\author{L.-G.~Strolger\altaffilmark{\affilref{STScI},\affilref{WKU}}}
\affilreftxt{STScI}{\STScI}
\affilreftxt{WKU}{\WKU}

\author{S.~H.~Suyu\altaffilmark{\affilref{ASIAA}}}
\affilreftxt{ASIAA}{\ASIAA}

\author{A.~V.~Filippenko\altaffilmark{\affilref{Berkeley}}}
\affilreftxt{Berkeley}{\Berkeley}

\author{R.~J.~Foley\altaffilmark{\affilref{IllinoisAstro},\affilref{IllinoisPhysics}}}
\affilreftxt{IllinoisAstro}{\IllinoisAstro}
\affilreftxt{IllinoisPhysics}{\IllinoisPhysics}

\author{S.~W.~Jha\altaffilmark{\affilref{Rutgers}}}
\affilreftxt{Rutgers}{\Rutgers}

\author{B.~Patel\altaffilmark{\affilref{Rutgers}}}
\affilreftxt{Rutgers}{\Rutgers}

\begin{abstract}
Supernova ``Refsdal,'' multiply imaged by cluster MACSJ1149.5+2223,
represents a rare opportunity to make a true blind test of model
predictions in extragalactic astronomy, on a time scale that is short
compared to a human lifetime. In order to take advantage of this
event, we produced seven gravitational lens models with five
independent methods, based on {\it Hubble Space Telescope (HST)}
Hubble Frontier Field images, along with extensive spectroscopic
follow-up observations by \HST, the Very Large and the Keck
Telescopes. We compare the model predictions and show that they agree
reasonably well with the measured time delays and magnification ratios
between the known images, even though these quantities were not used
as input. This agreement is encouraging, considering that the models
only provide statistical uncertainties, and do not include additional
sources of uncertainties such as structure along the line of sight,
cosmology, and the mass sheet degeneracy. We then present the model
predictions for the other appearances of SN ``Refsdal.'' A future
image will reach its peak in the first half of 2016, while another
image appeared between 1994 and 2004. The past image would have been
too faint to be detected in existing archival images. The future
image should be approximately one third as bright as the brightest
known image (i.e., $H_{\rm AB}\approx 25.7$ mag at peak and $H_{\rm
AB}\approx 26.7$ mag six months before peak), and thus detectable in
single-orbit \HST\ images. We will find out soon whether our
predictions are correct.
\end{abstract}


\section{Introduction}
\label{sec:intro}

In 1964 Sjur Resfdal speculated that a supernova (SN) multiply imaged 
by a foreground massive galaxy could be used to measure distances and,
therefore, the Hubble constant \citep{Ref64}. The basic physics behind
this phenomenon is very simple. According to Fermat's principle, in
gravitational optics as in standard optics, multiple images form at
the stationary points of the excess arrival time
\citep{1985A&A...143..413S,B+N86}. The excess arrival time is the
result of the competition between the geometric time delay and the
\citet{Sha64} delay. The arrival time thus depends on the apparent
position of the image in the sky as well as the gravitational
potential. Since the arrival time is measured in seconds, while all
the other lensing observables are measured in angles on the sky, their
relationship depends on the angular diameter distance $D$. In the
simplest case of single-plane lensing, the time delay between two
images is proportional to the so-called time-delay distance, $D_d D_s
(1+z_d)/D_{ds}$, where $d$ and $s$ represent the deflector and the
source, respectively
\citep[see, e.g., for definitions][]{2006glsw.conf.....M,Tre10,Suy++10}.

Over the past decades, many authors have highlighted the importance
and applications of identifying such events
\citep[e.g.,][]{1998MNRAS.296..763K,2001ApJ...556L..71H,2002A&A...393...25G,2003ApJ...592...17B,2003MNRAS.338L..25O}, computed rates and proposed search strategies \citep{1988ApJ...324..786L,2000MNRAS.319..549S,2003ApJ...583..584O,O+M10},
and identified highly magnified supernovae \citep{Qui++14}. 

\begin{figure*}
\begin{center}
  \includegraphics[width=\textwidth]{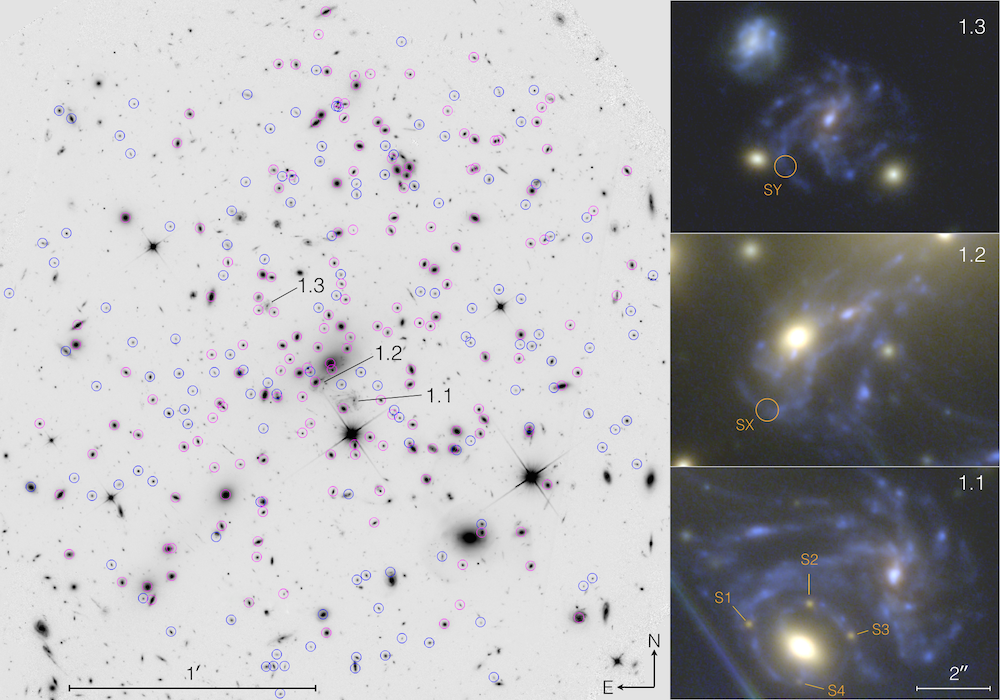}
  \caption{\label{fig:RefsdalHostImages} Multiple images of the SN
  Refsdal host galaxy behind \M. The left panel shows a wide
  view of the cluster, encompassing the entire footprint of the
  WFC3-IR camera. Spectroscopically confirmed cluster member galaxies
  are highlighted in magenta circles.  Cyan circles indicate those
  associated with the cluster based on their photometric properties.
  The three panels on the right show in more detail the multiple
  images of the SN Refsdal host galaxy (labeled 1.1, 1.2, and
  1.3). The positions of the known images of SN Refsdal are labeled as
  S1--S4, while the model-predicted locations of the future and past
  appearances are labeled as SX and SY, respectively.}
\end{center}    
\end{figure*}

Finally, 50 years after the initial proposal by Refsdal, the first
multiply imaged SN was discovered in November 2014 \citep{Kel++15} 
in {\it Hubble Space Telescope (\HST)} images of the cluster
\M\ \citep{Ebe++07,Smith++09,Zitrin:2009p33878}, taken as part of the
Grism Lens Amplified Survey from Space \citep[GLASS; GO-13459, PI
Treu;][]{Schmidt:2014p33661,Tre++15}, and aptly nicknamed ``SN Refsdal.''
SN Refsdal was identified in difference imaging as four
point sources that were not present in earlier images taken as part of
the CLASH survey \citep{Postman:2012p27556}. Luckily, the event was
discovered just before the beginning of an intensive imaging campaign
as part of the Hubble Frontier Field (HFF) initiative \citep[Lotz et
al. 2015, in prep.;][]{CBZ15}. Additional epochs were obtained
as part of the Frontier SN program (GO-13790, PI Rodney), and a
Director Discretionary Time program (GO/DD-14041, PI Kelly). The
beautiful images that have emerged
(Figure~\ref{fig:RefsdalHostImages}) are an apt celebration of the
international year of light and the 100-year anniversary of the
theory of general relativity
\citep[e.g.,][]{T+E15}.

The gravitational lensing configuration of SN Refsdal is very
remarkable. The SN exploded in one arm of an almost face-on
spiral galaxy that is multiply imaged and highly magnified by the
cluster gravitational potential \citep[$z_s=1.489$ and $z_d=0.542$;
redshifts from][]{Gri++15}. Furthermore, the spiral arm hosting
SN Refsdal happens to be sufficiently close to a cluster member galaxy
that four additional multiple images are formed with average
separation of order arcseconds, typical of galaxy-scale strong
lensing. This set of four images close together in an ``Einstein
cross'' configuration is where SN Refsdal has been detected so far
(labeled S1--S4 in Fig.~\ref{fig:RefsdalHostImages}). As we discuss
below, the cluster-scale images are more separated in terms of their
arrival time, with time delays that can be much longer than the
duration of the event, and therefore it is consistent with the lensing
interpretation that they have not yet been seen.

The original suggestion by \citet{Ref64} was to use such events to
measure distances and therefore cosmological parameters, starting from
the Hubble constant.  While distances with interesting accuracy and
precision have been obtained from gravitational time delays in 
galaxy-scale systems lensing quasars \citep[e.g.,][]{Suy++14}, it is
premature to attempt this in the case of SN Refsdal.  The time delay is
not yet known with precision comparable to that attained for lensed
quasars
\citep[e.g.,][]{Tew++13}, and the mass distribution of the cluster \M\ is
inherently much more complex than that of a single elliptical galaxy.

However, SN Refsdal gives us a unique opportunity to test the current
mass models of \M, by conducting a textbook-like falsifiable
experiment \citep{Pop92}. All of the models that have been published
after the discovery of SN Refsdal
\citep{Kel++15,Ogu2015,S+J15,Die++15,Jau++15} predict that an
additional image will form some time in the near future (close to
image 1.2 of the host galaxy, shown in
Figure~\ref{fig:RefsdalHostImages}). It could appear as early as
October 2015 or in a few years. The field of \M\ is unobservable with
\HST at the time of submission of this paper, but observations will
resume at the end of October 2015 as part of an approved Cycle 23
program (GO-14199, PI Kelly). We thus have the opportunity to carry
out a true blind test of the models, if we act sufficiently fast. This
test is similar in spirit to the test of magnification models using
singly imaged Type Ia supernovae
\citep{Pat+14,Nor+14,Rod++15}. The uniqueness of our test lies in the 
fact that it is based on the prediction of an event that has not yet 
happened, and it is thus intrinsically blind and immune to experimenter 
bias.

The quality and quantity of data available to lens modelers have
improved significantly since the discovery of SN Refsdal and the
publication of the first modeling papers. As part of the HFF and
follow-up programs, there are now significantly deeper \HST\
images. Spectroscopy of hundreds of sources in the field
(Fig.~\ref{fig:RefsdalHostImages}) is available from \HST\ grism data
obtained as part of GLASS and SN Refsdal follow-up campaign (PI
Kelly), from Multi-Unit Spectroscopic Explorer (MUSE) Very Large
Telescope (VLT) Director's Discretionary Time follow-up observations
(Prog. ID 294.A-5032, PI Grillo), and from follow-up observations
with the DEep Imaging Multi-Object Spectrograph
\citep[DEIMOS][]{Fab++03} on the Keck-II Telescope (PI Jha).

The timing is thus perfect to ask the question: ``Given
state-of-the-art data and models, how accurately can we predict the
arrival time and magnification of the next appearance of a multiply
imaged supernova?'' Answering this question will give us an absolute
measurement of the quality of present-day models, although one should
keep in mind that this is a very specific test. The arrival time and
especially the magnification of a point source depend strongly on the
details of the gravitational potential in the vicinity of the
images. Additional uncertainties in the time delay and magnification
arise from the inhomogeneous distribution of mass along the line of
sight \citep{Suy++10,Collett:2013p34320,Gre++13}, the mass-sheet
degeneracy and its generalizations
\citep{FGS85,SS13,SS14,Suy++14,Xu15}, and the residual uncertainties
in cosmological parameters, especially the Hubble constant
\citep{Rie++11,Fre++12}.  Average or global 
quantities of more general interest, such as the total volume behind
the cluster, or the average magnification, are much less sensitive to
the details of the potential around a specific point.

In order to answer this question in the very short amount of time
available, the SN Refsdal follow-up team worked hard to reduce and
analyze the new data. By May 2015 it was clear that the quality of the
follow-up data would be sufficient to make substantial improvements to
their lens models. Therefore, the follow-up team contacted the three
other groups who had by then published predictions for SN Refsdal, and
offered them the new datasets to update their models, as part of a
concerted comparison effort. The five groups worked together to
incorporate the new information into the lensing analysis, first by
identifying and rigorously vetting new sets of multiple images, and
then to promptly update their models to make a timely prediction. A
synopsis and comparison between the results and predictions of the
various models is presented in this paper. Companion papers by the
individual groups will describe the follow-up campaigns as well as the
details of each modeling effort.

This paper is organized as follows. In Section~\ref{sec:data}, we
briefly summarize the datasets and measurements that are used in this
comparison effort. In Section~\ref{sec:constraints}, we review the
constraints used by the modeling teams. Section~\ref{sec:model} gives
a concise description of each of the five lens modeling techniques
adopted. Section~\ref{sec:compare} presents the main results of this
paper --- a comparison of the predictions of the different
models. Section~\ref{sec:discussion} discusses the results, and
Section~\ref{sec:summary} concludes with a summary. To ensure
uniformity with the modeling effort for the Hubble Frontier Fields
clusters, we adopt a concordance cosmology with $h=0.7$,
$\Omega_m=0.3$, and $\Omega_\Lambda=0.7$. All magnitudes are given in
the AB system \citep{O+G83}.

\section{Summary of Datasets and Measurements}
\label{sec:data}

\begin{figure*}[]
\centerline{
\includegraphics[width=\textwidth]{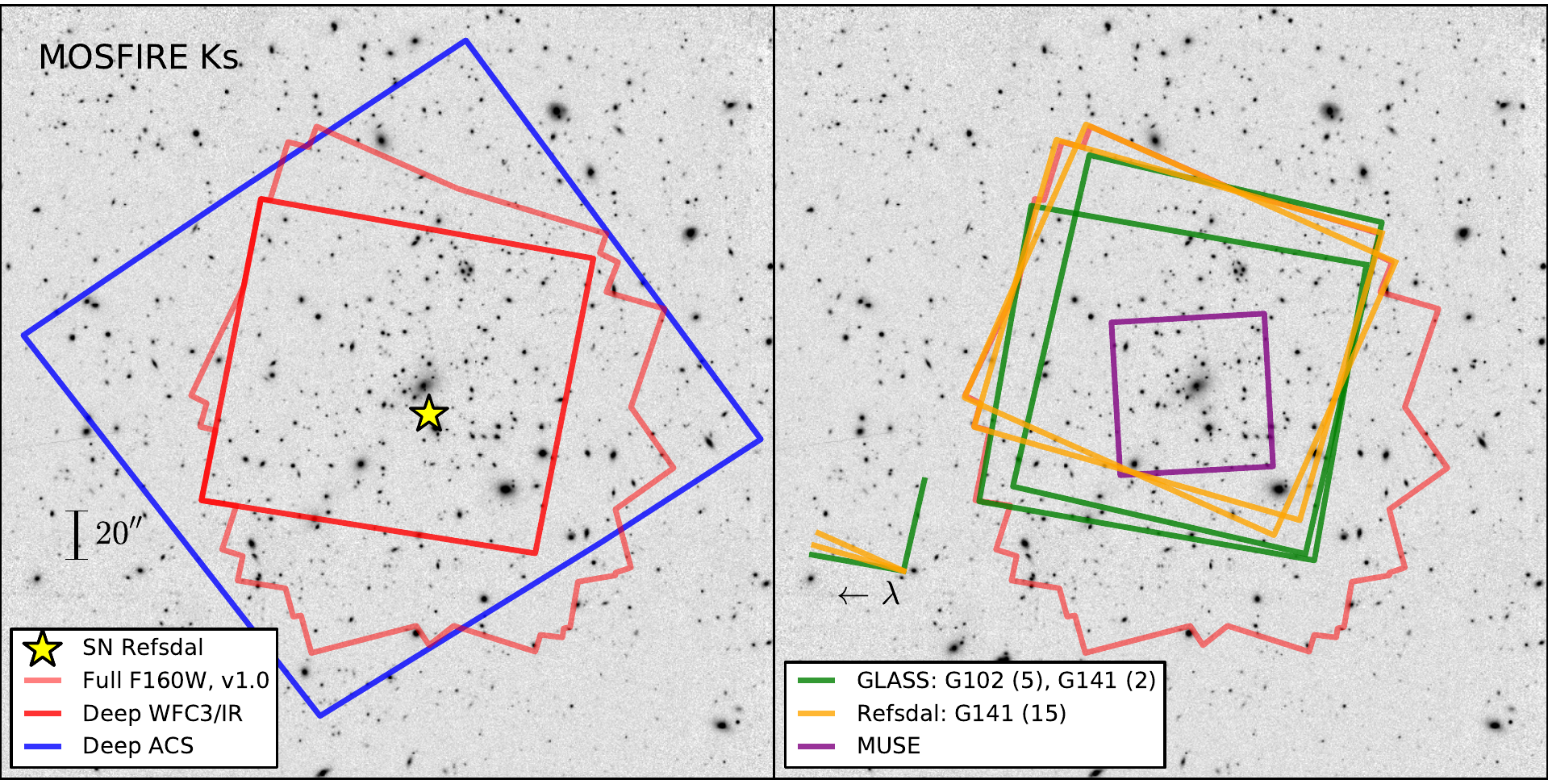}}
\caption{Observational layout of the MUSE and \HST\ spectroscopy in the context of existing imaging data for \M. The ``Full F160W" polygon is the full footprint of the F160W v1.0 FF release image.  The numbers in parentheses in the spectroscopy panel at the right are the number of orbits per grism in each of two orientations. The background image was taken with the MOSFIRE instrument on the Keck-I telescope (Brammer et al. 2015, in prep.).
\label{fig:layout}}
\end{figure*}

\begin{deluxetable*}{lcccccc} \tablecolumns{7}
\tablewidth{0pt} 
\tablecaption{Measured Time Delays and Magnification Ratios}
\tablehead{
\colhead{}  & \multicolumn{2}{c}{SN Template (prelim.)} & \multicolumn{2}{c}{SN Template} & \multicolumn{2}{c}{Polynomial} \\
\colhead{Image pair} & \colhead{$\Delta$t} & \colhead{$\mu$ ratio} & \colhead{$\Delta$t} & \colhead{$\mu$ ratio} & \colhead{$\Delta$t}     & \colhead{$\mu$ ratio}  \\
\colhead{}  & \colhead{(days)} & \colhead{} & \colhead{(days)} & \colhead{} & \colhead{(days)}  & \colhead{} }
S2 S1 & $-$2.1$\pm$1.0  &   1.09$\pm$0.01 & $-$0.8$\pm$1.1 &  1.13$\pm$0.01 &  8.0$^{+1.5}_{-1.4}$ &   1.17$\pm$0.01 \\
S3 S1 &  5.6$\pm$2.2  &   1.04$\pm$0.02 & $-$0.9$\pm$1.1 &  1.03$\pm$0.01 & $-$0.4$^{+1.9}_{-2.8}$ &   1.01$\pm$0.02 \\
S4 S1 &   22$\pm$11   &   0.35$\pm$0.01 & 14.9$\pm$2.4 &  0.34$\pm$0.03 & 30.7$^{+4.8}_{-4.3}$ &   0.38$\pm$0.01 \\
\enddata
\tablecomments{\label{tab:measurements} 
Observed delays and relative magnifications between images S1--S4
of SN~Refsdal.  For the values in columns 2 and 3, light curves
extending up until July 2015 were used by one of us (P.K.) to derive
time delays and magnification ratios using a range of templates. For
the values in columns 4 and 5, Rodney et al. (2015, in prep.)
matched the observed S1--S4 light curves with the best available SN
light-curve template, a model based on SN 1987A with corrections to
account for the bluer color of SN Refsdal.  The values in columns 6
and 7 were determined by fitting the {\it HST} photometry using a
second-order Chebyshev polynomial.}
\end{deluxetable*}
\clearpage

We summarize the datasets and measurements used in this paper.
An overview of the field of view and pointing of the instruments used
here is shown in Figure~\ref{fig:layout}.

\subsection{HST Imaging}

Different versions of the images were used at different stages of the
process. However, the final identifications of multiple images and
their positions were based on the HFF data release v1.0, and their
world coordinate system. The reader is referred to the HFF data
release
webpages\footnote{\url{http://www.stsci.edu/hst/campaigns/frontier-fields/}}
for more information on these data.

\subsubsection{The Light Curves of SN Refsdal}

Since the discovery of SN Refsdal on November 11, 2014, the \M\ field
has been observed in great detail, with {\it HST} imaging in optical
and infrared bands, and deep spectroscopy from {\it HST}, Keck, and
the VLT.  The main goal of the spectroscopic data is to determine the
spectral classification of the SN (Kelly et al. 2015, in prep.).
Photometry from the {\it HST} imaging provides well-sampled
multi-color light curves for SN Refsdal images S1--S4 that exhibit a
slow rise over $\sim$150 days, reaching a peak brightness in the F160W
band on approximately April 13, 2015, with an uncertainty of $\pm$20
days in the observer frame (Rodney et al. 2015, in prep.).

The S1--S4 light curves enable a first measurement of the relative
time delays and magnification ratios (Rodney et al. 2015, in prep.).
Preliminary results from that analysis, using light-curve data up
until the end of July 2015, were included in the first version of this
paper, which was posted to the arxiv before October 30 in order to
make a truly blind prediction before new observations could have
revealed the reappearance of SN Refsdal at position SX. Those
preliminary time delays were only used as a test of the models and not
as an input to the model, so they did not affect any of the
predictions given in this paper. Those preliminary measurements
obtained by one of us (P.K.) using a range of SN templates are listed
in Table~\ref{tab:measurements} for reference. The preliminary peak
date of S1 was found to be April 26 2015 ($\pm20$ days), i.e.
consistent with the final measurement given here (April 13) within the
uncertainties. The final analysis by Rodney et al. (2015) incorporates
Cycle 23 observations collected through Nov. 14, 2015. In
Section~\ref{ssec:cross} we compare our model predictions against all
three sets of available measurements of the S1--S4 time delays and
magnifications. However, to preserve the blind test, the mass models
in this work were not updated to accommodate new information from
those late epochs (e.g., the peak is still assumed to be April 26
2015 in the plots).

The Rodney et al. (2015) analysis uses two approaches for the 
time-delay measurements, first matching the best available SN template
(based on SN 1987A) to each of the S1--S4 light curves, and then using
a simple polynomial representation for the intrinsic light-curve
shape.  Uncertainties were derived using a mock light-curve algorithm
similar to those developed for lensed quasars \citep{TCM13}, although
both measurements ignore effects like microlensing fluctuations
\citep{doblerkeeton06}, and therefore this should be considered as a
lower limit to the total uncertainty.  

Results from these updated measurements are presented in
Table~\ref{tab:measurements} alongside the preliminary ones.  The
relative magnification ratios are measured to within 2\%, and
consistent values are derived from both methods.  However, the time
delays inferred from the two approaches do not agree within the
measured uncertainties.  
This is in part due to the
slowly evolving, comparatively featureless light curve of SN Refsdal, and
the fact that we were not able observe the \M\ field in August
through October, when the SN faded substantially from maximum
light. The differences in inferred time delays may also reflect
systematic biases inherent to one or both of the measurement methods,
as it is possible that none of the available SN light-curve templates
or the simple polynomial model are able to accurately represent the
intrinsic light-curve shape of this peculiar event. The difference
between the two sets of measurements provides an estimate of the
systematic uncertainties.
  
\subsection{Spectroscopy}
\label{ssec:spec}

\subsubsection{HST Spectroscopy}

The \HST\ grism spectroscopy comprises two datasets. The GLASS
data consist of 10 orbits of exposures taken through the G102 grism
and 4 orbits of exposures taken through the G141 grism, spanning the
wavelength range 0.81--1.69~$\mu$m. The GLASS data were taken at two
approximately orthogonal position angles (PAs) to mitigate contamination 
by nearby sources (the first one on 2014 February 23--25, the second PA on
2014 November 3--11). The SN Refsdal follow-up effort was focused on the
G141 grism, reaching a depth of 30 orbits. The pointing and PA
of the follow-up grism data were chosen to optimize the
spectroscopy of the SN itself, and are therefore different from
the ones adopted by GLASS. The SN Refsdal follow-up spectra were taken
between 2014 December 23 and 2015 January 4. Only a brief description
of the data is given here. For more details the reader is referred to
\citet{Schmidt:2014p33661} and \citet{Tre++15} for GLASS, and Brammer
et al.\ (2015, in prep.) and Kelly et al.\ (2015, in
prep.) for the deeper follow-up data.

The observing strategies and data-reduction schemes were very similar
for the two datasets, building on previous work by the 3D-HST survey
\citep{Brammer:2012p12977}. At least 4 subexposures were taken during 
each visit with semi-integer pixel offsets. This enables rejection of
defects and cosmic rays as well as recovery of some of the resolution
lost to undersampling of the point-spread function through interlacing.  
The data were reduced with an updated version of the 3D-HST reduction
pipeline\footnote{http://code.google.com/p/threedhst/} described by
\citet{Brammer:2012p12977} and \citet{Mom++15}. 
The pipeline takes care of alignment, defect removal, background
removal, image combination, and modeling of contamination by
nearby sources. One and two dimensional spectra are extracted for each
source.

The spectra were inspected independently by two of us (T.T. and G.B.)
using custom tools and the interfaces GiG and GiGz (available at
\url{https://github.com/kasperschmidt/GLASSinspectionGUIs}) developed
as part of the GLASS project. Information obtained from the multiband
photometry, continuum, and emission lines was combined to derive a
redshift and quality flag. The few discrepancies between redshifts and
quality flags were resolved by mutual agreement. In the end, we
determined redshifts for \Nzhst\ sources, with quality 3 or 4
\citep[probable or secure, respectively, as defined by][]{Tre++15}.

\begin{deluxetable}{lcccccccc} \tablecolumns{7}
\tablewidth{0pt} 
\tablecaption{Redshift Catalog}
\tablehead{
\colhead{ID$^\star$} & \colhead{$\alpha$}     & \colhead{$\delta$}     & \colhead{$z$} & \colhead{quality} & \colhead{source} & \colhead{Notes} \\
\colhead{}  & \colhead{(J2000)} & \colhead{(J2000)} & \colhead{}  & \colhead{}        & \colhead{} & \colhead{}}
1 & 177.397188 & 22.393744 & 0.0000 & 4 & 2 & \nodata \\ 
2 & 177.404017 & 22.403067 & 0.5660 & 4 & 2 & \nodata \\ 
3 & 177.394525 & 22.400653 & 0.5410 & 4 & 2 & \nodata \\ 
4 & 177.399663 & 22.399597 & 0.5360 & 4 & 2 & \nodata \\ 
5 & 177.404054 & 22.392108 & 0.0000 & 4 & 2 & \nodata \\ 
6 & 177.398554 & 22.389792 & 0.5360 & 4 & 2 & \nodata \\ 
7 & 177.393010 & 22.396799 & 2.9490 & 4 & 4 & 4.1 \\ 
8 & 177.394400 & 22.400761 & 2.9490 & 4 & 2 & 4.2 \\ 
9 & 177.404192 & 22.406125 & 2.9490 & 4 & 2 & 4.3 \\ 
10& 177.392904 & 22.404014 & 0.5140 & 4 & 2 & \nodata \\
\enddata
\tablecomments{\label{tab:catalog} First entries of the redshift catalog. 
The full catalog is given in its entirety in the electronic edition.
The column ``quality'' contains the quality flag (3=secure,
4=probable).  The column ``source'' gives the original source of the
redshift: 1={\it HST}, Brammer et al. (2015, in prep.); 2=MUSE,
\citep{Gri++15}; 3={\it HST}+MUSE, 4=MUSE+Keck.  The column ``note'' lists special comments
about the object, e.g., if the object is part of a known multiple-image
system.}
\end{deluxetable}
\vspace{1cm}

\subsubsection{VLT Spectroscopy}

Integral-field spectroscopy was obtained with the MUSE instrument on
the VLT between 2015 February 14 and 2015 April 12, as part of a
Director Discretionary Time program to observe SN Refsdal (PI 
Grillo). The main goal of the program was to facilitate the
computation of an accurate model to forecast the next appearance of
the lensed SN. MUSE covers the wavelength range 480--930~nm, with an
average spectral resolution of $R \approx 3000$, over a $1' \times 1'$ 
field of view, with a pixel scale of
0\farcs2~px$^{-1}$. Details of the data acquisition and processing are
given in a separate paper \citep{Gri++15}; only a brief summary of
relevant information is given here to guide the reader.

Twelve exposures were collected in dark time under clear conditions
and with an average seeing of $\sim$1\farcs0. Bias
subtraction, flatfielding, wavelength calibration, and flux calibration 
were obtained with the MUSE Data Reduction Software version 1.0, as
described by \citet{Kar++15,Kar2015}. The different exposures were
combined into a single datacube, with a spectral sampling of
1.25~\AA~px$^{-1}$, and a resulting total integration time of 4.8~hr. 
One-dimensional (1D) spectra within circular apertures of 0\farcs6\ radius 
were extracted for all the objects visible in the coadded image along 
the spectral direction. We also searched in the datacube for faint
emission-line galaxies that were not detected in the stacked
image. Redshifts were first measured independently by two of the
coauthors (W.K. and I.B.) and later reconciled in the very few cases
with inconsistent estimates. The analysis yielded secure redshift
values for \Nzmuse\ objects, of which 15 are multiple images of 6
different background sources.

\subsubsection{Keck Spectroscopy}
\label{ssec:Keck}

Spectroscopy of the field was obtained using the DEIMOS spectrograph
\citep{Fab++03} on the 10~m Keck-II telescope on 2014 December 20.
Conditions were acceptable, with photometric transparency and $1.3''$
seeing. The 600 line mm$^{-1}$ grating was used, set to a central
wavelength of 7200~\AA, resulting in a scale of 0.65~\AA\
pixel$^{-1}$. A multi-slit mask of the field with $1''$ wide slits was
observed for $7 \times 1600$~s exposures. The SN images were the main
targets, but a slit was also placed on image 4.1, yielding a
measurement of its redshift, $z = 2.951$, independent of but fully
consistent with the results from VLT-MUSE. A full analysis of these
data and subsequent Keck spectroscopy will be discussed elsewhere.

\subsubsection{Combined Redshift Catalog}

Redshifts for \Nzboth\ objects were measured independently using both
MUSE and GLASS data. We find that the redshifts of all objects in
common agree within the uncertainties, attesting to the excellent
quality of the data. The final redshift catalog, consisting of \Nztot\
entries, is given in electronic format in Table~\ref{tab:catalog},
and will be available through the GLASS public website at URL
\url{https://archive.stsci.edu/prepds/glass/} after the acceptance of this 
manuscript. We note that owing to the high resolution of the MUSE
data, we improved the precision of the redshift of the SN Refsdal host
galaxy to $z=1.489$ \citep[cf. 1.491 previously reported
by][]{Smith++09}. Also, we revise the redshift of the multiply imaged
source 3 with the new and reliable measurement $z=3.129$ based on
unequivocal multiple line identifications ([O~II] in the grism data,
plus Lyman-$\alpha$ in the MUSE data).

\section{Summary of Lens Modeling Constraints}
\label{sec:constraints}

\subsection{Multiple Images}

The strong lensing models that are considered in this paper use as
constraints sets of multiply imaged lensed galaxies, as well as knots
in the host galaxy of SN Refsdal.  The five teams independently
evaluated known sets of multiple images
\citep{Zitrin:2009p33878,Smith++09,Johnson++14,S+J15,Die++15}, and suggested new identifications of
images across the entire field of view, based on the new HFF data.  In
evaluating the image identifications, the teams relied on their
preliminary lens models and the newly measured spectroscopic redshifts
(Section~\ref{ssec:spec}).  Each team voted on known and new systems 
on a scale of\linebreak

\LongTables
\begin{deluxetable*}{llllllllllllll}
\tablecolumns{14}
\tablewidth{0pc}
\tablecaption{Multiply Imaged Systems \label{tab:arcs}}
\tablehead{        
\colhead{ID} & 
\colhead{$\alpha$} & 
\colhead{$\delta$} & 
\colhead{Z09} &
\colhead{S09} &
\colhead{R14,} & 
\colhead{D15} & 
\colhead{Spec-$z$} & 
\colhead{ref} & 
\colhead{Spec-$z$} & 
\colhead{source} & 
\colhead{Notes} & 
\colhead{Avg.} & 
\colhead{Category} \\
\colhead{} & 
\colhead{(J2000)} & 
\colhead{(J2000)} & 
\colhead{} &
\colhead{} &
\colhead{ J14} & 
\colhead{} & 
\colhead{} & 
\colhead{} & 
\colhead{} & 
\colhead{} & 
\colhead{} & 
\colhead{Score} & 
\colhead{} 
}
\startdata
1.1& 177.39700& 22.396000& 1.2& A1.1& 1.1&1.1&1.4906&S09&1.488&3&\nodata&1.0&Gold \\
1.2& 177.39942& 22.397439& 1.3& A1.2& 1.2&1.2&1.4906&S09&1.488&3&\nodata&1.0&Gold \\
1.3& 177.40342& 22.402439& 1.1& A1.3& 1.3&1.3&1.4906&S09&1.488&3&\nodata&1.0&Gold \\
1.5& 177.39986& 22.397133& 1.4& \nodata& \nodata&1.5&\nodata&\nodata&\nodata&\nodata&1&2.0&\nodata \\
2.1& 177.40242& 22.389750& 3.3& A2.1& 2.1&2.3&1.894&S09&1.891&3&\nodata&1.0&Gold \\
2.2& 177.40604& 22.392478& 3.2& A2.2& 2.2&2.2&1.894&S09&1.891&3&\nodata&1.0&Gold \\
2.3& 177.40658& 22.392886& 3.1& A2.3& 2.3&2.1&1.894&S09&1.891&3&\nodata&1.0&Gold \\
3.1& 177.39075& 22.399847& 2.1& A3.1& 3.1&3.1&2.497&S09&3.129&3&2&1.0&Gold \\
3.2& 177.39271& 22.403081& 2.2& A3.2& 3.2&3.2&2.497&S09&3.129&3&2&1.0&Gold \\
3.3& 177.40129& 22.407189& 2.3& A3.3& 3.3&3.3&\nodata&\nodata&3.129&3&2&1.1&Gold \\
4.1& 177.39300& 22.396825& 4.1& \nodata& 4.1&4.1&\nodata&\nodata&2.949&4&\nodata&1.0&Gold \\
4.2& 177.39438& 22.400736& 4.2& \nodata& 4.2&4.2&\nodata&\nodata&2.949&2&\nodata&1.0&Gold \\
4.3& 177.40417& 22.406128& 4.3& \nodata& 4.3&4.3&\nodata&\nodata&2.949&2&\nodata&1.0&Gold \\
5.1& 177.39975& 22.393061& 5.1& \nodata& 5.1&5.1&\nodata&\nodata&2.80&1&\nodata&1.0&Gold \\
5.2& 177.40108& 22.393825& 5.2& \nodata& 5.2&5.2&\nodata&\nodata&\nodata&\nodata&\nodata&1.0&Gold \\
5.3& 177.40792& 22.403553& 5.3& \nodata& 5.3&\nodata&\nodata&\nodata&\nodata&\nodata&\nodata&1.7&Silver \\
6.1& 177.39971& 22.392544& 6.1& \nodata& 6.1&6.1&\nodata&\nodata&\nodata&\nodata&\nodata&1.1&Gold \\
6.2& 177.40183& 22.393858& 6.2& \nodata& 6.2&6.2&\nodata&\nodata&\nodata&\nodata&\nodata&1.1&Gold \\
6.3& 177.40804& 22.402506& 5.4/6.3& \nodata& 6.3&\nodata&\nodata&\nodata&\nodata&\nodata&\nodata&1.7&Silver \\
7.1& 177.39896& 22.391339& 7.1& \nodata& 7.1&\nodata&\nodata&\nodata&\nodata&\nodata&\nodata&1.1&Gold \\
7.2& 177.40342& 22.394269& 7.2& \nodata& 7.2&\nodata&\nodata&\nodata&\nodata&\nodata&\nodata&1.1&Gold \\
7.3& 177.40758& 22.401242& \nodata& \nodata& 7.3&\nodata&\nodata&\nodata&\nodata&\nodata&\nodata&1.2&Gold \\
8.1& 177.39850& 22.394350& 8.1& \nodata& 8.1&8.1&\nodata&\nodata&\nodata&\nodata&\nodata&1.2&Gold \\
8.2& 177.39979& 22.395044& 8.2& \nodata& 8.2&8.2&\nodata&\nodata&\nodata&\nodata&\nodata&1.2&Gold \\
8.4& 177.40709& 22.404722& \nodata& \nodata& \nodata&\nodata&\nodata&\nodata&\nodata&\nodata&3&1.2&Gold \\
\nodata& 177.40704& 22.405553& \nodata& \nodata& \nodata&\nodata&\nodata&\nodata&2.78&1&3&3.0&Rejected \\
\nodata& 177.40517& 22.401563& 8.3& \nodata& \nodata&\nodata&\nodata&\nodata&\nodata&\nodata&3&3.0&Rejected \\
9.1& 177.40517& 22.426233& \nodata& A6.1& 9.1&\nodata&\nodata&\nodata&\nodata&\nodata&\nodata&1.8&\nodata \\
9.2& 177.40388& 22.427231& \nodata& A6.2& 9.2&\nodata&\nodata&\nodata&\nodata&\nodata&\nodata&1.8&\nodata \\
9.3& 177.40325& 22.427228& \nodata& A6.3& 9.3&\nodata&\nodata&\nodata&\nodata&\nodata&\nodata&1.8&\nodata \\
9.4& 177.40364& 22.426422& \nodata& A6.4?& \nodata&\nodata&\nodata&\nodata&\nodata&\nodata&\nodata&1.8&\nodata \\
10.1& 177.40450& 22.425514& \nodata& A7.1& 10.1&\nodata&\nodata&\nodata&\nodata&\nodata&\nodata&1.8&\nodata \\
10.2& 177.40362& 22.425636& \nodata& A7.2& 10.2&\nodata&\nodata&\nodata&\nodata&\nodata&\nodata&1.8&\nodata \\
10.3& 177.40221& 22.426625& \nodata& A7.3& 10.3&\nodata&\nodata&\nodata&\nodata&\nodata&\nodata&1.8&\nodata \\
12.1& 177.39857& 22.389356& \nodata& \nodata& \nodata&\nodata&\nodata&\nodata&1.020&3&4&2.6&Rejected \\
12.2& 177.40375& 22.392345& \nodata& \nodata& \nodata&\nodata&\nodata&\nodata&0.929&2&4&2.9&Rejected \\
12.3& 177.40822& 22.398801& \nodata& \nodata& \nodata&\nodata&\nodata&\nodata&1.118&3&4&2.6&Rejected \\
13.1& 177.40371& 22.397786& \nodata& \nodata& 13.1&\nodata&\nodata&\nodata&1.23&1&\nodata&1.0&Gold \\
13.2& 177.40283& 22.396656& \nodata& \nodata& 13.2&\nodata&\nodata&\nodata&1.25&1&\nodata&1.0&Gold \\
13.3& 177.40004& 22.393858& \nodata& \nodata& 13.3&\nodata&\nodata&\nodata&1.23&1&\nodata&1.3&Gold \\
14.1& 177.39167& 22.403489& \nodata& \nodata& 14.1&\nodata&\nodata&\nodata&3.703&2&\nodata&1.3&Gold \\
14.2& 177.39083& 22.402647& \nodata& \nodata& 14.2&\nodata&\nodata&\nodata&3.703&2&\nodata&1.3&Gold \\
110.1& 177.40014& 22.390162& \nodata& \nodata& \nodata&\nodata&\nodata&\nodata&3.214&2&\nodata&1.0&Gold \\
110.2& 177.40402& 22.392894& \nodata& \nodata& \nodata&\nodata&\nodata&\nodata&3.214&2&\nodata&1.0&Gold \\
110.3& 177.40907& 22.400242& \nodata& \nodata& \nodata&\nodata&\nodata&\nodata&\nodata&\nodata&\nodata&2.0&\nodata \\
21.1& 177.40451& 22.386704& \nodata& \nodata& \nodata&\nodata&\nodata&\nodata&\nodata&\nodata&\nodata&1.8&Silver \\
21.2& 177.40800& 22.389057& \nodata& \nodata& \nodata&\nodata&\nodata&\nodata&\nodata&\nodata&\nodata&1.6&Silver \\
21.3& 177.40907& 22.390407& \nodata& \nodata& \nodata&\nodata&\nodata&\nodata&\nodata&\nodata&\nodata&1.6&Silver \\
22.1& 177.40370& 22.386838& \nodata& \nodata& \nodata&\nodata&\nodata&\nodata&\nodata&\nodata&\nodata&1.8&\nodata \\
22.2& 177.40791& 22.389232& \nodata& \nodata& \nodata&\nodata&\nodata&\nodata&\nodata&\nodata&\nodata&1.8&\nodata \\
22.3& 177.40902& 22.391053& \nodata& \nodata& \nodata&\nodata&\nodata&\nodata&\nodata&\nodata&\nodata&1.8&\nodata \\
23.1& 177.39302& 22.411428& \nodata& A5& \nodata&\nodata&\nodata&\nodata&\nodata&\nodata&\nodata&1.8&\nodata \\
23.2& 177.39308& 22.411455& \nodata& A5& \nodata&\nodata&\nodata&\nodata&\nodata&\nodata&\nodata&1.8&\nodata \\
23.3& 177.39315& 22.411473& \nodata& A5& \nodata&\nodata&\nodata&\nodata&\nodata&\nodata&\nodata&1.8&\nodata \\
24.1& 177.39285& 22.412872& \nodata& \nodata& \nodata&\nodata&\nodata&\nodata&\nodata&\nodata&\nodata&1.7&\nodata \\
24.2& 177.39353& 22.413071& \nodata& \nodata& \nodata&\nodata&\nodata&\nodata&\nodata&\nodata&\nodata&1.7&\nodata \\
24.3& 177.39504& 22.412697& \nodata& \nodata& \nodata&\nodata&\nodata&\nodata&\nodata&\nodata&\nodata&1.8&\nodata \\
\pagebreak
25.1& 177.40428& 22.398782& \nodata& \nodata& \nodata&\nodata&\nodata&\nodata&\nodata&\nodata&\nodata&2.0&\nodata \\
25.2& 177.40411& 22.398599& \nodata& \nodata& \nodata&\nodata&\nodata&\nodata&\nodata&\nodata&\nodata&2.0&\nodata \\
25.3& 177.39489& 22.391796& \nodata& \nodata& \nodata&\nodata&\nodata&\nodata&\nodata&\nodata&\nodata&2.3&\nodata \\
26.1& 177.41035& 22.388749& 9.1& \nodata& \nodata&\nodata&\nodata&\nodata&\nodata&\nodata&\nodata&1.8&Silver \\
26.2& 177.40922& 22.387697& 9.2& \nodata& \nodata&\nodata&\nodata&\nodata&\nodata&\nodata&\nodata&1.8&Silver \\
26.3& 177.40623& 22.385369& \nodata& \nodata& \nodata&\nodata&\nodata&\nodata&\nodata&\nodata&\nodata&1.8&Silver \\
27.1& 177.40971& 22.387665& \nodata& \nodata& \nodata&\nodata&\nodata&\nodata&\nodata&\nodata&\nodata&1.8&Silver \\
27.2& 177.40988& 22.387835& \nodata& \nodata& \nodata&\nodata&\nodata&\nodata&\nodata&\nodata&\nodata&1.8&Silver \\
27.3& 177.40615& 22.385142& \nodata& \nodata& \nodata&\nodata&\nodata&\nodata&\nodata&\nodata&\nodata&2.5&\nodata \\
28.1& 177.39531& 22.391809& \nodata& \nodata& \nodata&\nodata&\nodata&\nodata&\nodata&\nodata&\nodata&2.0&\nodata \\
28.2& 177.40215& 22.396750& \nodata& \nodata& \nodata&\nodata&\nodata&\nodata&\nodata&\nodata&\nodata&2.2&\nodata \\
28.3& 177.40562& 22.402434& \nodata& \nodata& \nodata&\nodata&\nodata&\nodata&\nodata&\nodata&\nodata&2.0&\nodata \\
200.1& 177.40875& 22.394467& \nodata& \nodata& \nodata&\nodata&\nodata&\nodata&2.32&1&\nodata&2.6&\nodata \\
200.2& 177.40512& 22.391261& \nodata& \nodata& \nodata&\nodata&\nodata&\nodata&\nodata&\nodata&\nodata&2.6&\nodata \\
200.3& 177.40256& 22.389233& \nodata& \nodata& \nodata&\nodata&\nodata&\nodata&\nodata&\nodata&\nodata&2.8&\nodata \\
201.1& 177.40048& 22.395444& \nodata& \nodata& \nodata&\nodata&\nodata&\nodata&\nodata&\nodata&5&1.6&\nodata \\
201.2& 177.40683& 22.404517& \nodata& \nodata& \nodata&\nodata&\nodata&\nodata&\nodata&\nodata&5&1.6&\nodata \\
202.1& 177.40765& 22.396789& \nodata& \nodata& \nodata&\nodata&\nodata&\nodata&\nodata&\nodata&\nodata&2.0&\nodata \\
202.2& 177.40224& 22.391489& \nodata& \nodata& \nodata&\nodata&\nodata&\nodata&\nodata&\nodata&\nodata&2.0&\nodata \\
202.3& 177.40353& 22.392586& \nodata& \nodata& \nodata&\nodata&\nodata&\nodata&\nodata&\nodata&\nodata&2.0&\nodata \\
203.1& 177.40995& 22.387244& \nodata& \nodata& \nodata&\nodata&\nodata&\nodata&\nodata&\nodata&\nodata&1.8&Silver \\
203.2& 177.40657& 22.384511& \nodata& \nodata& \nodata&\nodata&\nodata&\nodata&\nodata&\nodata&\nodata&2.0&Silver \\
203.3& 177.41123& 22.388461& \nodata& \nodata& \nodata&\nodata&\nodata&\nodata&\nodata&\nodata&\nodata&1.8&Silver \\
204.1& 177.40961& 22.386661& \nodata& \nodata& \nodata&\nodata&\nodata&\nodata&\nodata&\nodata&\nodata&1.8&Silver \\
204.2& 177.40668& 22.384322& \nodata& \nodata& \nodata&\nodata&\nodata&\nodata&\nodata&\nodata&\nodata&1.8&Silver \\
204.3& 177.41208& 22.389056& \nodata& \nodata& \nodata&\nodata&\nodata&\nodata&\nodata&\nodata&\nodata&1.8&Silver \\
205.1& 177.40520& 22.386042& \nodata& \nodata& \nodata&\nodata&\nodata&\nodata&\nodata&\nodata&\nodata&2.0&\nodata \\
205.2& 177.40821& 22.388119& \nodata& \nodata& \nodata&\nodata&\nodata&\nodata&\nodata&\nodata&\nodata&2.0&\nodata \\
205.3& 177.41038& 22.390625& \nodata& \nodata& \nodata&\nodata&\nodata&\nodata&\nodata&\nodata&\nodata&2.0&\nodata \\
206.1& 177.40764& 22.385647& \nodata& \nodata& \nodata&\nodata&\nodata&\nodata&\nodata&\nodata&\nodata&2.2&\nodata \\
206.2& 177.40863& 22.386453& \nodata& \nodata& \nodata&\nodata&\nodata&\nodata&\nodata&\nodata&\nodata&2.2&\nodata \\
206.3& 177.41133& 22.388997& \nodata& \nodata& \nodata&\nodata&\nodata&\nodata&\nodata&\nodata&\nodata&2.2&\nodata \\
207.1& 177.40442& 22.397303& \nodata& \nodata& \nodata&\nodata&\nodata&\nodata&\nodata&\nodata&\nodata&2.2&\nodata \\
207.2& 177.40397& 22.396039& \nodata& \nodata& \nodata&\nodata&\nodata&\nodata&\nodata&\nodata&\nodata&2.2&\nodata \\
208.1& 177.40453& 22.395761& \nodata& \nodata& \nodata&\nodata&\nodata&\nodata&\nodata&\nodata&\nodata&2.0&\nodata \\
208.2& 177.40494& 22.396397& \nodata& \nodata& \nodata&\nodata&\nodata&\nodata&\nodata&\nodata&\nodata&2.0&\nodata \\
209.1& 177.38994& 22.412694& \nodata& \nodata& \nodata&\nodata&\nodata&\nodata&\nodata&\nodata&\nodata&3.0&\nodata \\
209.2& 177.39055& 22.413408& \nodata& \nodata& \nodata&\nodata&\nodata&\nodata&\nodata&\nodata&\nodata&3.0&\nodata \\
210.1& 177.39690& 22.398061& \nodata& \nodata& \nodata&\nodata&\nodata&\nodata&0.702&2&\nodata&3.0&\nodata \\
210.2& 177.39505& 22.397497& \nodata& \nodata& \nodata&\nodata&\nodata&\nodata&0.702&2&\nodata&3.0&\nodata \\
\enddata
 \tablecomments{Coordinates and ID notations of multiply imaged
 families of lensed galaxies. The labels in previous publications are
 indicated for Zitrin et al. (2009; Z09), Smith et al. (2009; S09),
 Richard et al. (2014; R14), Johnson et al. (2014; J14), and Diego et
 al. (2015; D15). New identifications were made by Sharon, Oguri, and
 Hoag.  Each modeling team used a modified version or subset of the
 list above, with the coordinates of each knot varying slightly between
 modelers. The source of the new spectroscopic redshift is as in
 Table~\ref{tab:catalog}: 1={\it HST}, Brammer et al. (2015, in prep.); 2=MUSE,
 \citep{Gri++15}; 3={\it HST}+MUSE; 4=MUSE+Keck. The redshift of image 4.1
 was measured independently at Keck (\S~\ref{ssec:Keck}). The average
 score among the team is recorded; ``1'' denotes secure
 identification, ``2'' is a possible identification, and higher scores
 are considered unreliable by the teams. \\ 
$^1$See Table~\ref{tab:knots} for information on all the knots in
source 1.\\
 $^2$We revise the redshift of source 3 with the new and reliable
 measurement from MUSE (see \S~\ref{ssec:spec}).  \\
$^3$We revise the identification of a counterimage of 8.1 and 8.2,
and determine that it is at a different position compared to previous
publications. To limit confusion we label the newly identified 
counterimage 8.4.\\
 $^4$The identification of source 12 was ruled out in HFF work prior
 to the 2014 publications; we further reject this set with
 spectroscopy.\\
$^5$This image is identified as part of the same source as source 8;
the third image is buried in the light of a nearby star.  }
\end{deluxetable*}


\clearpage
\noindent 1--4, where 1 denotes secure identification, 2 is a possible
identification, and higher values are considered
unreliable. Images that had large variance in their scores were
discussed and reevaluated, and the final score was then recorded.
The list of multiple images considered in this work is given in
Table~\ref{tab:arcs}. For each system we give coordinates, the average
score, and the redshift if available. We also indicate the labels given 
to known images that were previously identified in the literature,
previously published redshifts, and references to these publications.

We define three samples of image sets (``gold,'' ``silver,'' and
``all'') based on the voting process. Following the approach of
\citet{Wan+15}, we conservatively include in our gold sample only
the systems about which every team was confident. The silver sample
includes images that were considered secure by most teams, or are
outside the MUSE field of view. The ``all'' sample includes all of the
images that were not rejected as false identifications, based on
imaging and/or spectroscopy.  In order to facilitate the comparison,
most teams produced baseline models based on the gold sample of
images, and some of the teams produced additional models based on
larger sets of images. However, owing to differences in investigators'
opinions and specifics of each code, small differences between the
constraints adopted by each team persist. They are described below for
each of the teams. The reader is referred to the publications of each
individual team for more details.

\begin{figure*}[]
\centerline{
\includegraphics[width=1\textwidth]{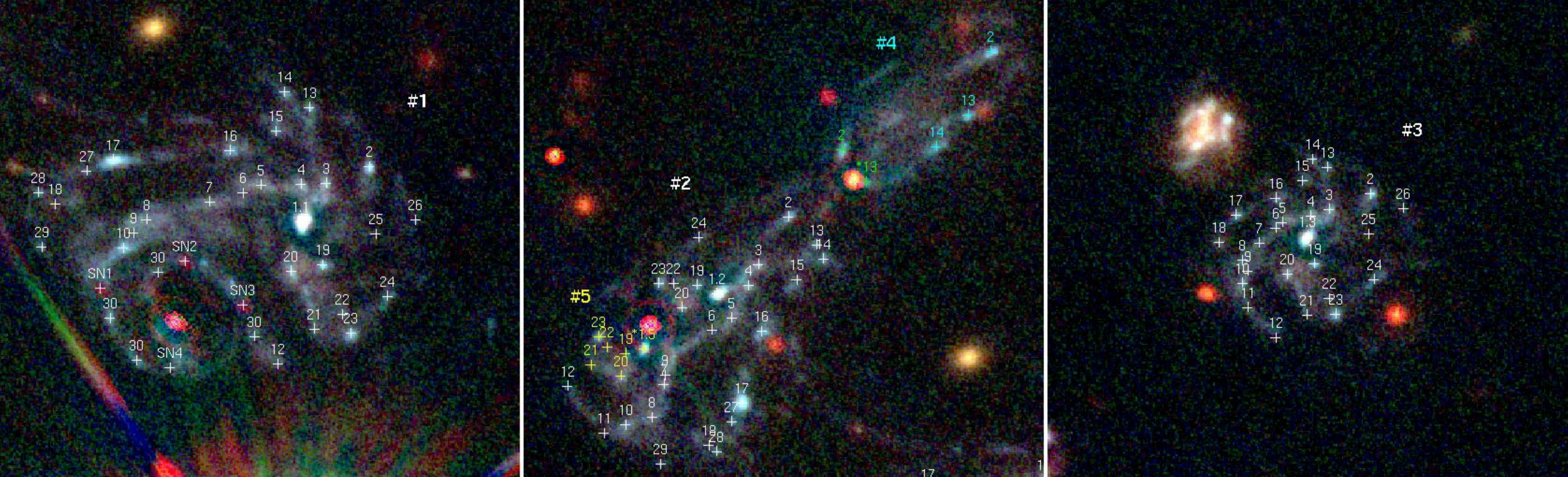}}
\caption{Knots and morphological features in the host galaxy of SN
  Refsdal at $z=1.489$. The color composite on which the regions are
  overplotted is generated by scaling and subtracting the F814W image
  from the F435W, F606W, and F105W images, in order to suppress the
  light from the foreground cluster galaxies. The left panel shows
  image 1.1, and the right panel shows image 1.3. In the middle panel,
  the complex lensing potential in the central region is responsible
  for one full image, 1.2, and additional partial images of the
  galaxy, 1.4, and 1.5 (see also Smith et al. 2009, Zitrin et
  al. 2009, and Sharon \& Johnson 2015). To guide the eye, we label
  knots that belong to 1.4 and 1.5 in cyan and yellow, respectively. A
  possible sixth image of a small region of the galaxy is labeled in
  green.  The two features marked with an asterisk in this panel,
  *1.5 and *13, are the only controversial identifications. We could
  not rule out the identification of *1.5 (knot 1.1.5 in
  Table~\ref{tab:knots}) as a counterpart of the bulge of the galaxy;
  however, it is likely only partly imaged. Image *13 (1.13.6 in
  Table~\ref{tab:knots}) is suggested by some of the models, but hard
  to confirm, and is thus not used as a constraint in the gold lens
  models considered here. We note that the exact coordinates of each
  feature may vary slightly between modelers, and we refer the reader to
  detailed publications (in preparation) by each modeling team for
  exact positions and features used.
\label{fig:knots}}
\end{figure*}

We also evaluated the identification of knots in the 
spiral galaxy hosting SN Refsdal.  Table~\ref{tab:knots} and
Figure~\ref{fig:knots} list the emission knots and features in the
host galaxy of SN Refsdal that were considered in this work.  Not
all knots were used in all models, and again, there are slight
differences between the teams as the implementation of these
constraints vary among lensing algorithms. Nevertheless, the overall
mapping of morphological features between the images of this galaxy
was in agreement between the modeling teams.

\subsection{Time Delays}

The time delay and magnification ratios between the known images were
not yet measured at the time when the models were being
finalized. Therefore, they were not used as input and they can be
considered as a valuable test of the lens model.

\subsection{Cluster Members}
\label{sssec:mem}

Cluster member galaxies were selected based on their redshifts in the
combined redshift catalog and their photometry, as follows.  In order
to account for the cluster velocity dispersion, as well as the
uncertainty in the grism-based redshifts, we define cluster membership
loosely as galaxies with spectroscopic redshift in the range
$0.520<z<0.570$, within a few thousand kilometers per second of
the fiducial cluster redshift ($z_d=0.542$). This is sufficiently
precise for the purpose of building lens models, even though not all
the cluster members are necessarily physically bound to the cluster,
from a dynamical point of view. Naturally, these cluster members still
contribute to the deflection field as the dynamically bound cluster
members. The spectroscopic cluster-member catalog comprises \Nzmem\
galaxies.

To obtain a more complete member catalog, the
spectroscopically confirmed members were supplemented by
photometrically selected galaxies. This list includes galaxies down to 
the limit (F814W $\approx$ 25 mag) of spectroscopically confirmed
members. It consists mostly of galaxies belonging to the last
two-magnitude bins of the luminosity distribution, for which the
spectroscopic sample is significantly incomplete. The missing galaxies from the spectroscopic catalog are the brightest ones that fall outside the MUSE field of view or the ones that are
contaminated in the \HST\ grism data.  The photometric analysis is
restricted to the WFC3-IR area, in order to exploit the full
multi-band photometric catalog from CLASH. The method is briefly
described by \citet{Gri2015}, and it uses a Bayesian technique to
compute the probability for a galaxy to be a member from the
distribution in color space of all spectroscopic galaxies (from 13
bands --- i.e., not including the 3 in the UV). For the photometric
selection, we started from spectroscopically confirmed members, with
redshift within $0.520< z <0.570$, and provided a catalog with only
the objects having measured F160W magnitudes.  The total catalog of
cluster members comprises
\Nzmem\ galaxies with spectroscopically determined membership, and \Npmem\
galaxies with photometrically determined membership.

\LongTables
\begin{deluxetable*}{lllllll}
\tablecolumns{7}
\tablewidth{0pc}
\tablecaption{Knots in the Host Galaxy of SN Refsdal \label{tab:knots} }
\tablehead{        
\colhead{ID} & 
\colhead{$\alpha$ (J2000)} & 
\colhead{$\delta$ (J2000)} & 
\colhead{ID Smith et al. (2009)} &
\colhead{ID Sharon et al. (2015)} & 
\colhead{ID Diego et al. (2015)} & 
\colhead{Notes} 
}
\startdata
1.1.1& 177.39702& 22.396003& 2& 1.1& 1.1.1& 1 \\
1.1.2& 177.39942& 22.397439& 2& 1.2& 1.2.1& 1 \\
1.1.3& 177.40341& 22.402444& 2& 1.3& 1.3.1& 1 \\
1.*1.5& 177.39986& 22.397133& \nodata& \nodata& 1.5.1& 1,2 \\
1.2.1& 177.39661& 22.396308& 19& 23.1& 1.1.8& \nodata \\
1.2.2& 177.39899& 22.397867& 19& 23.2& 1.2.8& \nodata \\
1.2.3& 177.40303& 22.402681& 19& 23.3& 1.3.8& \nodata \\
1.2.4& 177.39777& 22.398789& 19& 23.4& 1.4.8a& \nodata \\
1.2.6& 177.39867& 22.398242& \nodata& \nodata& 1.4.8b& \nodata \\
1.3.1& 177.39687& 22.396219& 16& 31.1& 1.1.15& \nodata \\
1.3.2& 177.39917& 22.397600& 16& 31.2& 1.2.15& \nodata \\
1.3.3& 177.40328& 22.402594& 16& 31.3& 1.3.15& \nodata \\
1.4.1& 177.39702& 22.396214& 11& 32.1& \nodata& \nodata \\
1.4.2& 177.39923& 22.397483& 11& 32.2& \nodata& \nodata \\
1.4.3& 177.40339& 22.402558& 11& 32.3& \nodata& \nodata \\
1.5.1& 177.39726& 22.396208& 18& 33.1& \nodata& \nodata \\
1.5.2& 177.39933& 22.397303& 18& 33.2& \nodata& \nodata \\
1.5.3& 177.40356& 22.402522& 18& 33.3& \nodata& \nodata \\
1.6.1& 177.39737& 22.396164& \nodata& \nodata& 1.1.13& \nodata \\
1.6.2& 177.39945& 22.397236& \nodata& \nodata& 1.2.13& \nodata \\
1.6.3& 177.40360& 22.402489& \nodata& \nodata& 1.3.13& \nodata \\
1.7.1& 177.39757& 22.396114& \nodata& 40.1& \nodata& \nodata \\
1.7.2& 177.39974& 22.396933& \nodata& 40.2& \nodata& \nodata \\
1.7.3& 177.40370& 22.402406& \nodata& 40.3& \nodata& \nodata \\
1.8.1& 177.39795& 22.396014& \nodata& \nodata& \nodata& \nodata \\
1.8.2& 177.39981& 22.396750& \nodata& \nodata& \nodata& \nodata \\
1.8.3& 177.40380& 22.402311& \nodata& \nodata& \nodata& \nodata \\
1.9.1& 177.39803& 22.395939& \nodata& \nodata& 1.1.9& \nodata \\
1.9.2& 177.39973& 22.396983& \nodata& \nodata& 1.2.9& \nodata \\
1.9.3& 177.40377& 22.402250& \nodata& \nodata& 1.3.9& \nodata \\
1.10.1& 177.39809& 22.395856& \nodata& \nodata& \nodata& \nodata \\
1.10.2& 177.39997& 22.396708& \nodata& 36.2& \nodata& \nodata \\
1.10.3& 177.40380& 22.402183& \nodata& 36.3& \nodata& \nodata \\
1.11.2& 177.40010& 22.396661& \nodata& \nodata& 1.2.3& \nodata \\
1.11.3& 177.40377& 22.402047& \nodata& \nodata& 1.3.3& \nodata \\
1.12.1& 177.39716& 22.395211& \nodata& \nodata& 1.1.14& \nodata \\
1.12.2& 177.40032& 22.396925& \nodata& \nodata& 1.2.14& \nodata \\
1.12.3& 177.40360& 22.401878& \nodata& \nodata& 1.3.14& \nodata \\
1.13.1& 177.39697& 22.396639& 7& 24.1& 1.1.19& \nodata \\
1.13.2& 177.39882& 22.397711& 7& 24.2& 1.2.19& \nodata \\
1.13.3& 177.40329& 22.402828& 7& 24.3& 1.3.19& \nodata \\
1.13.4& 177.39791& 22.398433& 7& 24.4& 1.4.19& \nodata \\
1.*13.6& 177.39852& 22.398061& \nodata& \nodata& \nodata& 3 \\
1.14.1& 177.39712& 22.396725& 6& 25.1& 1.1.7& \nodata \\
1.14.2& 177.39878& 22.397633& 6& 25.2& 1.2.7& \nodata \\
1.14.3& 177.40338& 22.402872& 6& 25.3& 1.3.7& \nodata \\
1.14.4& 177.39810& 22.398256& \nodata& 25.4& 1.4.7& \nodata \\
1.15.1& 177.39717& 22.396506& \nodata& 41.1& 1.1.20& \nodata \\
1.15.2& 177.39894& 22.397514& \nodata& 41.2& 1.2.20& \nodata \\
1.15.3& 177.40344& 22.402753& \nodata& 41.3& 1.3.20& \nodata \\
1.16.1& 177.39745& 22.396400& 4& 26.1& 1.1.6& \nodata \\
1.16.2& 177.39915& 22.397228& 4& 26.2& 1.2.6& \nodata \\
1.16.3& 177.40360& 22.402656& 4& 26.3& 1.3.6& \nodata \\
1.17.1& 177.39815& 22.396347& 3& 11.1& 1.1.5& \nodata \\
1.17.2& 177.39927& 22.396831& 3& 11.2& 1.2.5& \nodata \\
1.17.3& 177.40384& 22.402564& 3& 11.3& 1.3.5& \nodata \\
1.18.1& 177.39850& 22.396100& \nodata& \nodata& 1.1.11& \nodata \\
1.18.2& 177.39947& 22.396592& \nodata& \nodata& 1.2.11& \nodata \\
1.18.3& 177.40394& 22.402408& \nodata& \nodata& 1.3.11& \nodata \\
1.19.1& 177.39689& 22.395761& \nodata& 21.1& 1.1.17& \nodata \\
1.19.2& 177.39954& 22.397486& \nodata& 21.2& 1.2.17& \nodata \\
1.19.3& 177.40337& 22.402292& \nodata& 21.3& 1.3.17& \nodata \\
1.19.5& 177.39997& 22.397106& \nodata& 21.4& 1.5.17& \nodata \\
1.20.1& 177.39708& 22.395728& \nodata& 27.1& 1.1.16& \nodata \\
1.20.2& 177.39963& 22.397361& \nodata& \nodata& 1.2.16& \nodata \\
1.20.3& 177.40353& 22.402233& \nodata& 27.3& 1.3.16& \nodata \\
1.20.5& 177.40000& 22.396981& \nodata& 27.2& 1.5.16& \nodata \\
1.21.1& 177.39694& 22.395406& \nodata& \nodata& 1.1.18& \nodata \\
1.21.3& 177.40341& 22.402006& \nodata& \nodata& 1.3.18& \nodata \\
1.21.5& 177.40018& 22.397042& \nodata& \nodata& 1.5.18& \nodata \\
1.22.1& 177.39677& 22.395487& \nodata& \nodata& \nodata& \nodata \\
1.22.2& 177.39968& 22.397495& \nodata& \nodata& \nodata& \nodata \\
1.22.3& 177.40328& 22.402098& \nodata& \nodata& \nodata& \nodata \\
1.22.5& 177.40008& 22.397139& \nodata& \nodata& \nodata& \nodata \\
1.23.1& 177.39672& 22.395381& 15& 22.1& 1.1.2& \nodata \\
1.23.2& 177.39977& 22.397497& 15& 22.2& 1.2.2& \nodata \\
1.23.3& 177.40324& 22.402011& 15& 22.3& 1.3.2& \nodata \\
1.23.5& 177.40013& 22.397200& \nodata& 22.2& 1.5.2& \nodata \\
1.24.1& 177.39650& 22.395589& \nodata& 28.1& 1.1.4& \nodata \\
1.24.2& 177.39953& 22.397753& \nodata& 28.2& 1.2.4& \nodata \\
1.24.3& 177.40301& 22.402203& \nodata& 28.3& 1.3.4& \nodata \\
1.25.1& 177.39657& 22.395933& \nodata& \nodata& 1.1.21& \nodata \\
1.25.3& 177.40304& 22.402456& \nodata& \nodata& 1.3.21& \nodata \\
1.27.1& 177.39831& 22.396285& \nodata& 37.1& \nodata& \nodata \\
1.27.2& 177.39933& 22.396725& \nodata& 37.2& \nodata& \nodata \\
1.26.1& 177.39633& 22.396011& \nodata& \nodata& 1.1.12& \nodata \\
1.26.3& 177.40283& 22.402600& \nodata& \nodata& 1.3.12& \nodata \\
1.28.1& 177.39860& 22.396166& \nodata& 38.1& \nodata& \nodata \\
1.28.2& 177.39942& 22.396559& \nodata& 38.2& \nodata& \nodata \\
1.29.1& 177.39858& 22.395860& \nodata& 39.1& \nodata& \nodata \\
1.29.2& 177.39976& 22.396490& \nodata& 39.2& \nodata& \nodata \\
1.30.1& 177.39817& 22.395465& \nodata& 35.1& \nodata& \nodata \\
1.30.2& 177.39801& 22.395230& \nodata& 35.2& \nodata& \nodata \\
1.30.3& 177.39730& 22.395364& \nodata& 35.3& \nodata& \nodata \\
1.30.4& 177.39788& 22.395721& \nodata& 35.4& \nodata& \nodata \\
SN1& 177.39823& 22.395631& \nodata& 30.1& 1.1.3a& \nodata \\
SN2& 177.39772& 22.395783& \nodata& 30.2& 1.1.3b& \nodata \\
SN3& 177.39737& 22.395539& \nodata& 30.3& 1.1.3c& \nodata \\
SN4& 177.39781& 22.395189& \nodata& 30.4& 1.1.3d& \nodata \\
\enddata
\tablecomments{Coordinates and ID notations of emission knots in the
   multiply imaged host of SN Refsdal, at $z=1.489$. The labels in previous
   publications are indicated. New identifications were made by
   C.G., K.S., and J.D. 
 Each modeling team used a
   modified version or subset of the list above, with the coordinates of each knot
   varying slightly between modelers.  Nevertheless, there is consensus among
   the modelers on the identification and mapping of the
   different features between the multiple images of the same
   source.\\
$^1$Images 1.1, 1.2, 1.3, and 1.5 were labeled by \citet{Zitrin:2009p33878} as
1.2, 1.3, 1.1, and 1.4, respectively. The labels of other knots were not
given in that publication.\\
 $^2$This knot was identified as a counterimage of the bulge of the
 galaxy by \citet{Zitrin:2009p33878}, but rejected by \citet{Smith++09}. As in the paper by \citet{S+J15}, the modelers' consensus is that this
 knot is likely at least a partial image of the bulge.  \\
$^3$Image 1.13.6 is predicted by some models to be a counterimage of
1.13, but its identification is not sufficiently confident to be used as constraint.}
\end{deluxetable*}

\clearpage

\section{Brief Description of Modeling Techniques and their Inputs}
\label{sec:model}

For convenience to the reader, in this section we give a brief
description of each of the modeling techniques compared in this work
(summarized briefly in Table~\ref{tab:models}). We note that the five
models span a range of very different assumptions. Three of the teams
(Grillo et al., Oguri et al., Sharon et al.) used an approach based on
modeling the mass distribution with a set of physically motivated
components, described each by a small number of parameters,
representing the galaxies in the cluster and the overall cluster
halo. We refer to these models as ``simply-parametrized.'' One of the
approaches (Diego et al.)  describes the mass distribution with a
larger number of components. The components are not associated with
any specific physical object and are used as building blocks, allowing
for significant flexibility, balanced by regularization. We refer to
this model as ``free-form''\footnote{These models are sometimes
described incorrectly as ``nonparametric,'' even though they
typically have more parameters than the so-called parametric models.}.
The fifth approach (Zitrin et al.) is based on the assumption that
light approximately traces mass, and the mass components are built by
smoothing and rescaling the observed surface brightness of the cluster
members. We refer to this approach as ``light-traces-mass.''  All of 
the models considered here are single-plane lens models. As we will
discuss in Section~\ref{sec:discussion}, each type of model uses a
different approach to account for the effects of structure along the
line of sight, and to break the mass-sheet degeneracy. All model
outputs will be made available through the HFF website after the
acceptance of the individual modeling papers.

\begin{deluxetable}{lllll} \tablecolumns{5}
\tablewidth{0pt} 
\tablecaption{Summary of models}
\tablehead{
\colhead{Short name}  & \colhead{Team} & \colhead{Type}     & \colhead{RMS}& \colhead{Images}}
Die-a              & Diego et al.   & Free-form             & 0.78         & gold+sil                         \\
Gri-g              & Grillo et al.  & Simply-param          & 0.26         & gold                        \\
Ogu-g              & Oguri  et al.  & Simply-param          & 0.43         & gold                        \\
Ogu-a              & Oguri  et al.  & Simply-param          & 0.31         & all                         \\
Sha-g              & Sharon  et al. & Simply-param          & 0.16         & gold                        \\
Sha-a              & Sharon  et al. & Simply-param          & 0.19         & gold+sil                         \\
Zit-g              & Zitrin  et al. & Light-tr-mass         & 1.3          & gold                        \\
\enddata
\tablecomments{\label{tab:models} For each model we provide a short name as well as basic features and inputs. The column RMS lists the root-mean-square scatter of the observed vs. predicted image positions in arcseconds.
}
\end{deluxetable}
\vspace{1cm}

We note that members of our team have developed another complementary
``free-form'' approach, based on modeling the potential in pixels on
an adaptive grid \citep{Bra++04,Bra++09}. However, given the
pixelated nature of the reconstruction and the need to compute
numerical derivatives and interpolate from noisy pixels in order to
compute time delays and magnifications at the location of SN Refsdal, 
we did not expect this method to be competitive for this specific
application. Therefore, in the interest of time we did not construct
this model. A pre-HFF model of \M\ using this approach is available
through the HFF website and will be updated in the future.

When appropriate, we also describe additional sets of constraints used
by each modeler.

\subsection{Diego et al.}
A full description of the modeling technique used by this team (J.D.,
T.B.) and the various improvements implemented in the code can be
found in the literature
\citep{Diego2005,Diego2007,Sendra2014,Die++15}. Here is a brief
summary of the basic steps.

\subsubsection{Definition of the Mass Model} 

The algorithm (WSLAP+) relies on a division of the mass distribution
in the lens plane into two components. The first is compact and
associated with the member galaxies (mostly red ellipticals). The
second is diffuse and distributed as a superposition of Gaussians on a
regular (or adaptive) grid. In this specific case, a grid of
$512\times512$ pixels $0\farcs1875$ on a side was used. For the
compact component, the mass associated with the galaxies is assumed to
be proportional to their luminosity. If all the galaxies are assumed
to have the same mass-to-light ($M/L$) ratio, the compact component
(galaxies) contributes with just one ($N_g=1$) extra free parameter
which corresponds to the correction that needs to be applied to the
fiducial $M/L$ ratio. In a few particular cases, some galaxies (like the
brightest cluster galaxy [BCG] or massive galaxies very close to an 
arclet) are allowed to have their own $M/L$ ratio, adding additional 
free parameters to the lens model but typically no more than a few 
($N_g \approx$ O(1)). For this component associated with the galaxies, 
the total mass is assumed to follow either a 
\citet[][hereafter NFW]{NFW97} profile (with a fixed
concentration, and scale radius scaling with the fiducial halo mass)
or be proportional to the observed surface brightness. For this work
the team adopted $N_g = 2$ or $N_g=3$. The case $N_g=2$ considers one
central BCG and the elliptical galaxy near
image 1.2 to have the same $M/L$ ratio, while the remaining galaxies
have a different one. In the case $N_g=3$, the BCG and the galaxy near
image 1.2 each have their own $M/L$ ratio, and the remaining galaxies
are assumed to have a third independent value. In all cases, it is
important to emphasize that the member galaxy between the 4 observed
images of SN Refsdal was not allowed to have its own independent $M/L$
ratio. This results in a model that is not as accurate on the smallest
scales around this galaxy as other models that allow this galaxy to
vary.

The diffuse or ``soft'' component is described by as many free
parameters as grid (or cell) points. This number ($N_c$) varies but is
typically between a few hundred to one thousand ($N_c \approx$
O(100)-O(1000)) depending on the resolution and/or use of the adaptive
grid. In addition to the free parameters describing the lens model,
the problem includes as unknowns the original positions of the lensed
galaxies in the source plane. For the clusters included in the HFF
program the number of background sources, $N_s$, is typically a few
tens ($N_s \approx$ O(10)), each contributing with two unknowns ($\beta_x$
and $\beta_y$). All of the unknowns are then combined into a single array
$X$ with $N_x$ elements ($N_x \approx O(1000)$). 

\subsubsection{Definition of the Inputs} 

The inputs are the pixel position of the strongly lensed galaxies (not
just the centroids) for all the multiple images listed in
Tables~\ref{tab:arcs} and \ref{tab:knots}. In the case of elongated
arcs near the critical curves with no features, the entire arc is
mapped and included as a constraint. If the arclets have individual
features, these can be incorporated as semi-independent constraints
but with the added condition that they need to form the same source in
the source plane. The following inputs are added to the default set of
image and knot centers listed in Section~\ref{sec:constraints}:

\begin{enumerate}

\item Shape of the arclets. This is particularly useful for long
elongated arcs (with no counterimages) which lie in the regime between
the weak and strong lensing. These arcs are still useful constraints
that add valuable information beyond the Einstein radius.

\item Shape and morphology of arcs. By including this information one can
account (at least partially) for the magnification at a given
position.

\item Resolved features in individual systems. This new addition to the
code is motivated by the host galaxy of SN Refsdal, where multiple
features can be identified in the different counter images. In
addition, the counterimage in the North, when relensed, offers a
robust picture of the original source morphology (size, shape,
orientation). This information acts as an anchor, constraining the
range of possible solutions.

\end{enumerate}

Weak lensing shear measurements can also be used as input to the
inference. For the particular case of \M\ the weak lensing
measurements are not used, to ensure homogeneity with the other
methods.

\subsubsection{Description of the Inference Process and Error Estimation} 

The array of best-fit parameters, $X$, is obtained after solving the
system of linear equations
\begin{equation}
\Theta = \Gamma X,
\label{eq_WSLAP}
\end{equation}
where the $N_o$ observations (strong lensing, weak lensing, time
delays) are included in the array $\Theta$, and the matrix $\Gamma$ is
known and has dimension $N_o \times (N_c+N_g+2N_s)$.

In practice, $X$ is obtained by solving the set of linear equations
described in Eq.~\ref{eq_WSLAP} via a fast biconjugate algorithm, or
inverted with a singular value decomposition (after setting a
threshold for the eigenvalues) or solved with a more robust but slower
quadratic algorithm. The quadratic algorithm is the preferred method,
as it imposes the physical constraint that the solution $X$ must be
positive. This eliminates unphysical solutions with negative masses
and reduces the space of possible solutions. Like in the case of the
biconjugate gradient, the quadratic programming algorithm solves the
system of linear equations by finding the minimum of the associated
quadratic function. Errors in the solution are derived by minimizing
the quadratic function multiple times, after varying the initial
conditions of the minimization process, and/or varying the grid
configuration.

\subsection{Grillo et al.}

The software used by this team (C.G., S.H.S., A.H., P.R., W.K., I.B.,
A.M., G.B.C.) is {\sc Glee} (\citealt{S+H10,Suy++12}). The strong
lensing analysis performed here follows very closely the one presented
by \citet{Gri2015} for another HFF target,
MACSJ0416.1$-$2403. Cosmological applications of {\sc Glee} can be
found in the papers by \citet{Suy++13,Suy++14}, and further details on
the strong lensing modeling of \M\ are provided in a dedicated paper
\citep{Gri++15}.

\subsubsection{Definition of the Mass Model}

Different mass models have been explored for this galaxy cluster, but
only the best-fitting one is discussed here. The projected
dimensionless total surface mass density of 300 cluster members within
the WFC3 field of view of the CLASH observations is modeled as a dual
pseudoisothermal elliptical mass distribution (dPIE;
\citealt{Eli2007}) with vanishing ellipticity and core radius. The
zero-core dPIE profile corresponds to the three-dimensional mass
density profile:
\begin{equation}
\rho(r) \propto \frac{1}{r^2(r^2+r^2_{t})}.
\label{eq:grillo}
\end{equation}

\noindent
The galaxy luminosity values in the F160W band are used to assign the
relative weights to their total mass profile. The galaxy total $M/L$
ratio is scaled with luminosity as $M_{\rm T}/L \propto
L^{0.2}$, thus mimicking the so-called ``tilt'' of the Fundamental
Plane. The values of axis ratio, position angle, effective velocity
dispersion, and truncation radius of the two cluster members closest in
projection to the central and southern images of the SN Refsdal host
are left free. To complete the total mass distribution of the galaxy
cluster, three additional mass components are added to describe the
cluster dark matter halo on physical scales larger than those typical
of the individual cluster members. These cluster halo components are
parametrized as two-dimensional pseudo-isothermal elliptical mass
profiles (PIEMD as defined by \citealt{Kas1993}; see also
\citealt{Gri++15}).

No external shear or higher-order perturbations are included in the
model. The number of free parameters associated with the model of the
cluster total mass distribution is 28.

\subsubsection{Definition of the Inputs}

The positions of the multiple images belonging to the 10 systems of
the gold sample and to 18 knots of the SN Refsdal host are the
observables over which the values of the model parameters are
optimized. The adopted positional uncertainty of each image is
0\farcs065. The redshift values of the 7 spectroscopically confirmed
gold systems are fixed, while the remaining 3 systems are included
with a uniform prior on the value of $D_{ds}/D_{s}$, where $D_{ds}$
and $D_{s}$ are the deflector-source and observer-source angular
diameter distances, respectively. In total, 88 observed image
positions are used to reconstruct the cluster total mass distribution.

\subsubsection{Description of the Inference Process and Error Estimation}

The best-fitting, minimum-$\chi^{2}$ model is obtained by minimizing
the distance between the observed and model-predicted positions of the
multiple images in the lens plane. A minimum $\chi^{2}$ value of 1441,
corresponding to an RMS offset between the image observed and
reconstructed positions of 0\farcs26, is found. To sample the
posterior probability distribution function of the model parameters,
the image positional uncertainty is increased until the value of the
$\chi^{2}$ is comparable to the number of the degrees of freedom (89),
and standard Markov chain Monte Carlo (MCMC) methods are used. The
quantities shown in Figures~\ref{fig:tdcross} to~\ref{fig:mulong} are
for the model-predicted images of SN Refsdal and are obtained from 100
different models extracted from an MCMC chain with $10^{6}$ samples
and an acceptance rate of approximately 0.13.

\subsection{Oguri et al.}

\subsubsection{Definition of the Mass Model}
This team (M.O., M.I., R.K.) uses the public software {\sc glafic}
\citep{Ogu2010}. This ``simply-parametrized'' method assumes that the lens
potential consists of a small number of components describing dark
halos, cluster member galaxies, and perturbations in the lens
potential. The dark halo components are assumed to follow the
elliptical NFW mass density profile. In contrast, the elliptical
pseudo-Jaffe profile is adopted to describe the mass distribution of
cluster member galaxies. In order to reduce the number of free
parameters, the velocity dispersion $\sigma$ and the truncation radius
$r_{\rm cut}$ for each galaxy are assumed to scale with the
($F814W$-band) luminosity of the galaxy as $\sigma\propto L^{1/4}$ and
$r_{\rm cut}\propto L^\eta$, with $\eta$ being a free parameter. In
addition, the second-order (external shear) and third-order perturbations
are included so as to account for asymmetry of the overall lens
potential. Interested readers are referred to
\citet{Ogu2010,Ogu2015},
\citet{Ogu2012,Ogu2013}, and
\citet{Ishigaki2014} for more detailed descriptions and examples of
cluster mass modeling with {\sc glafic}. Additional details are given
in a dedicated paper \citep{Kaw++15}.

\subsubsection{Definition of the inputs}

The positions of multiple images and knots listed in
Section~\ref{sec:constraints} are used as constraints.  Image 1.5 was
not used as a constraint. To accurately recover the position of
SN Refsdal, different positional uncertainties are assumed for different
multiple images. Specifically, while the positional uncertainty of
$0\farcs4$ in the image plane is assumed for most of the multiple images,
smaller positional uncertainties of $0\farcs05$ and $0\farcs2$ are
assumed for SN Refsdal and knots of the SN host galaxy, respectively
\citep[see also][]{Ogu2015}. When spectroscopic redshifts are
available, their redshifts are fixed to the spectroscopic
redshifts. Otherwise source redshifts are treated as model parameters
and are optimized simultaneously with the other model parameters. For
a subsample of multiple image systems for which photometric redshift
estimates are secure and accurate, a conservative Gaussian prior with
a dispersion of $\sigma_z=0.5$ for the source redshift is
added. While {\sc glafic} allows one to include other types of
observational constraints, such as flux ratios, time delays, and weak
lensing shear measurements, those constraints are not used in the mass
modeling of \M.

\subsubsection{Description of the Inference Process and Error Estimation}

The best-fit model is obtained simply by minimizing $\chi^2$.  The
so-called source plane $\chi^2$ minimization is used for an efficient
model optimization \citep[see Appendix 2 of][]{Ogu2010}. A standard
MCMC approach is used to estimate errors on model
parameters and their covariance.

The predicted time delays and magnifications are computed at the
model-predicted positions.  For each mass model (chain), the best-fit
source position of the SN is derived. From that, the corresponding SN
image positions in the image plane (which can be slightly different
from observed SN positions) are obtained for that model, and finally
the time delays and magnifications of the images are calculated.

\subsection{Sharon et al.}

The approach of this team (K.S., T.J.) was based on the publicly
available software Lenstool \citep{Jul++07}. Lenstool is a
``simply-parametrized'' lens modeling code. In practice, the code assumes
that the mass distribution of the lens can be described by a
combination of mass halos, each of them taking a functional form whose
properties are defined by a set of parameters. The method assumes that
mass generally follows light, and assigns halos to individual galaxies
that are identified as cluster members. Cluster- or group-scale halos
represent the cluster mass components that are not directly related to
galaxies. The number of cluster or group-scale halos is determined by
the modeler. Typically, the positions of the cluster-scale halos are
not fixed and are left to be determined by the modeling algorithms. A
hybrid ``simply-parametrized''/``free-form'' approach has also been
implemented in Lenstool \citep{J+K09}, where numerous halos are placed
on a grid, representing the overall cluster component. This hybrid
method is not implemented in this work.

\subsubsection{Definition of the Mass Model} 

The halos are represented by elliptical mass distributions
corresponding to a spherical density profile $\rho(r)$ described by the equation
\begin{equation}
\rho(r) = \frac{\rho_0}{(1+r^2/r^2_{\rm core})(1+r^2/r^2_{\rm cut})}.
\label{eq:sharon}
\end{equation}
\noindent
These halos are isothermal at intermediate radii, i.e., $\rho \propto
r^{-2}$ at $r_{\rm core}\lesssim r \lesssim r_{\rm cut}$, and they have a 
flat core internal to $r_{\rm core}$. The profile is equivalent to that 
given in Eq.~\ref{eq:grillo} for $r_{\rm core}=0$.  It is sometimes known 
as dPIE or ``truncated PIEMD'' \citep{Eli2007}, although it differs from the
original PIEMD profiled defined by \citet{Kas1993}. The transition
between the different slopes is smooth. The quantity $\sigma_0$ defines the 
overall normalization as a fiducial velocity dispersion. In Lenstool, each 
of these halos has seven free parameters: centroid position ($x$,$y$);
ellipticity $e=(a^2-b^2)/(a^2+b^2)$ where $a$ and $b$ are the semimajor 
and semiminor axes, respectively; position angle $\theta$; and
$r_{\rm core}$, $r_{\rm cut}$, and $\sigma_0$ as defined above.

The selection of cluster member galaxies is described in
Section~\ref{sssec:mem}. In this model, 286 galaxies were selected from
the cluster member catalog, by a combination of their luminosity 
and projected distance from the cluster center, such that the
deflection caused by an omitted galaxy is much smaller than the
typical uncertainty caused by unseen structure along the line of
sight. This selection criterion results in removal of faint galaxies
at the outskirts of the cluster, and inclusion of all the galaxies
that pass the cluster-member selection in the core.

Cluster member galaxies are also modeled with the profile given by
Eq.~\ref{eq:sharon}. Their positional parameters are fixed on
their observed properties as measured with SExtractor \citep{B+A96}
for $x$, $y$, $e$, and $\theta$. The other parameters --- $r_{\rm core}$,
$r_{\rm cut}$, and $\sigma_0$ --- are linked to their luminosity in the
$F814W$ band through scaling relations
\citep[e.g.,][]{LKN05} assuming a constant $M/L$ ratio
for all galaxies,
\begin{equation}
\sigma _0 = \sigma_0^*\Big(\frac{L}{L^*}\Big)^{1/4} {\rm  ~~~~ and ~~~~}
r_{\rm cut} = r_{\rm cut}^*\Big(\frac{L}{L^*}\Big)^{1/2}.
\end{equation}
\subsubsection{Definition of the Inputs}
The lensing constraints are the positions of multiple images of each
lensed source, plus those of the knots in the host galaxy of SN
Refsdal, as listed in Section~\ref{sec:constraints}. In cases where
the lensed image is extended or has substructure, the exact positions
were selected to match similar features within multiple images of the
same galaxy with each other, thus obtaining more constraints, a better
local sampling of the lensing potential, and a better handle on the
local magnification.  Where available, spectroscopic redshifts are
used as fixed constraints.  For sources with no spectroscopic
redshift, the redshifts are considered as free parameters with
photometric redshifts informing their Bayesian priors. The
uncertainties of the photometric redshifts are relaxed in order to
allow for outliers (to an interval of approximately $\delta z=\pm2$
around the photo-$z$). We present two models here: Sha-g uses as
constraints the gold sample of multiply imaged galaxies, and Sha-a
uses gold, silver, and secure arcs outside the MUSE field of view,
to allow better coverage of lensing evidence in the outskirts of the
cluster and in particular to constrain the subhalos around \M.

\subsubsection{Description of the Inference Process and Error Estimation} 

The parameters of each halo are allowed to vary under Bayesian priors,
and the parameter space is explored in an MCMC process to identify the
set of parameters that provide the best fit. The quality of the lens
model is measured either in the source plane or in the image
plane. The latter requires significantly longer computation time.  In
source-plane minimization, the source positions of all the images of
each set are computed, by ray tracing the image-plane positions
through the lens model to the source plane. The best-fit model is the
one that results in the smallest scatter in the source positions of
multiple images of the same source.  In image-plane minimization, the
model-predicted counterimages of each of the multiple images of the
same source are computed. This results in a set of predicted images
near the observed positions. The best-fit model is the one that
minimizes the scatter among these image-plane positions.
The MCMC sampling of the parameter space is used to estimate the
statistical uncertainties that are inherent to the modeling
algorithm. 
In order to estimate the uncertainties on the magnification and time
delay, potential maps are generated from sets of
parameters from the MCMC chain that represent 1$\sigma$ in the
parameter space.

\subsection{Zitrin et al.}

\subsubsection{Definition of the Mass Model} 

The method used by this team (A.Z.) is a Light Traces Mass (LTM)
method, so that both the galaxies $\emph{and}$ the dark matter follow
the light distribution. The method is described in detail by
\citet{Zitrin2009a,Zitrin2013}, and it is inspired by the LTM assumptions
outlined by \citet{Broadhurst2005}. The model consists of two main
components. The first component is a mass map of the cluster galaxies,
chosen by following the red sequence. Each galaxy is represented with
a power-law surface mass density distribution, where the surface
density is proportional to its surface brightness. The power law is a
free parameter of the model and is iterated (all galaxies are
forced to have the same exponent). The second component is a smooth
dark matter map, obtained by smoothing (with a spline polynomial or
with a Gaussian kernel) the first component (i.e., the superposed red
sequence galaxy mass distribution). The smoothing degree is the second
free parameter of the model. The two components are then added with a
relative weight which is a free parameter, along with the overall
normalization. 

Next, a two-component external shear can be included to add
flexibility and generate ellipticity in the magnification map. Lastly,
individual galaxies can be assigned with free masses to be optimized
by the minimization procedure, to allow more degrees of freedom
deviating from the initial imposed LTM.  This procedure has been shown
to be very effective in locating multiple images in many clusters
\citep[e.g.,][]{Zitrin2009a,Zitrin2012b,Zitrin2013,Zitrin2015}, even
without any multiple images initially used as input
\citep{Zitrin2012a}. Most of the multiple images in \M\ that were
found by \citet{Zitrin2009b} and \citet{2012Natur.489..406Z} were
identified with this method.

\subsubsection{Definition of the Inputs} 

All sets of multiple images in the gold list were used except system
14. Most knots were used except those in the fifth radial BCG image.
All systems listed with spec-$z$ (aside for system 5) were kept fixed at
that redshift, while all other gold systems were left to be freely
optimized with a uniform flat prior.  Image position uncertainties
were adopted to be $0\farcs5$, aside for the four SN images for which
$0\farcs15$ was used.

\subsubsection{Description of the Inference Process and Error Estimation} 

The best-fit solution and uncertainties are obtained via converged MCMC
chains. 

\section{Comparison of Lens Models}
\label{sec:compare}

In this section we carry out a comparison of the 7 models, focusing
specifically on the quantities that are relevant for SN Refsdal. We
start in Section~\ref{ssec:map} by presenting the two-dimensional maps
of convergence, magnification, and time delay, for a deflector at the
redshift of the cluster and a source at the redshift of SN Refsdal
($z=1.489$; we note that assuming $z=1.491$, the redshift published by
\citet{Smith++09}, would not have made any significant
difference). Then, in Section~\ref{ssec:cross}, we compare
quantitatively the predicted time delays and magnification ratios of
the known images with their measured values.  Finally, in
Section~\ref{ssec:long} we present the forecast for the future (and
past) SN images. All of the lens models predict the appearance of an image
of the SN in the two other images of the host galaxy. In the following
sections, we refer to the predicted SN in image 1.2 of the host galaxy
as SX, and the one in image 1.3 of the host as SY, following the
labeling of previous publications. The predicted time delays and
magnification ratios are given in Table~\ref{tab:predictions}.

\subsection{Convergence, Magnification, and Time-Delay Maps}
\label{ssec:map}

\begin{figure*}
\begin{center}
\includegraphics[width=0.31\textwidth]{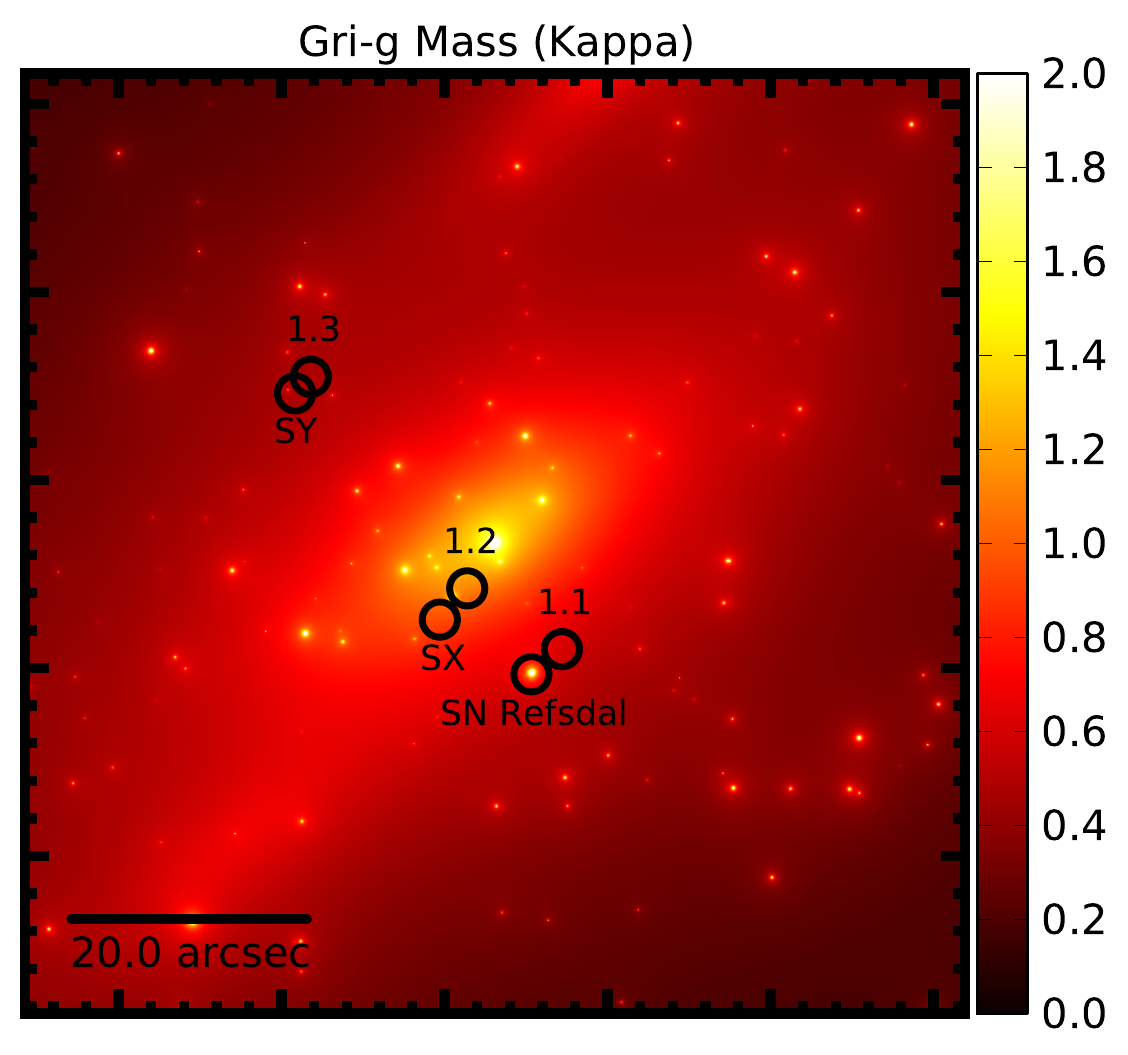}
\includegraphics[width=0.31\textwidth]{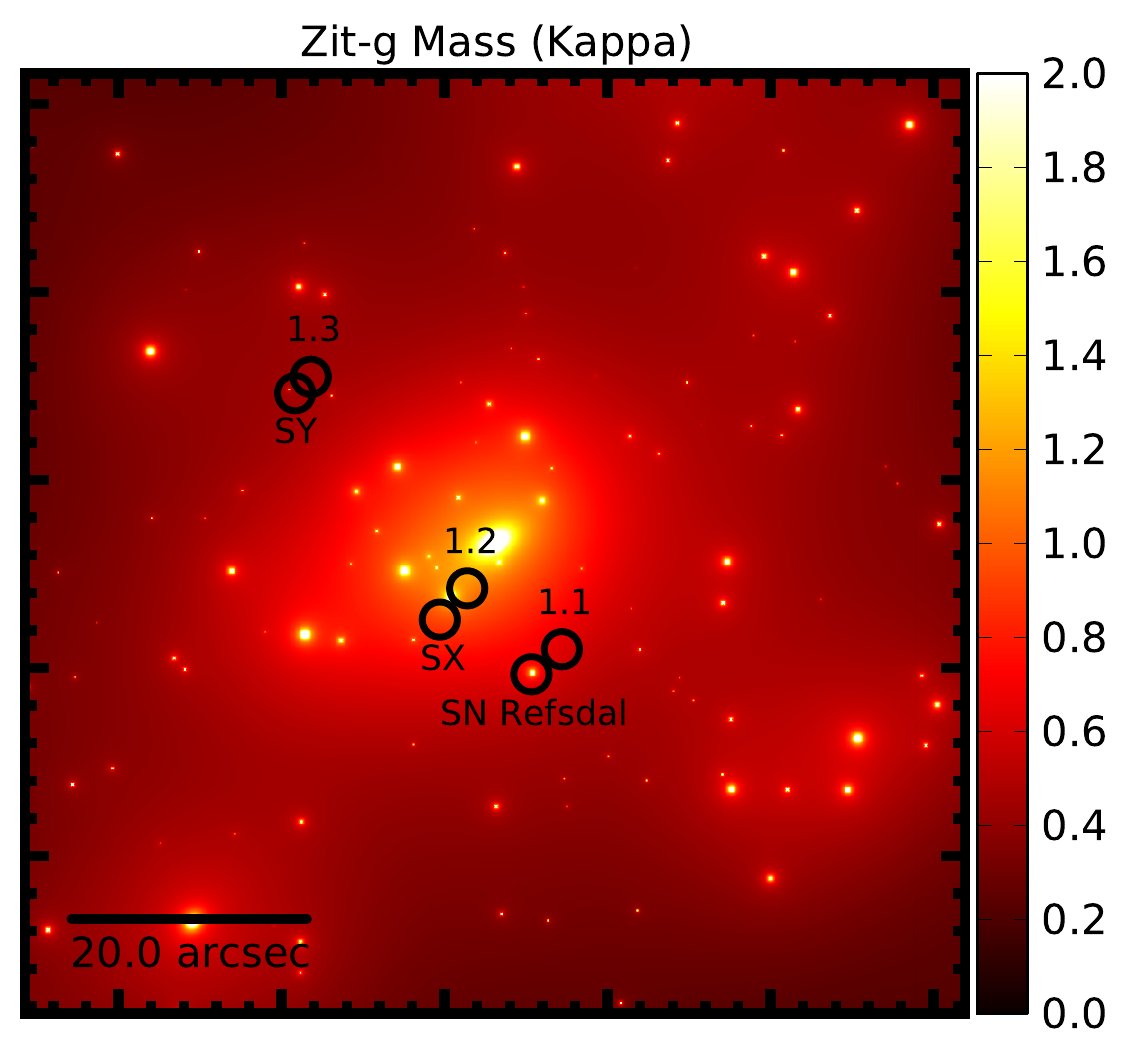}\\ 
\includegraphics[width=0.31\textwidth]{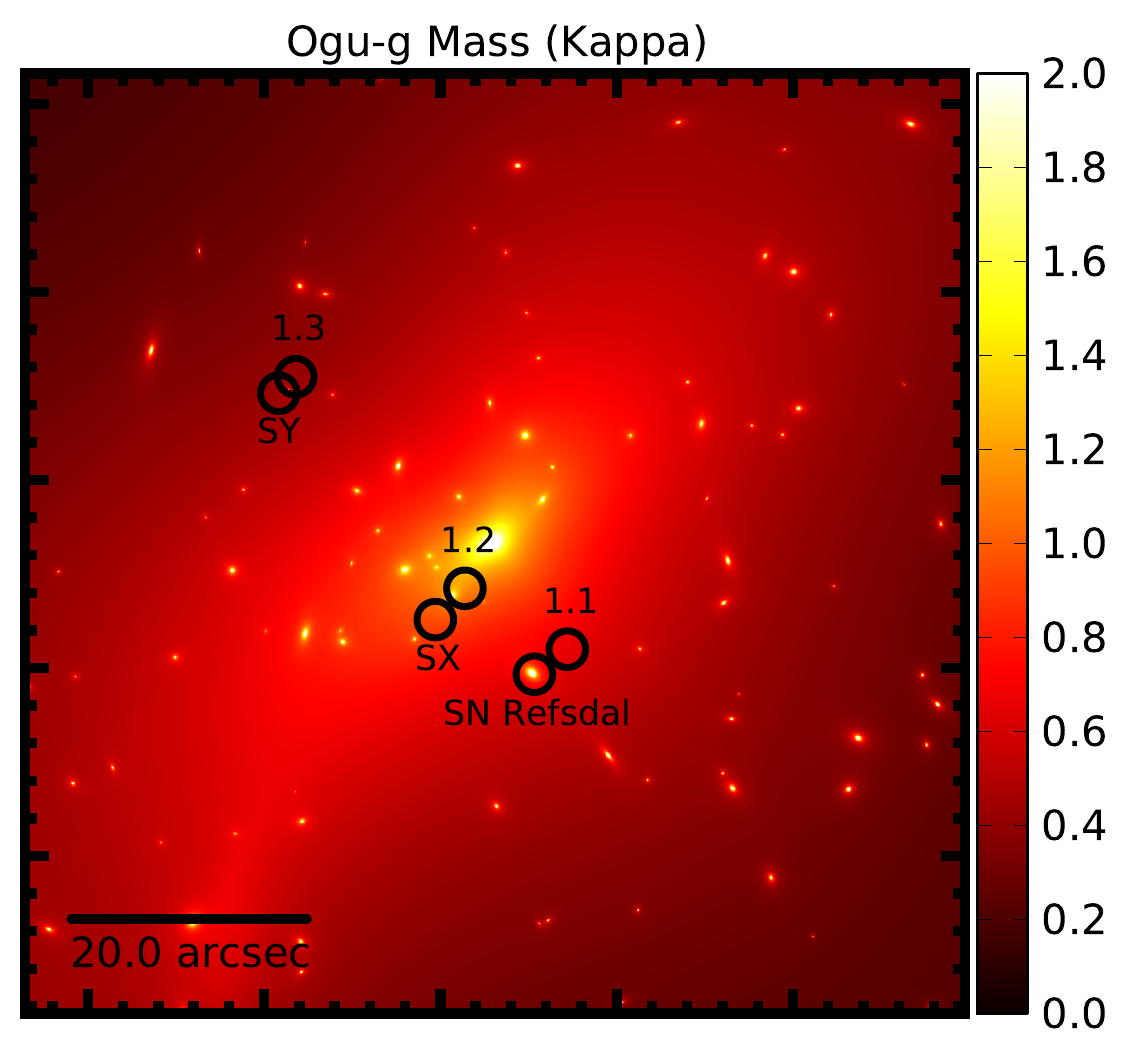}
\includegraphics[width=0.31\textwidth]{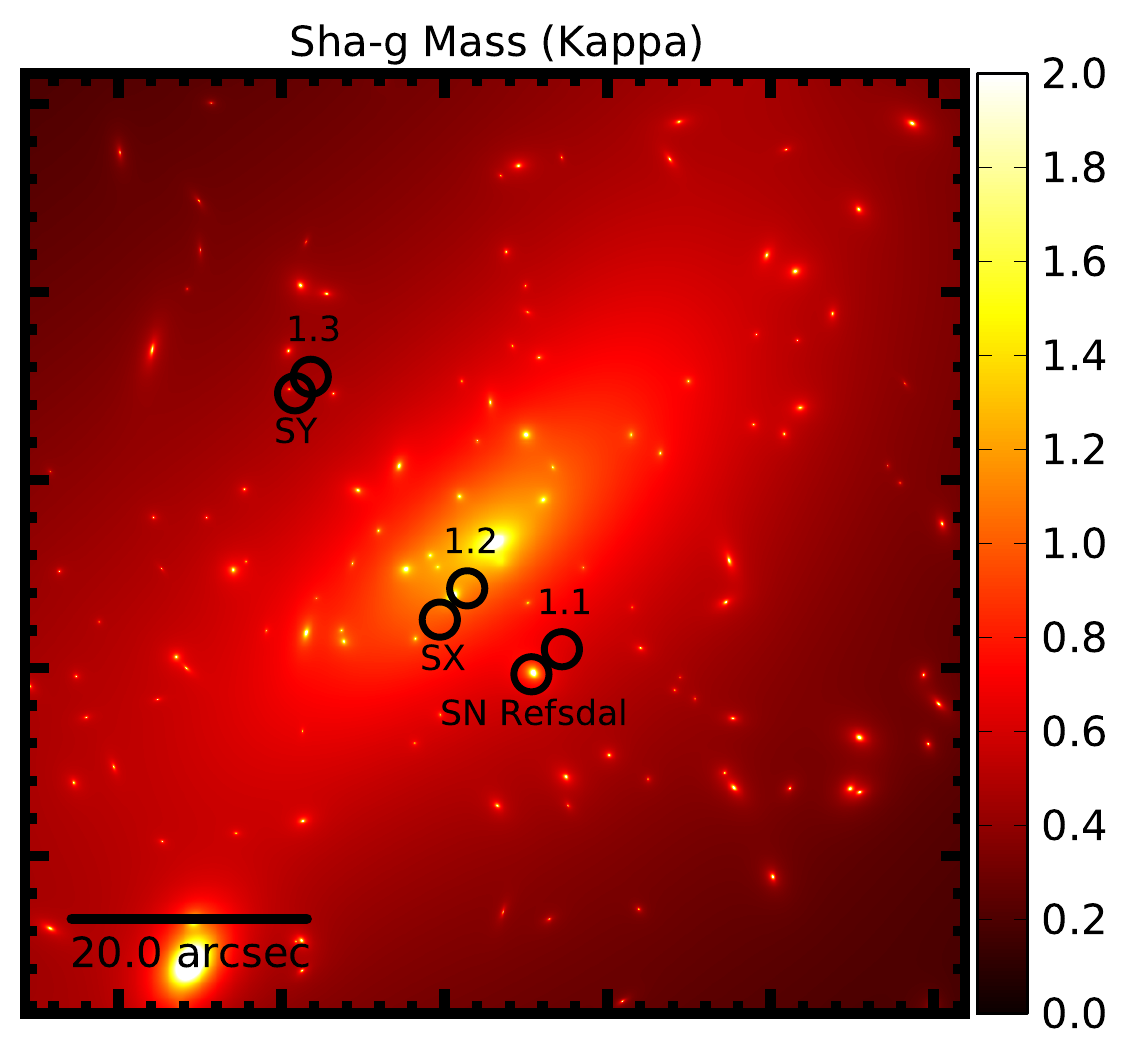}\\
\includegraphics[width=0.31\textwidth]{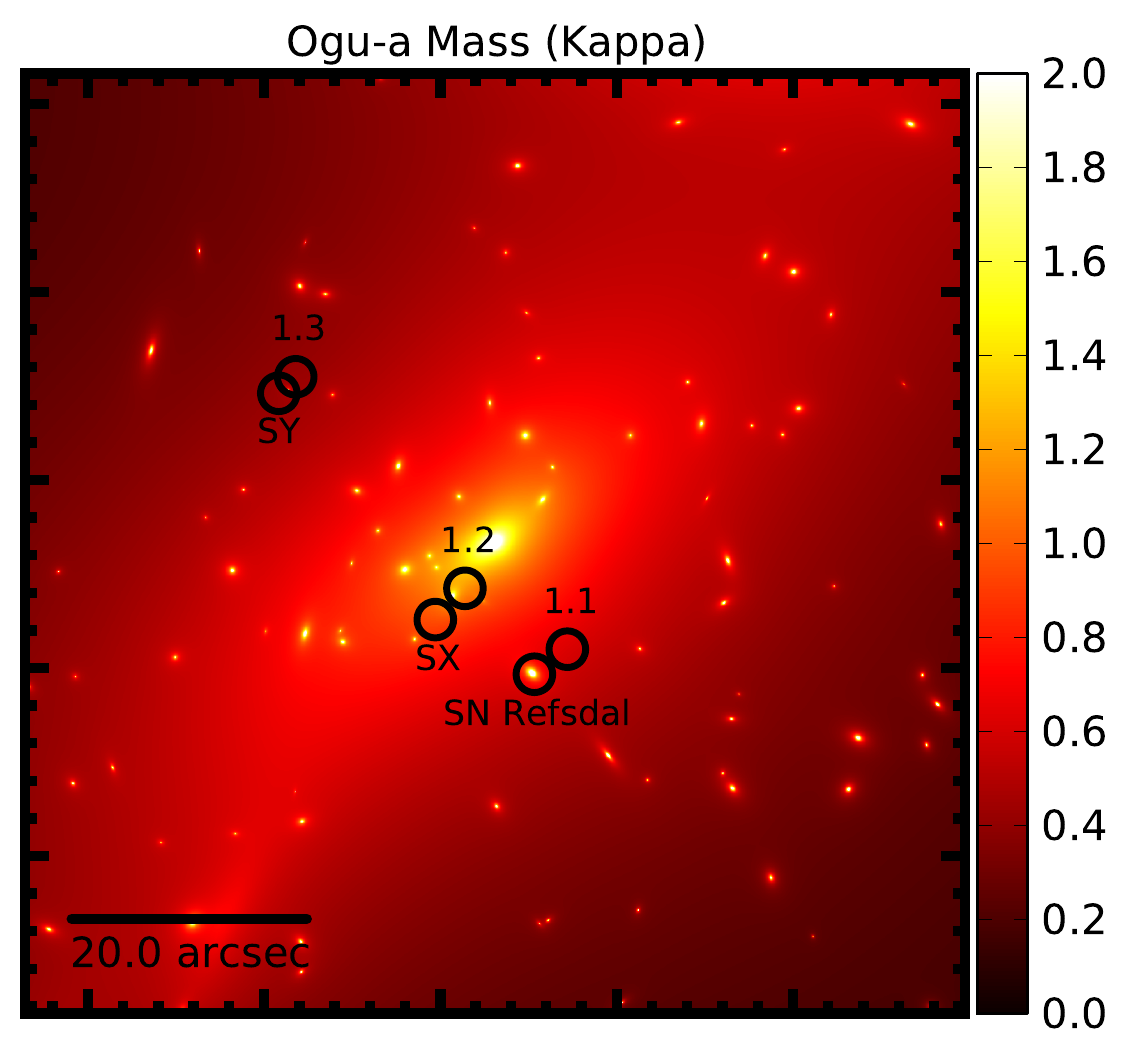}   
\includegraphics[width=0.31\textwidth]{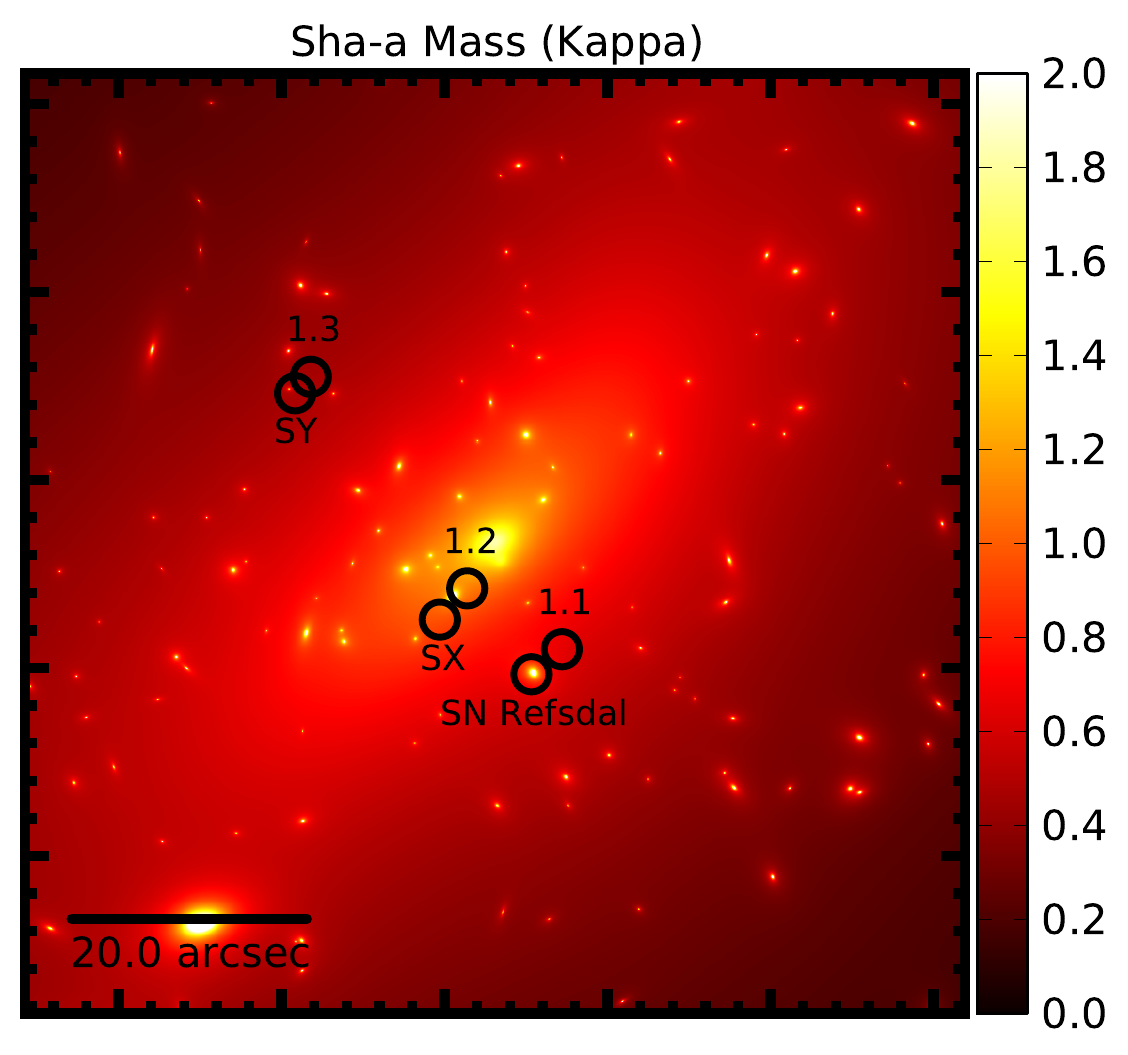}
\includegraphics[width=0.31\textwidth]{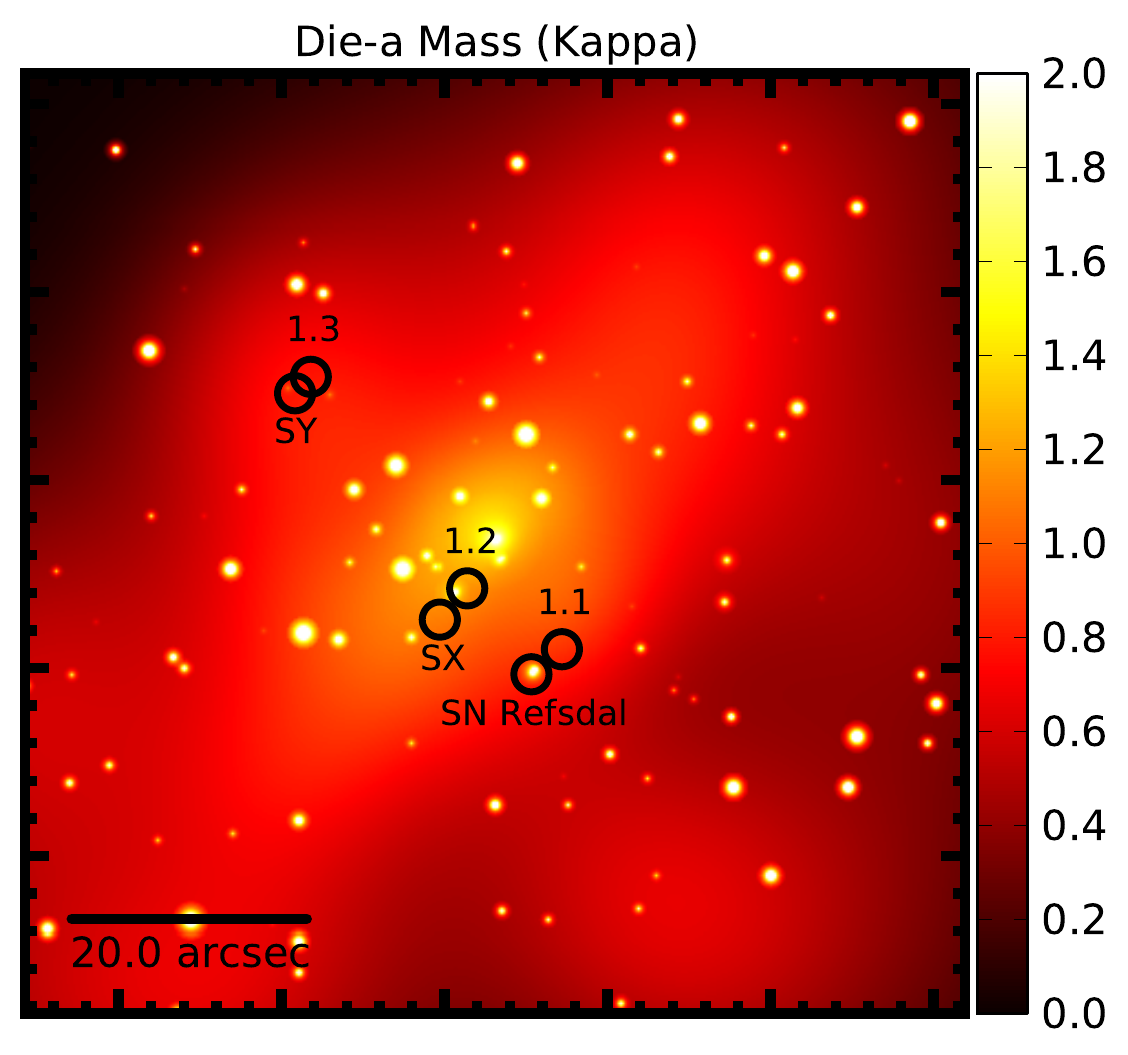}  
\caption{Comparing the mass distributions for the models, labeled as in Table~\ref{tab:models}. Convergence is computed relative to the critical density with the deflector at the redshift of the cluster and the source at the redshift of the SN. The circles identify the positions of the observed and predicted images of SN Refsdal and those of the multiple images of its host galaxy.
The top four panels are models including only the gold sample of images as constraints.}
\label{fig:compareMASS}
\end{center}
\end{figure*}

\begin{figure*}
\begin{center}
\includegraphics[width=0.31\textwidth]{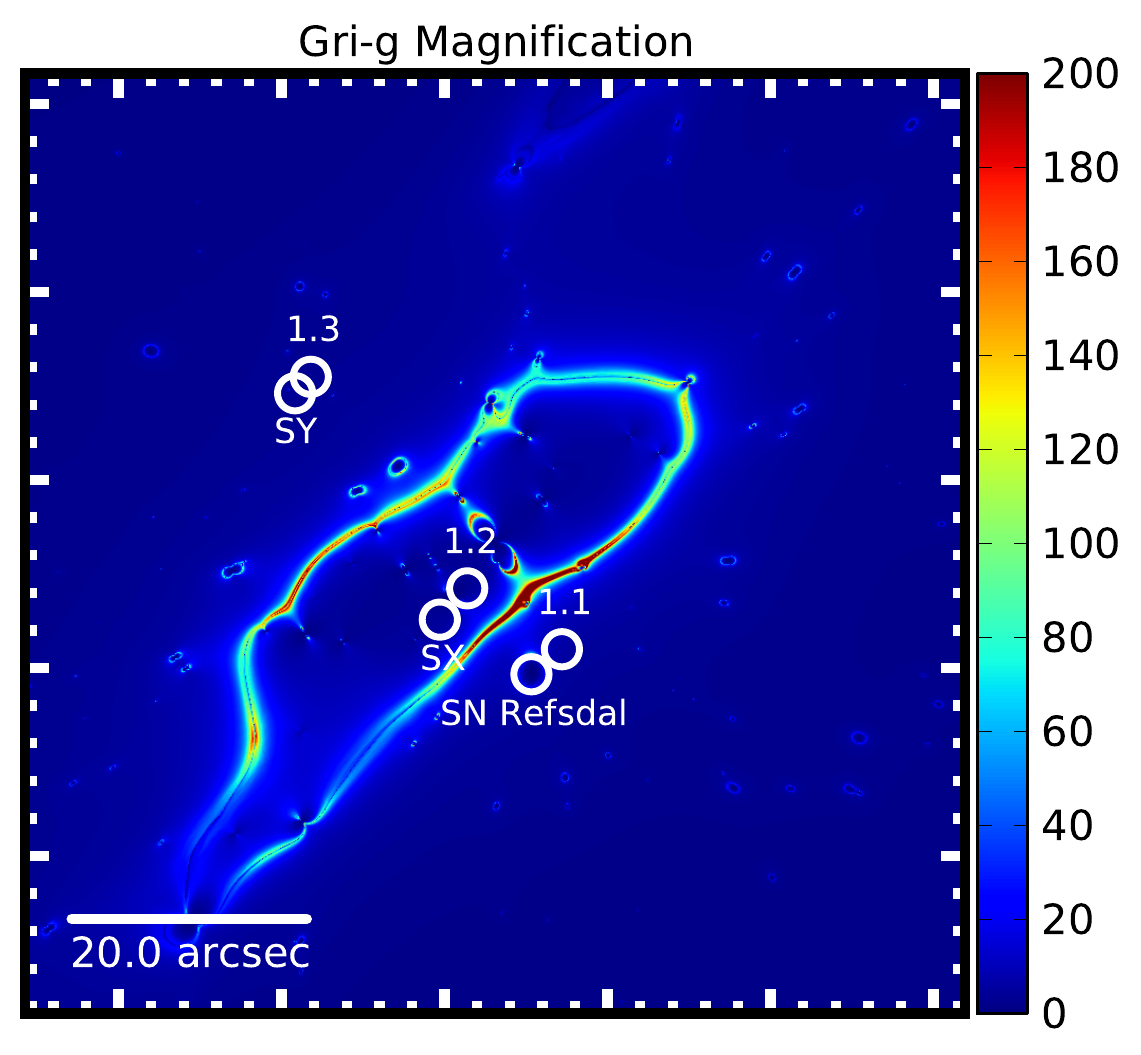} 
\includegraphics[width=0.31\textwidth]{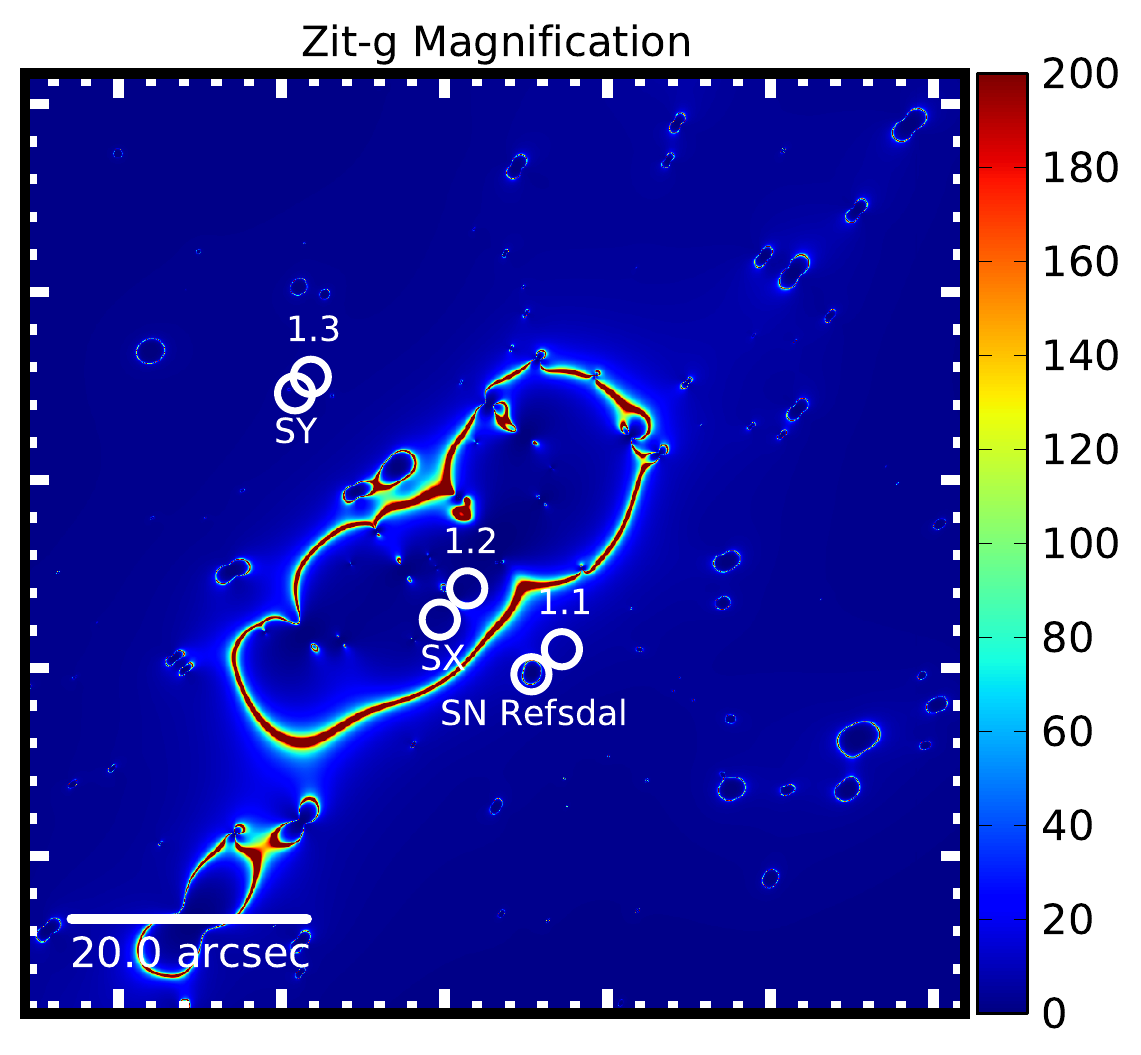}\\
\includegraphics[width=0.31\textwidth]{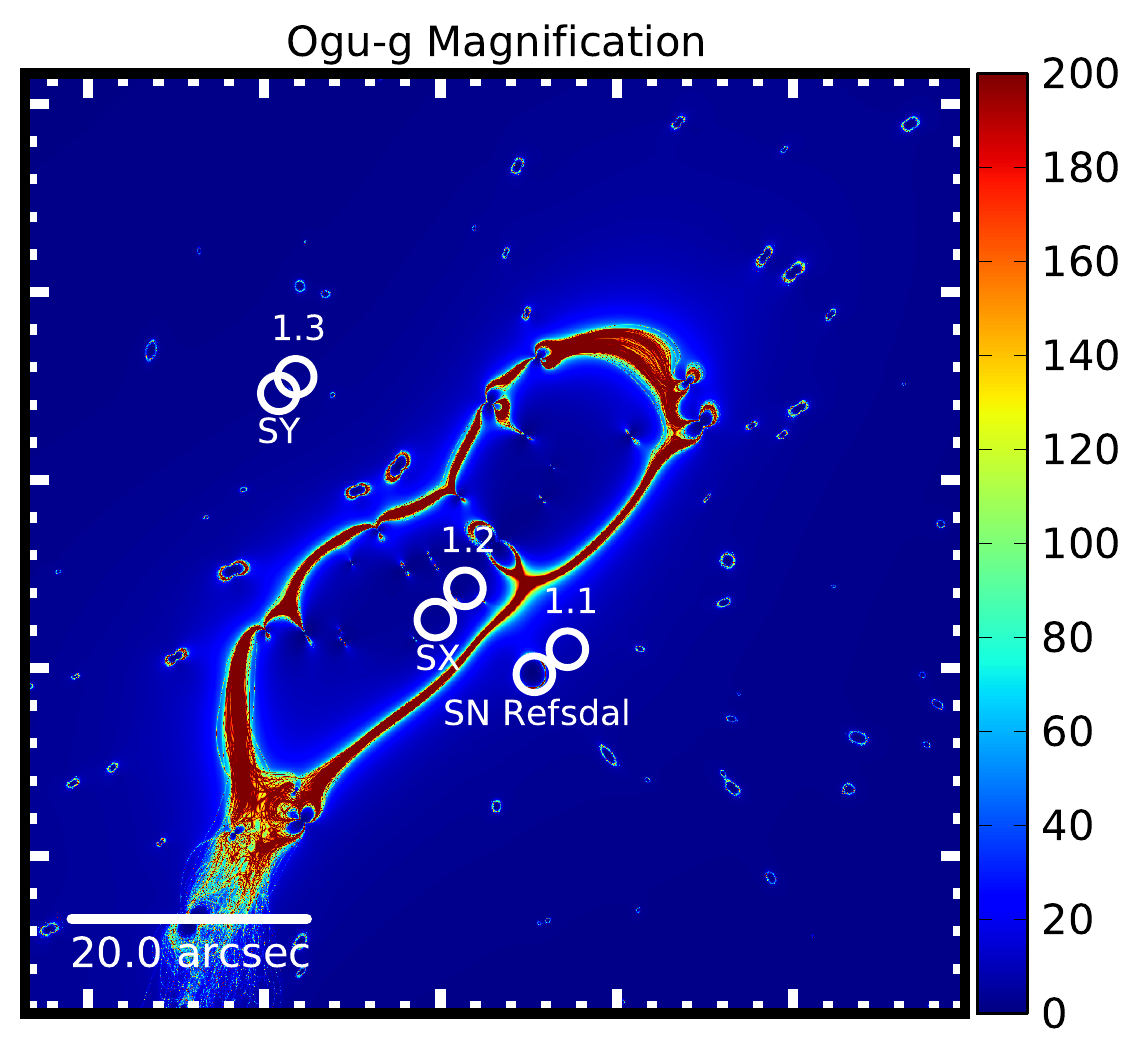}
\includegraphics[width=0.31\textwidth]{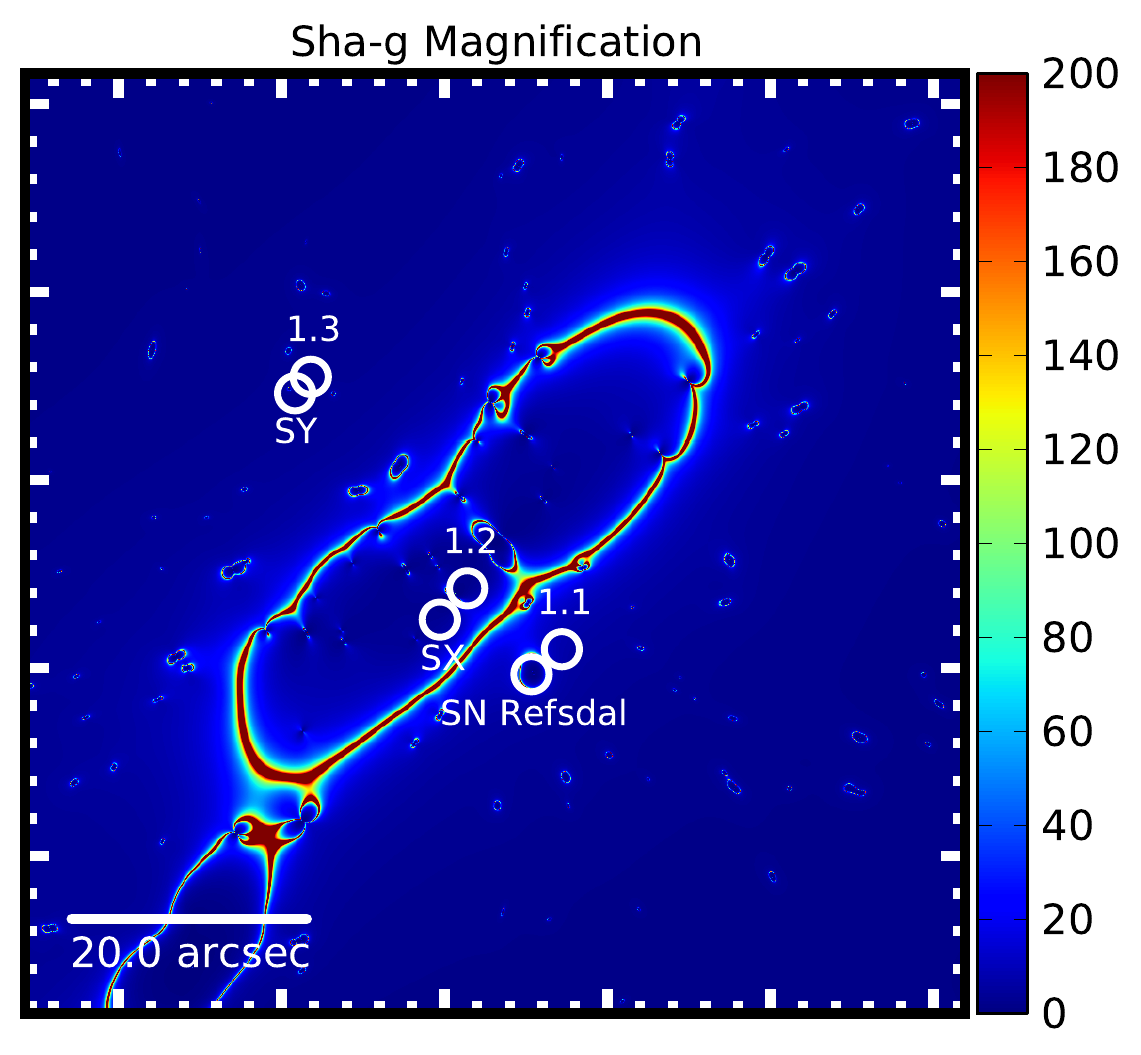}\\
\includegraphics[width=0.31\textwidth]{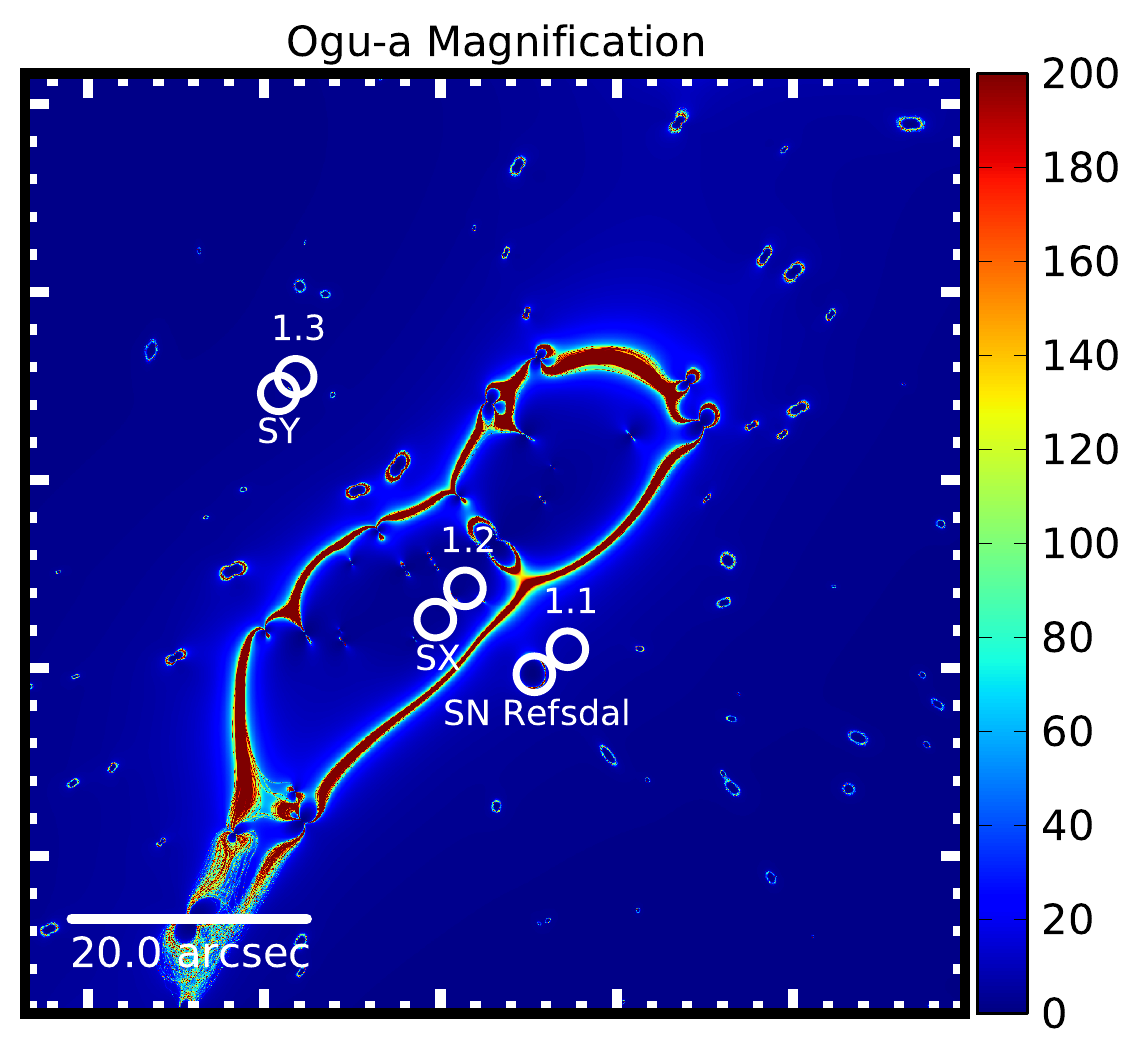}   
\includegraphics[width=0.31\textwidth]{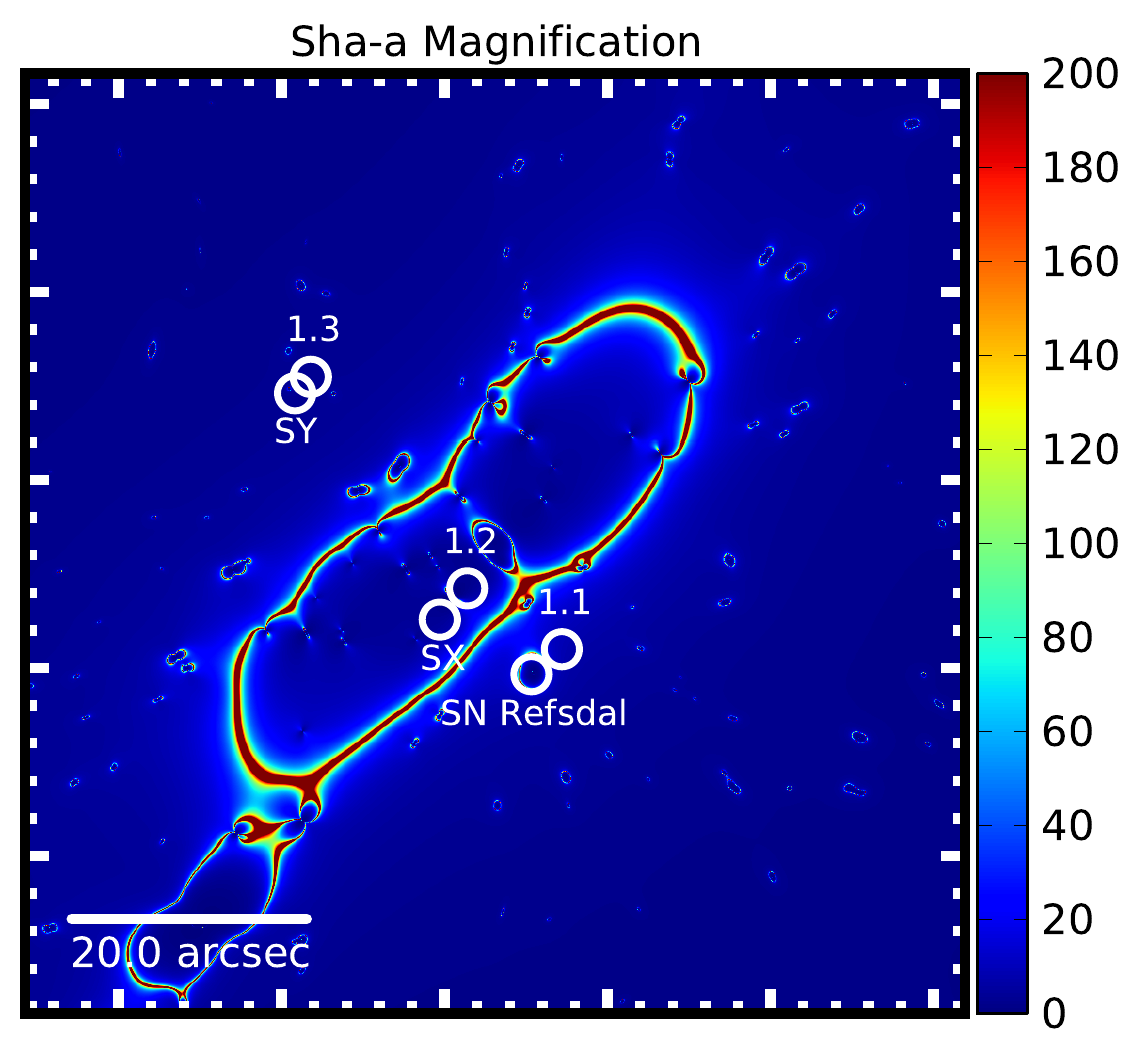}
\includegraphics[width=0.31\textwidth]{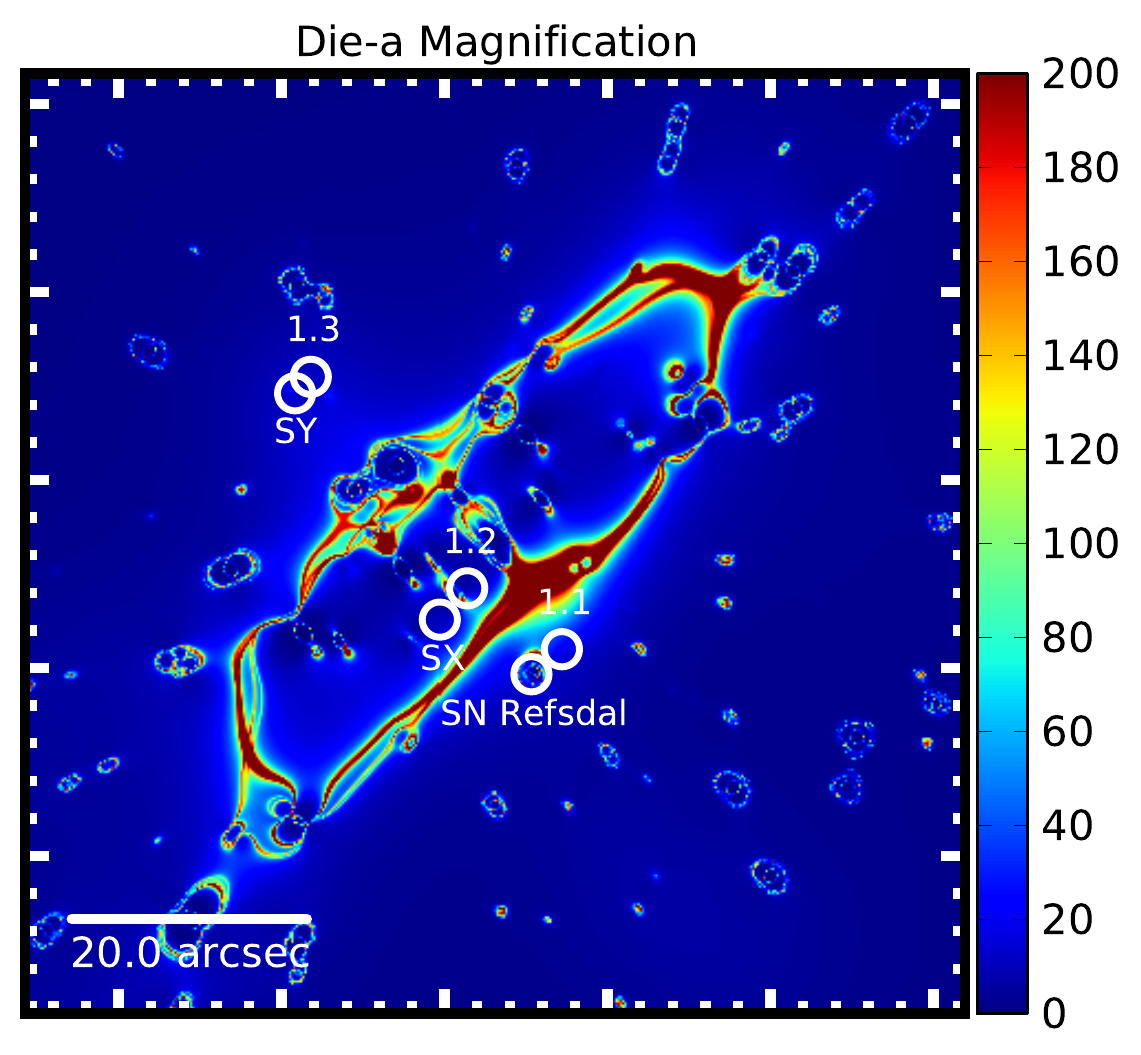} 
\caption{As in Figure~\ref{fig:compareMASS} for magnification.}
\label{fig:compareMAG}
\end{center}
\end{figure*}

\begin{figure*}
\begin{center}
\includegraphics[width=0.31\textwidth]{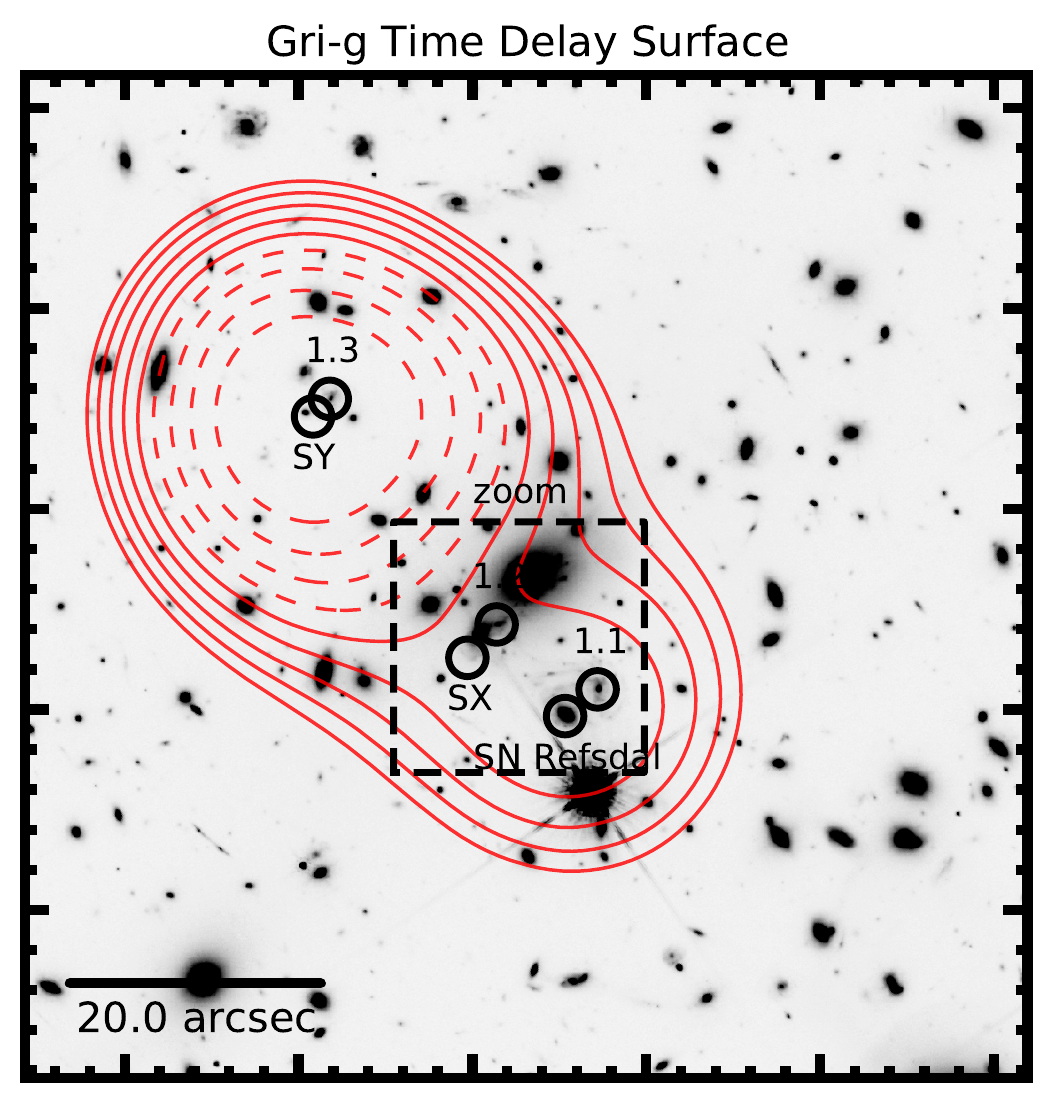}
\includegraphics[width=0.31\textwidth]{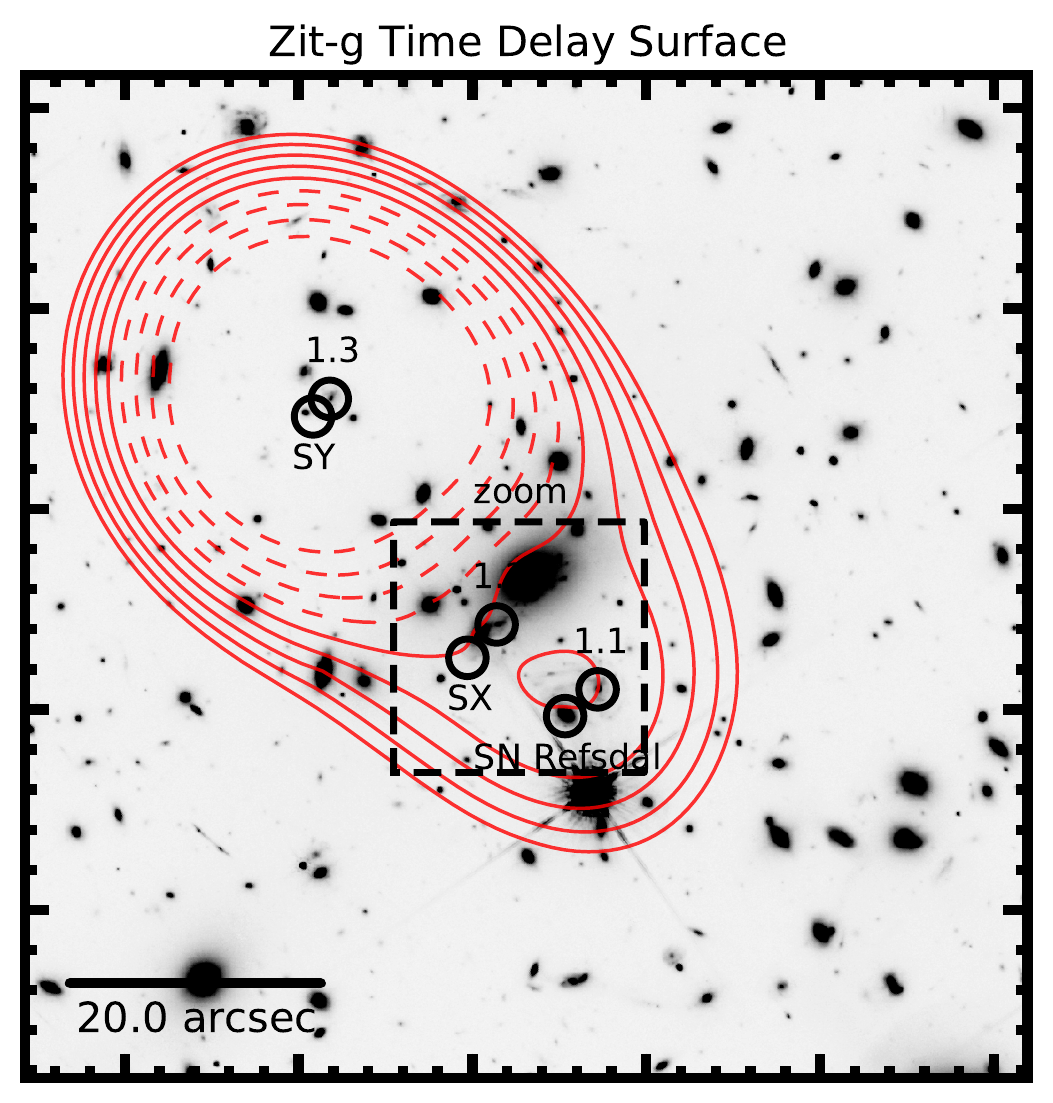}\\
\includegraphics[width=0.31\textwidth]{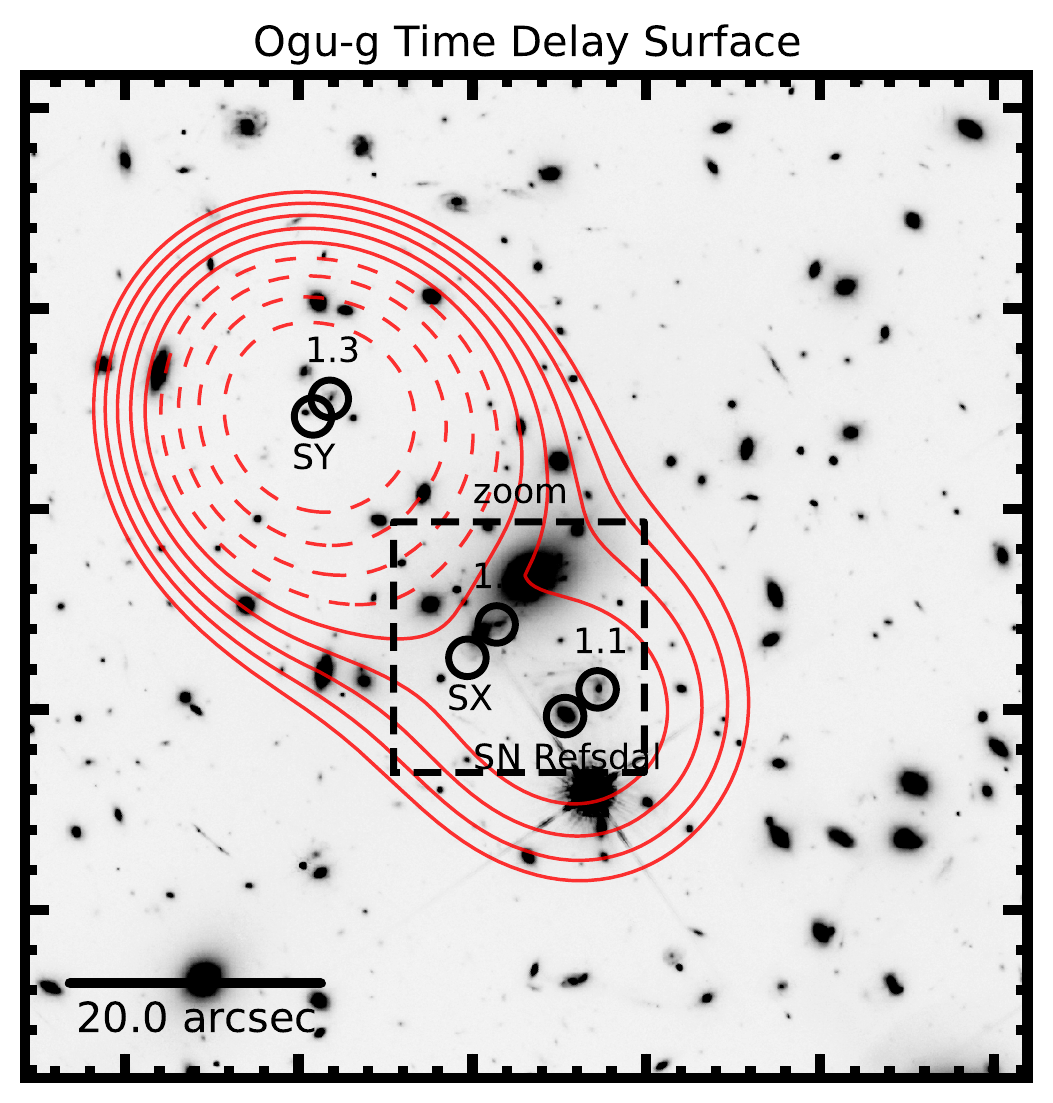}       
\includegraphics[width=0.31\textwidth]{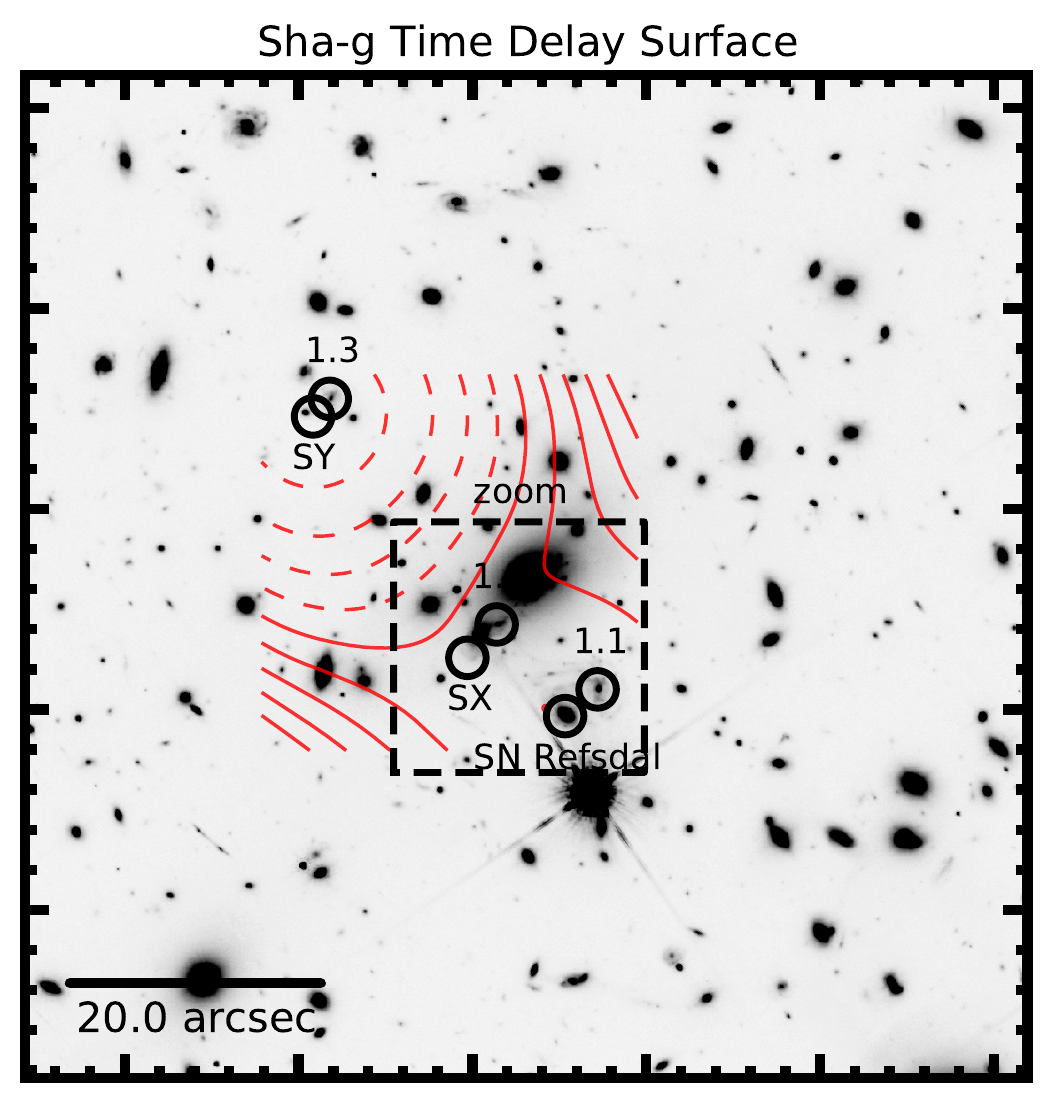}\\
\includegraphics[width=0.31\textwidth]{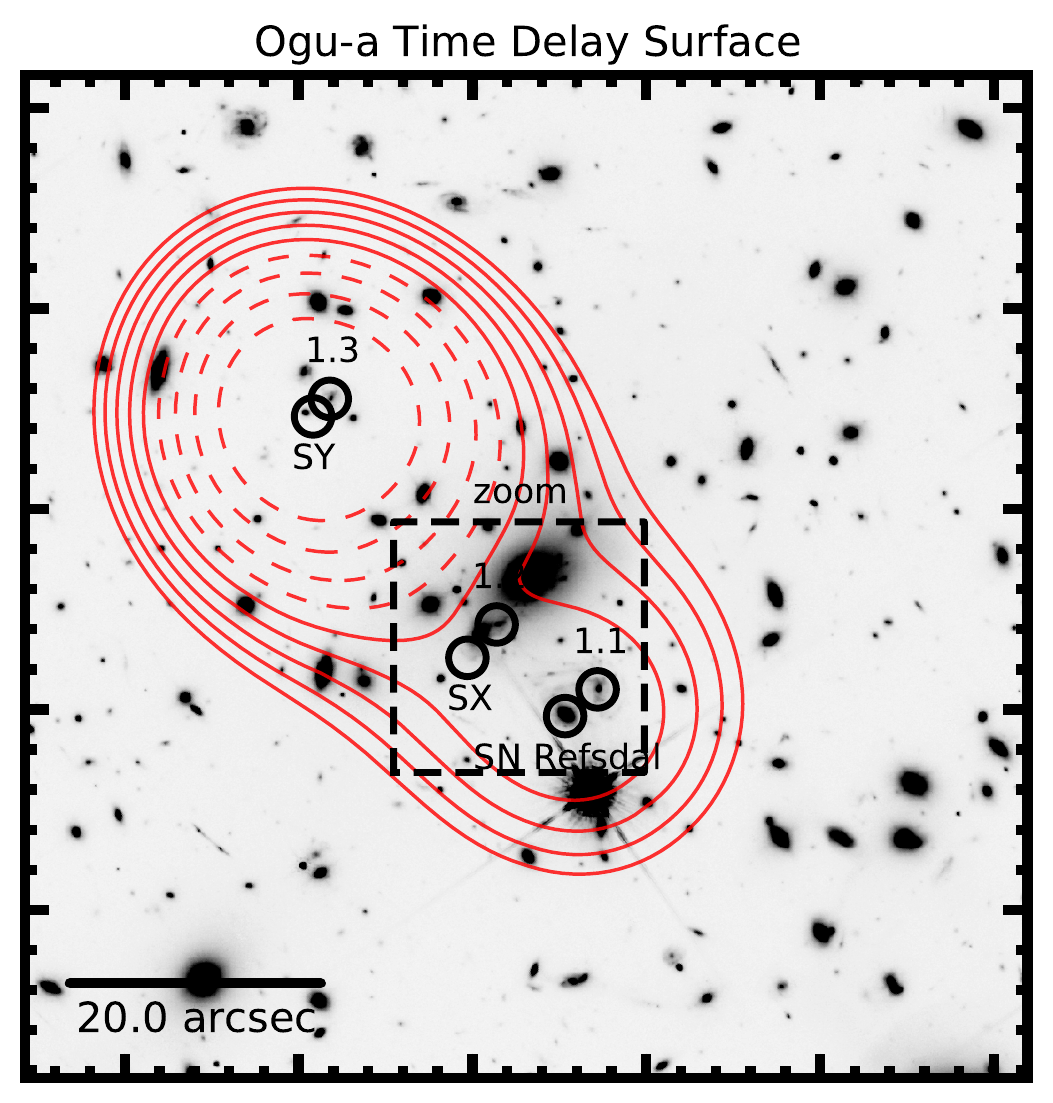}
\includegraphics[width=0.31\textwidth]{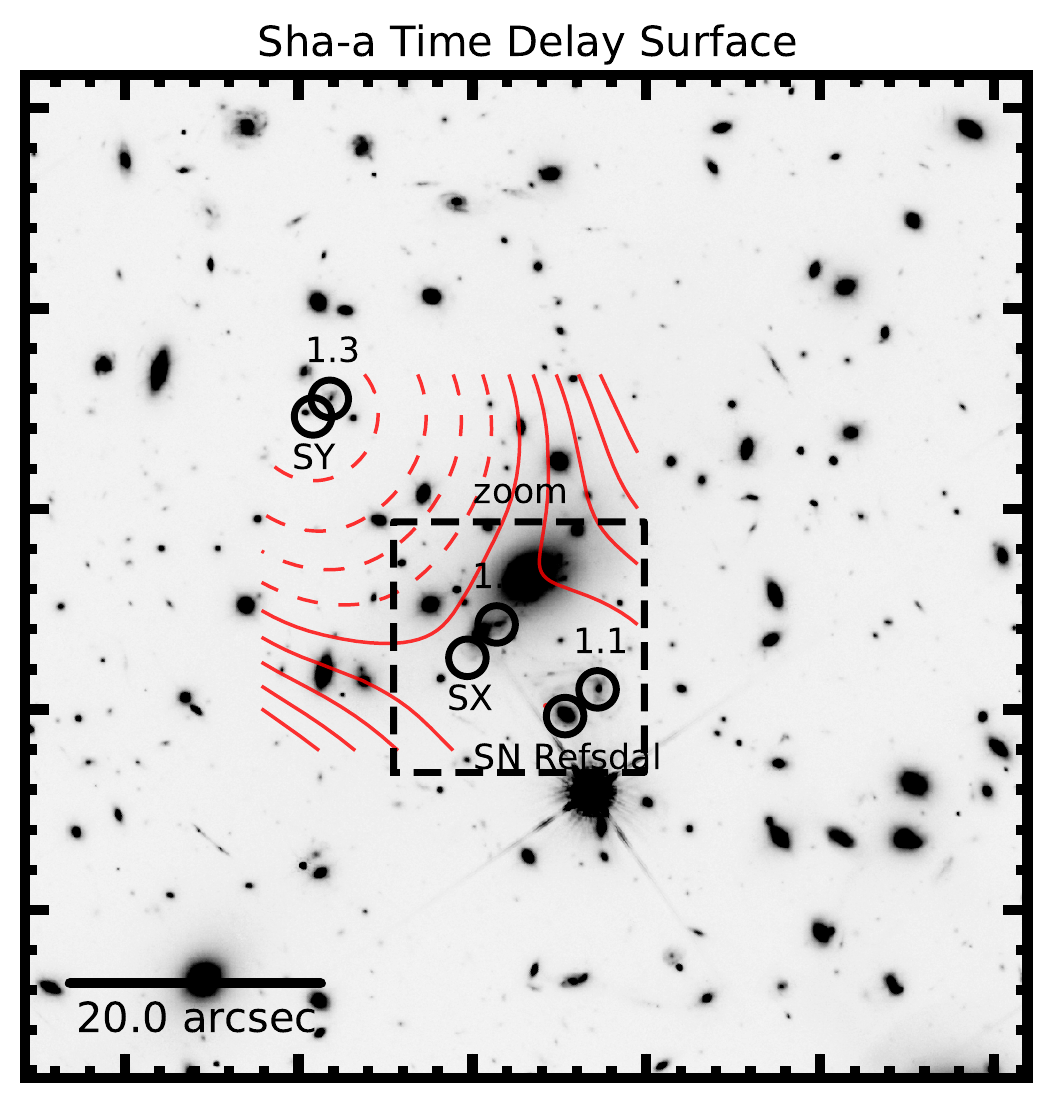}  
\includegraphics[width=0.31\textwidth]{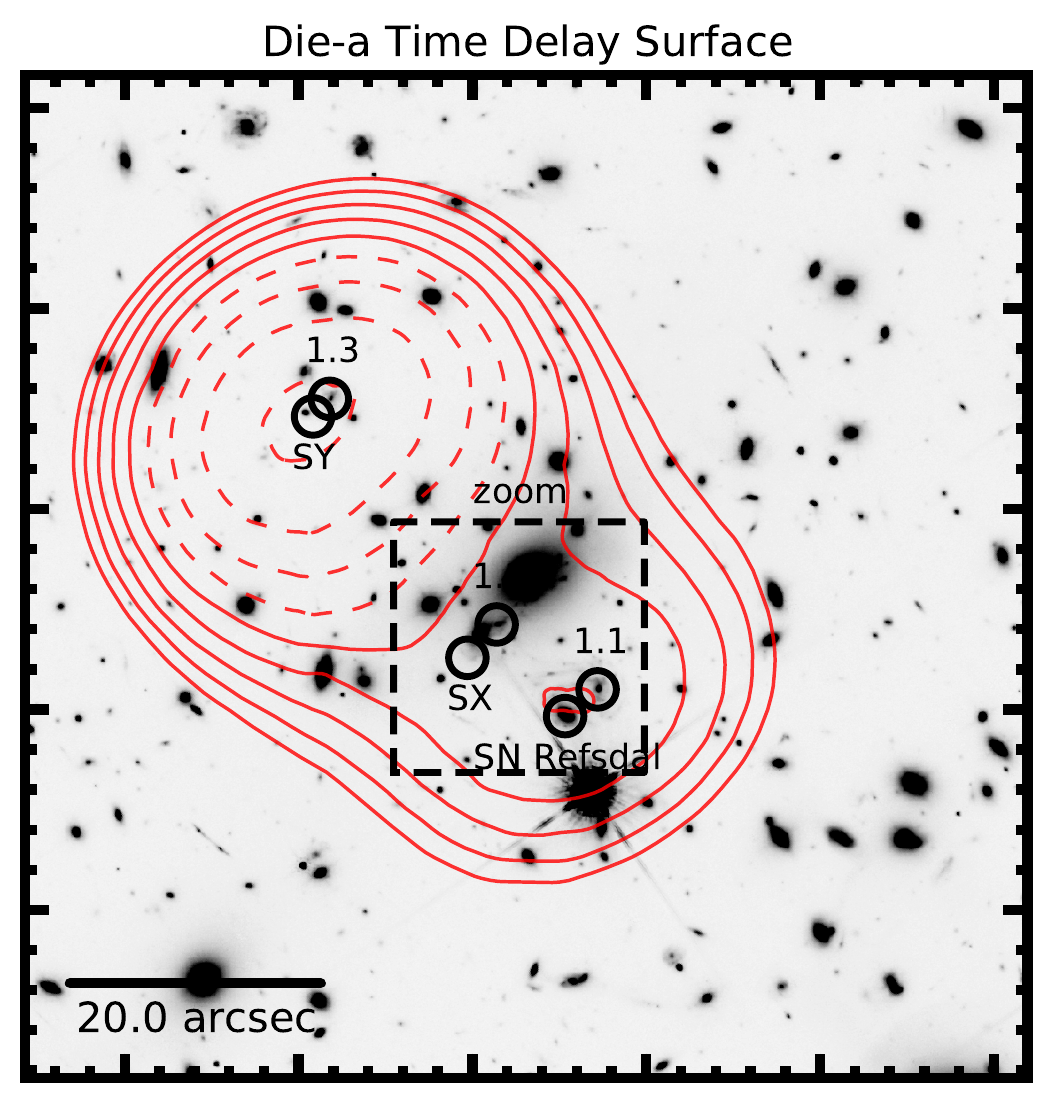}
\caption{As in Figure~\ref{fig:compareMASS} for time-delay surfaces.
The dashed boxes mark the location of the zoom-in regions shown in Figure~\ref{fig:compareTDzoom}. 
Contour levels indicate the time delay from $-12$ to 12 years in increments of 3 years, relative to S1. For the Sha-a and Sha-g models the time-delay surfaces were only calculated in the region shown.
Negative levels are marked by dashed contours.
The gray-scale background image shows the HFF F140W epoch2 version 1.0 mosaic.}
\label{fig:compareTD}
\end{center}
\end{figure*}

\begin{figure*}
\begin{center}
\includegraphics[width=0.31\textwidth]{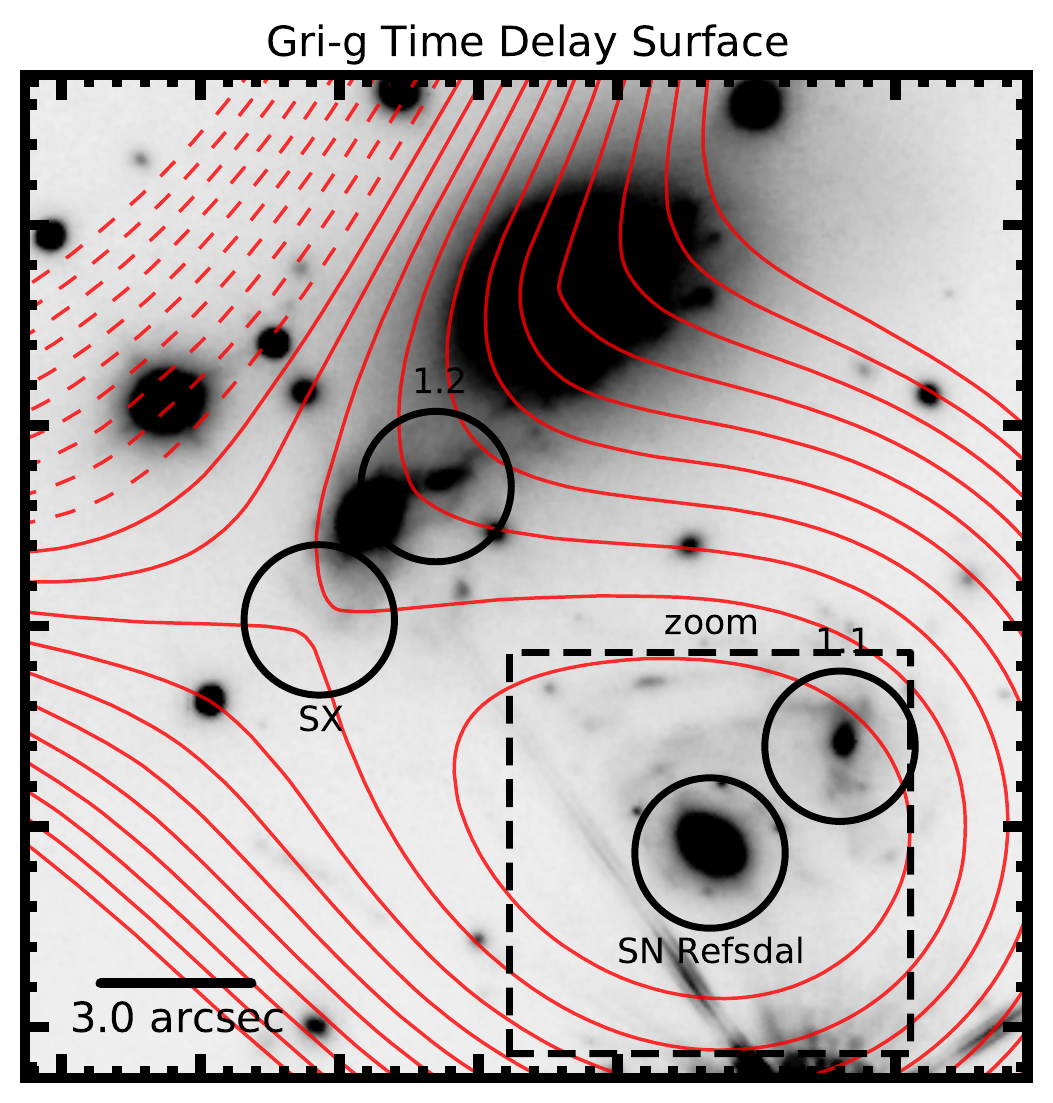}
\includegraphics[width=0.31\textwidth]{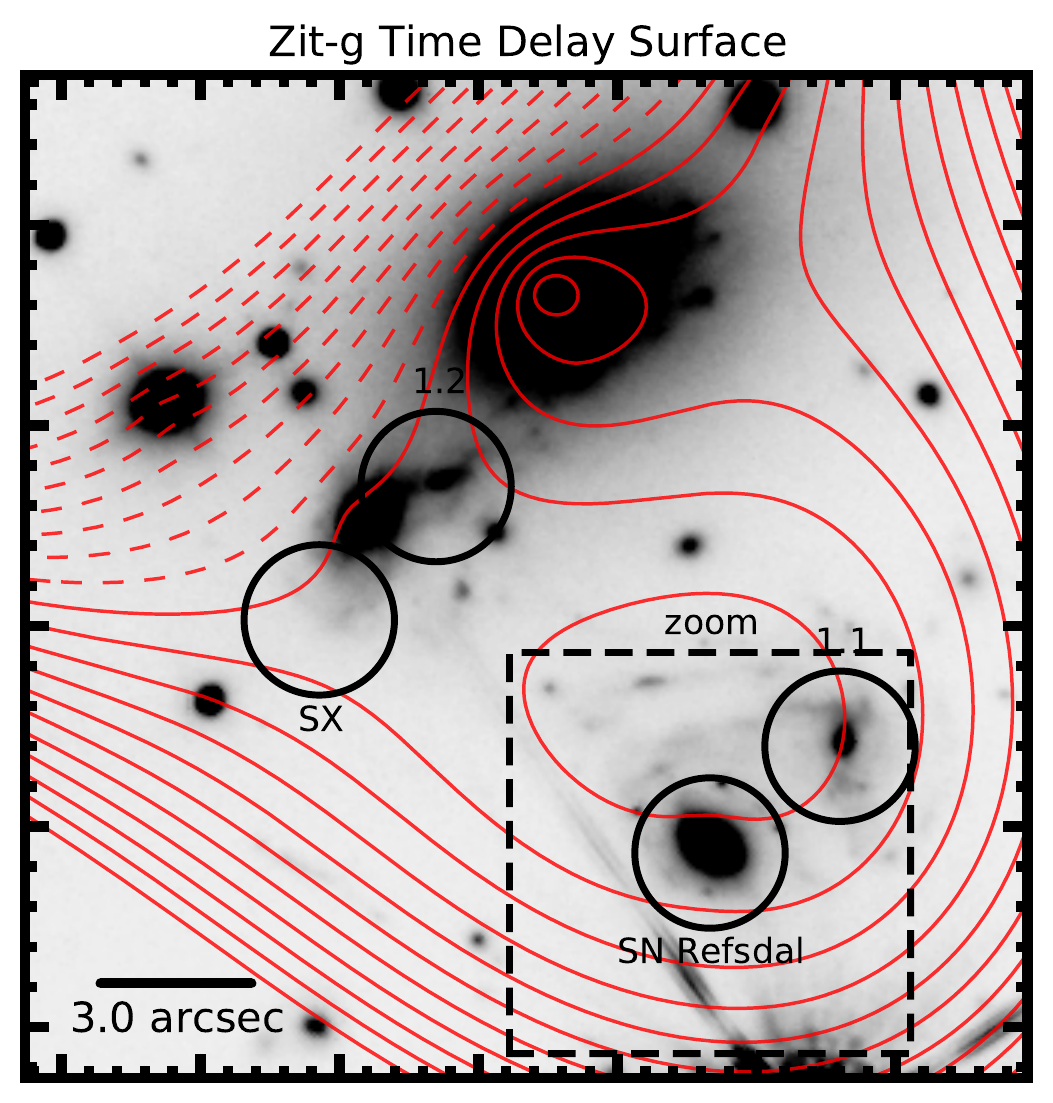}\\
\includegraphics[width=0.31\textwidth]{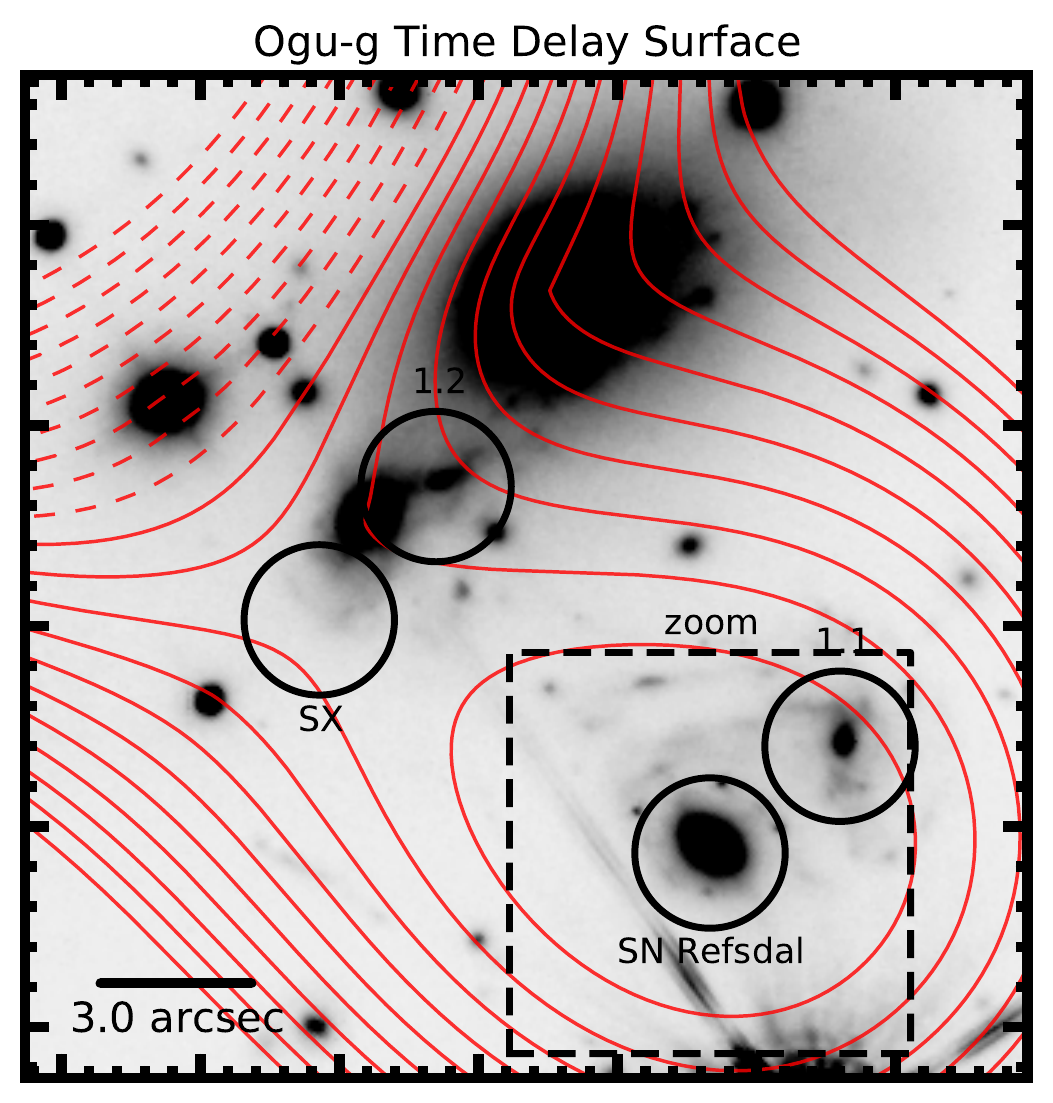} 
\includegraphics[width=0.31\textwidth]{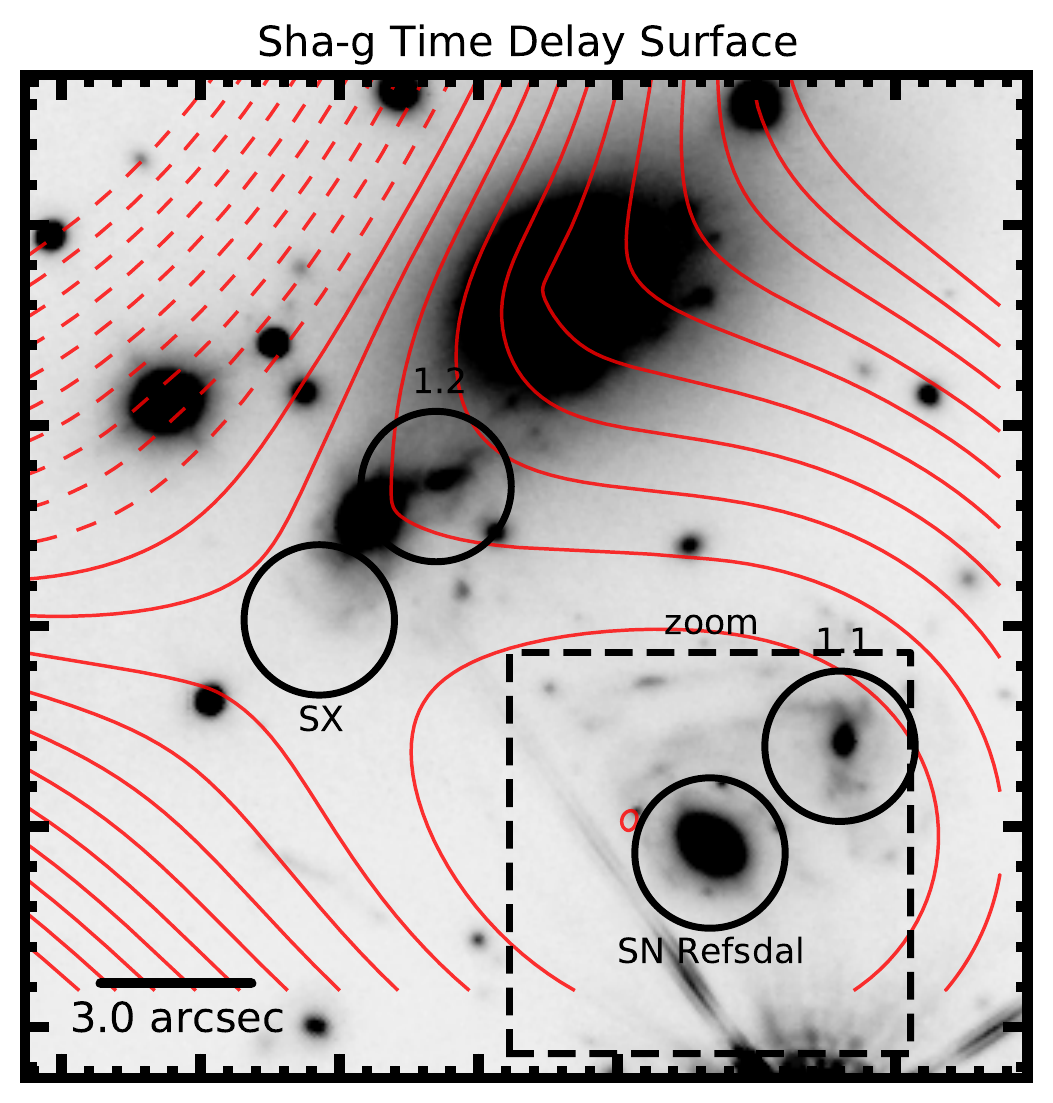}\\	
\includegraphics[width=0.31\textwidth]{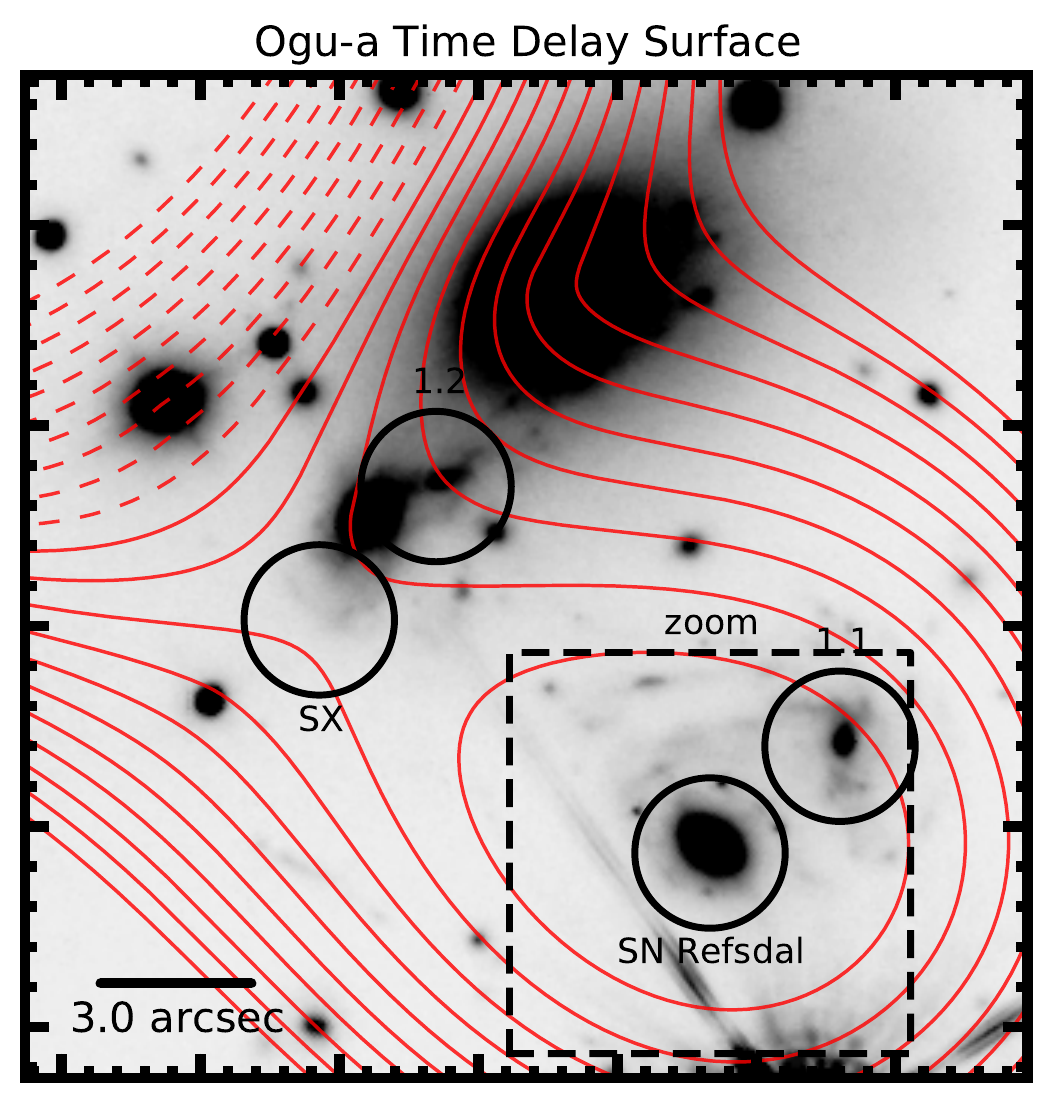}   
\includegraphics[width=0.31\textwidth]{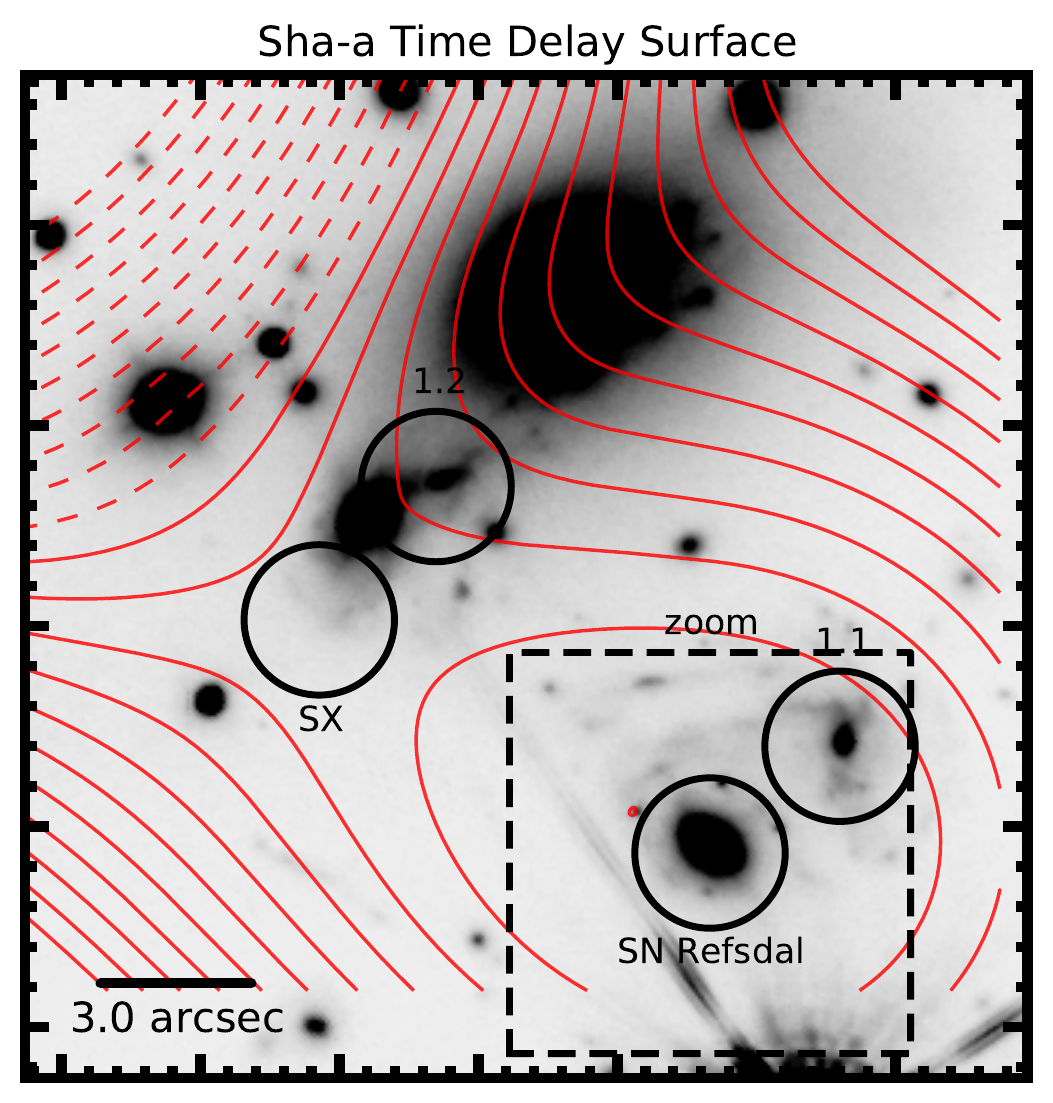}   
\includegraphics[width=0.31\textwidth]{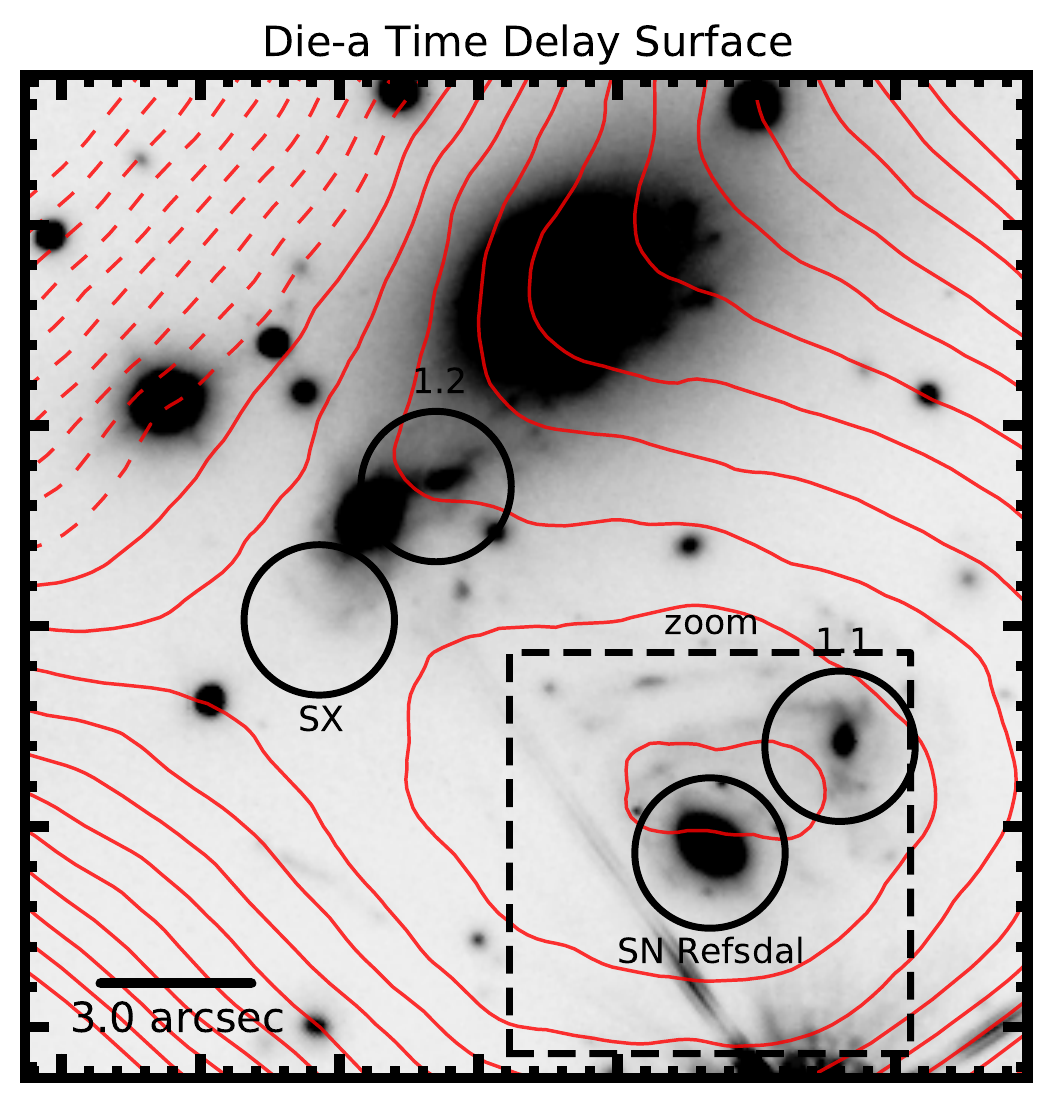}
\caption{The time-delay surface details in the region marked in Figure~\ref{fig:compareTD}. 
The dashed boxes mark the location of the zoom-in regions shown in Figure~\ref{fig:compareTDzoomrefsdal}. 
Contour levels indicate the time delay from $-5$ to 5 years in increments of 0.5 years, relative to S1.
Negative levels are marked by dashed contours.
The gray-scale background image shows the HFF F140W epoch2 version 1.0 mosaic.}
\label{fig:compareTDzoom}
\end{center}
\end{figure*}

\begin{figure*}
\begin{center}
\includegraphics[width=0.31\textwidth]{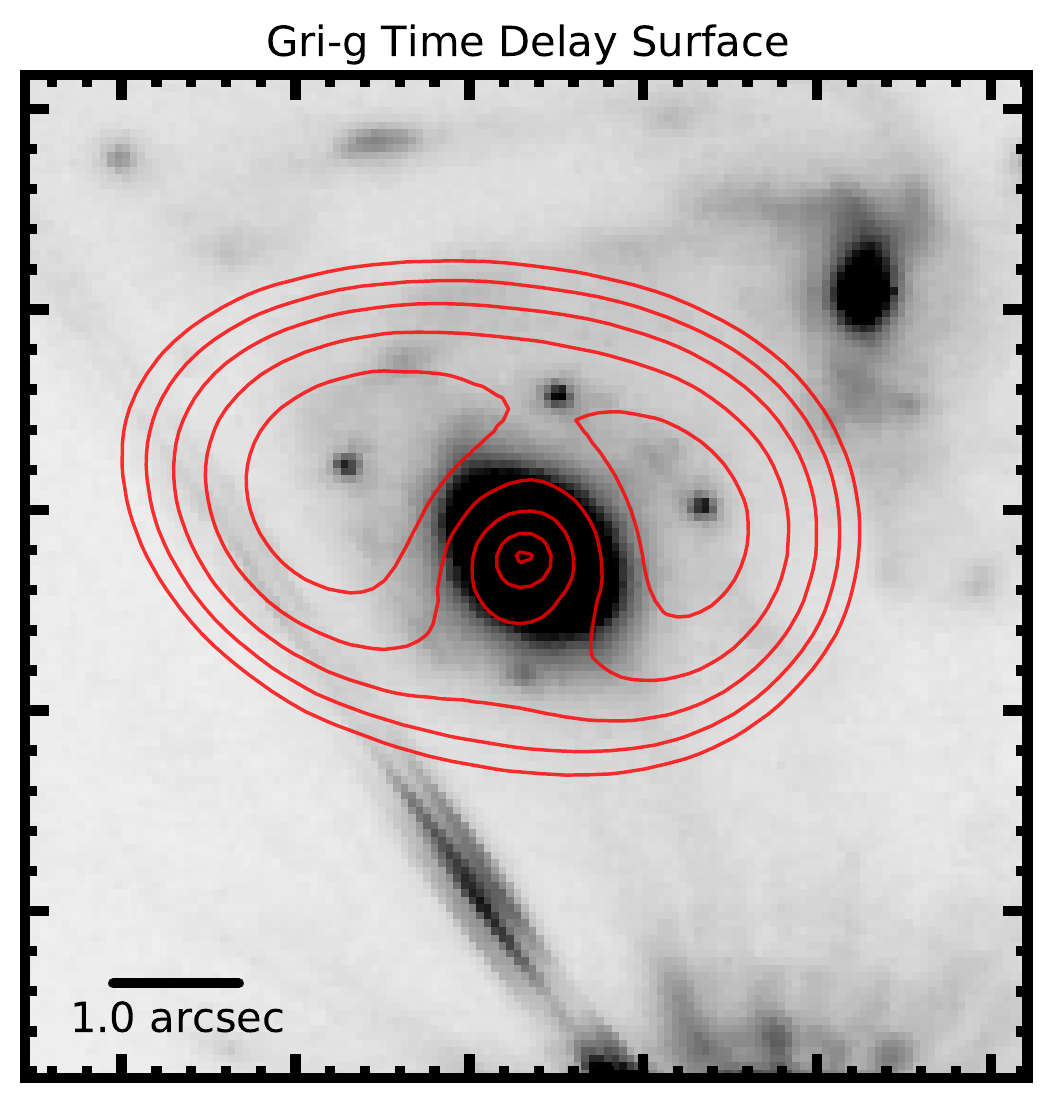}
\includegraphics[width=0.31\textwidth]{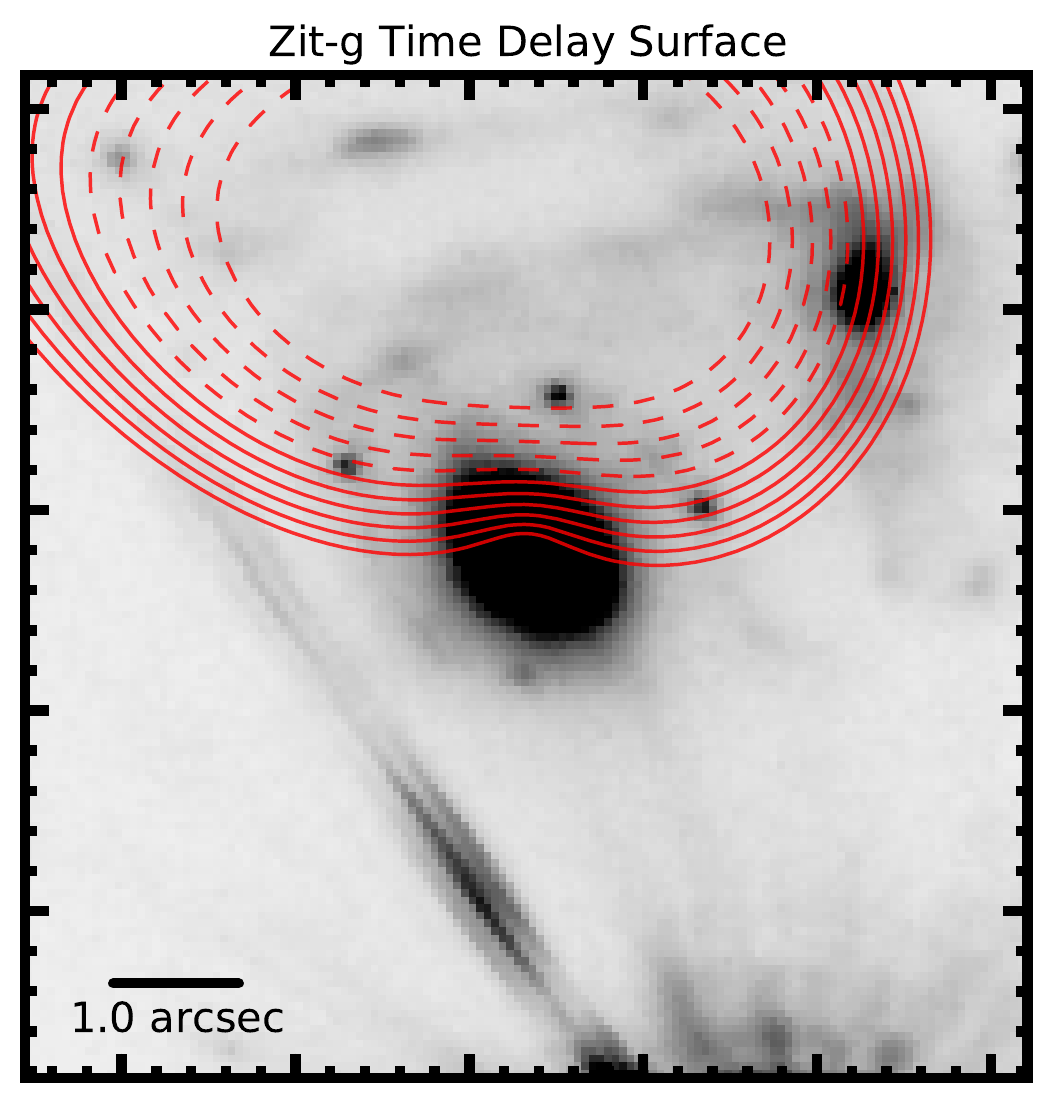}\\
\includegraphics[width=0.31\textwidth]{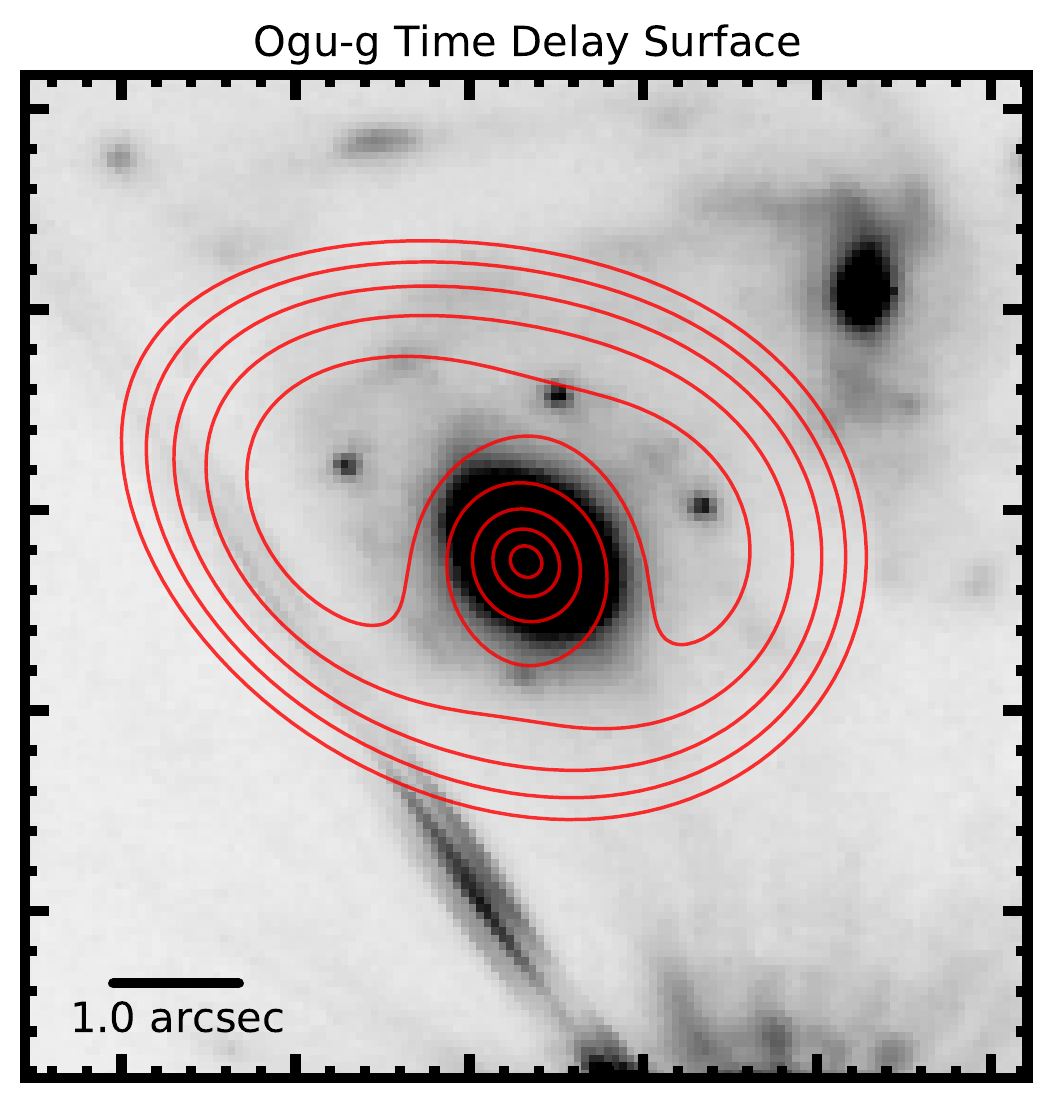} 
\includegraphics[width=0.31\textwidth]{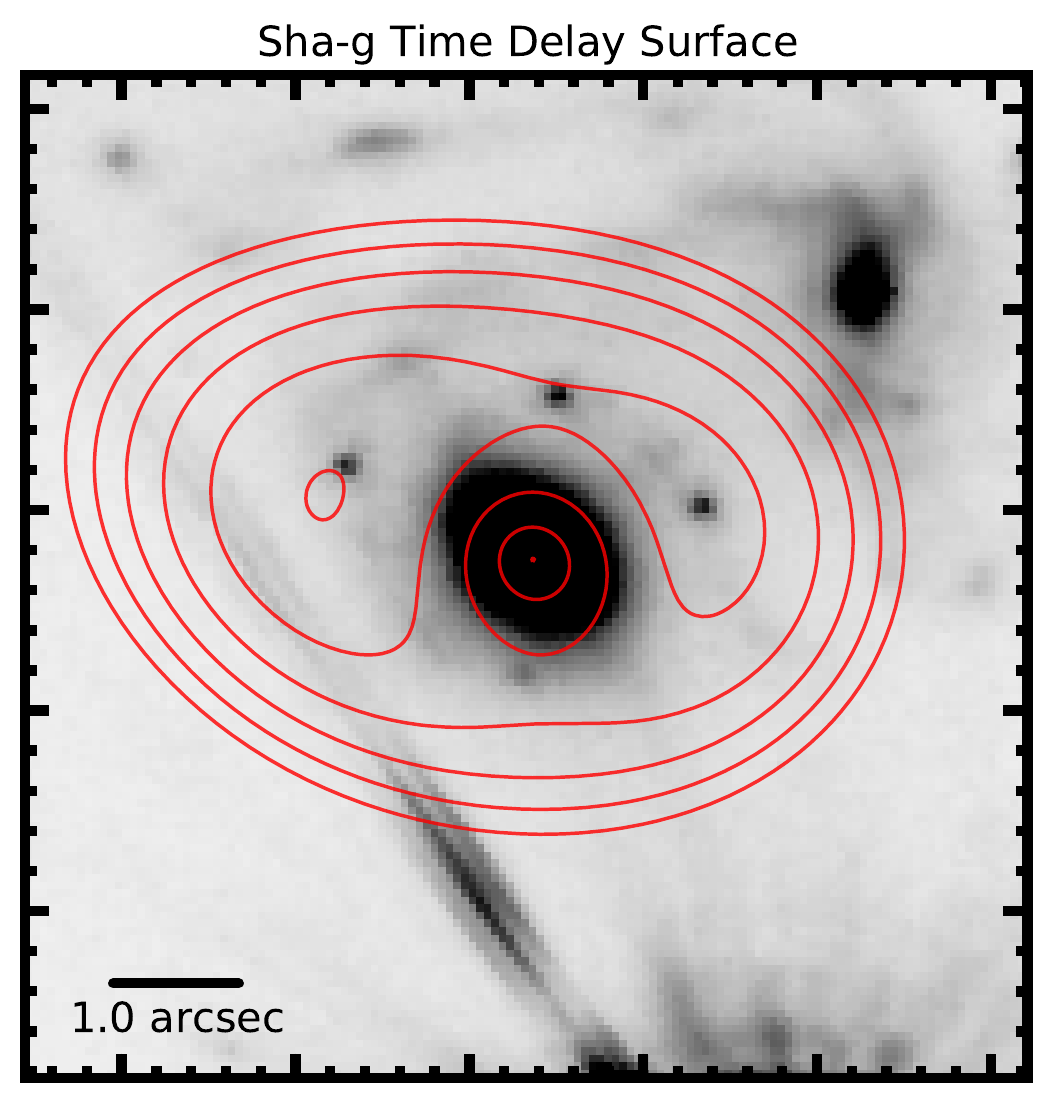}\\	
\includegraphics[width=0.31\textwidth]{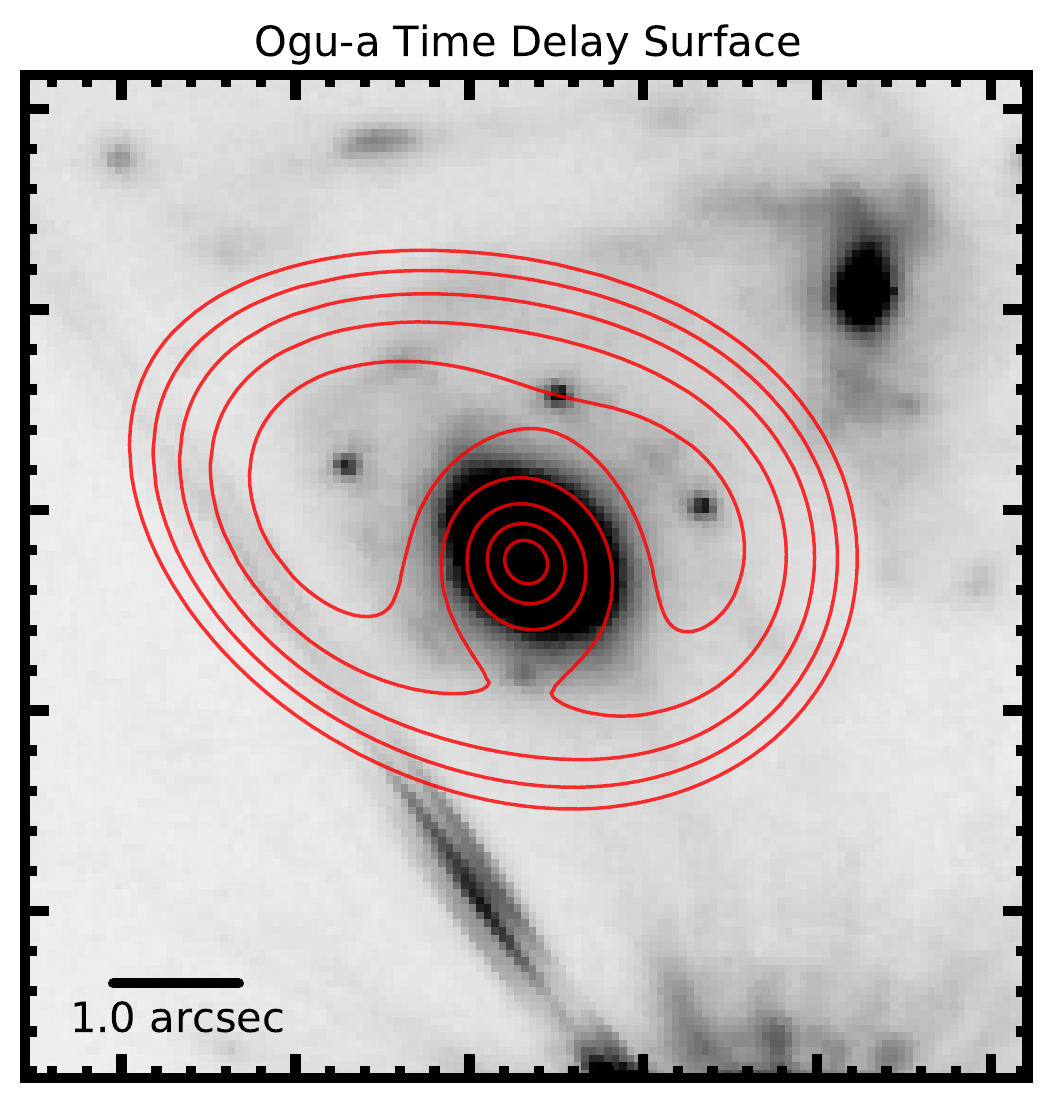}   
\includegraphics[width=0.31\textwidth]{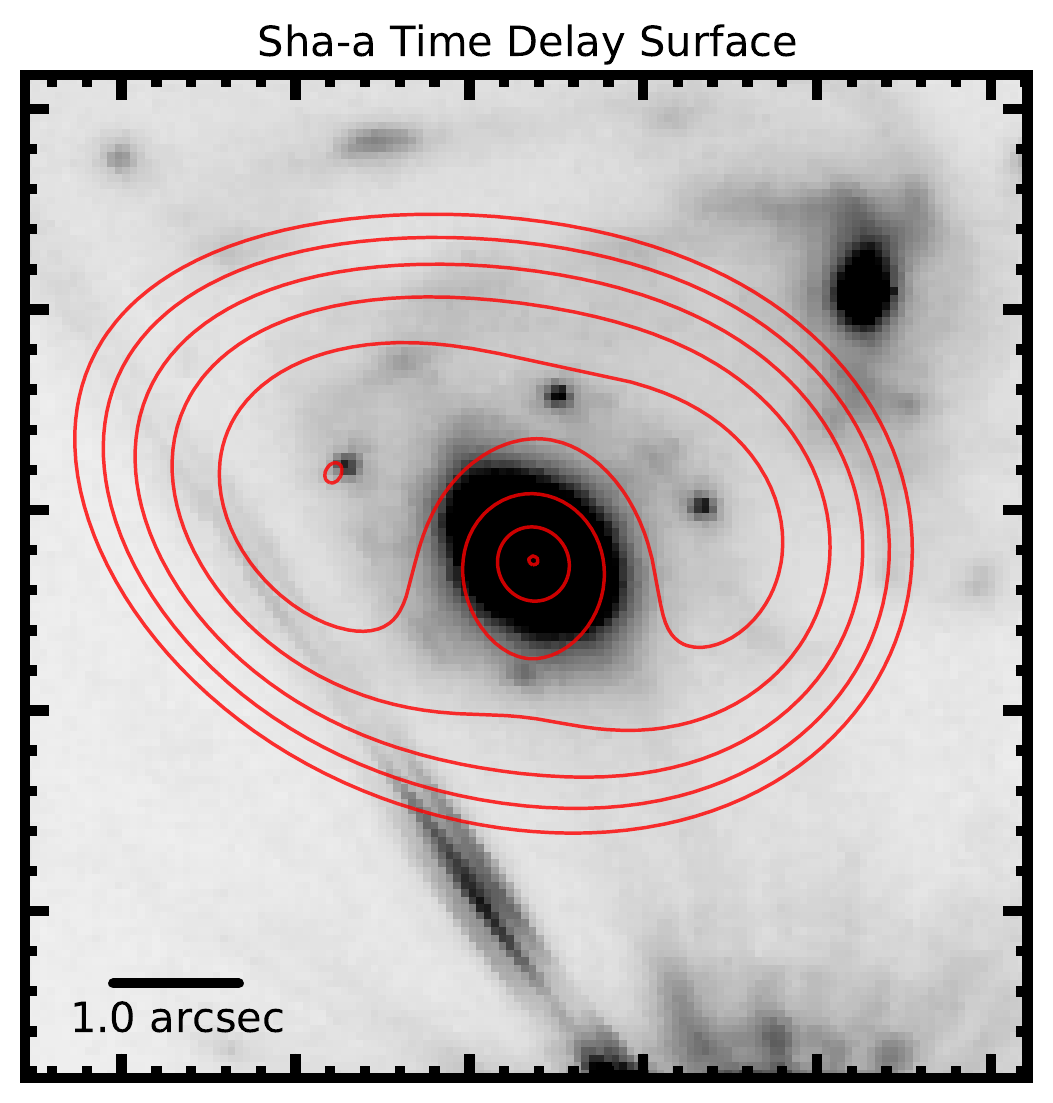}   
\includegraphics[width=0.31\textwidth]{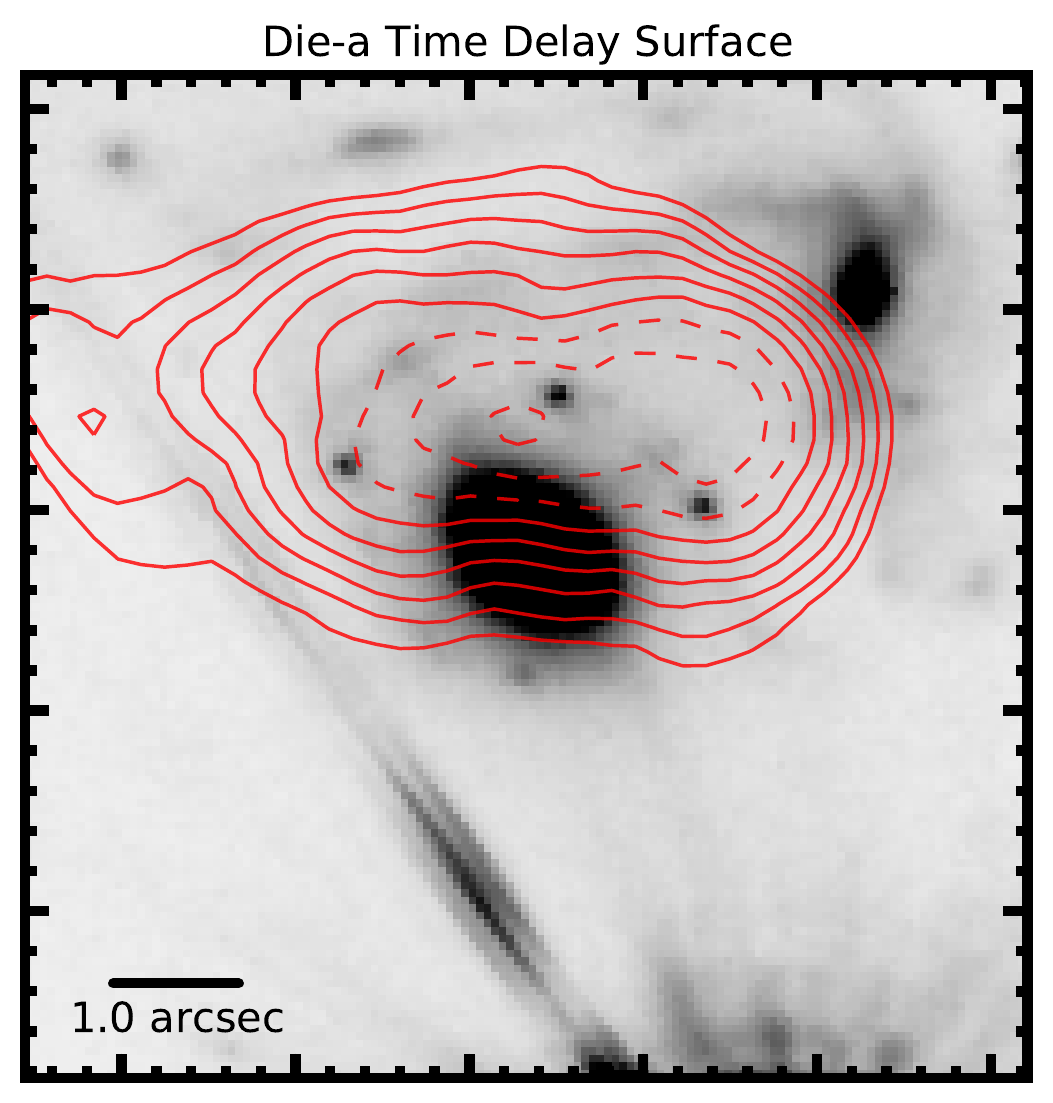}
\caption{The time-delay surface details in the region marked in Figure~\ref{fig:compareTDzoom}. 
Contour levels indicate the time delay from $-50$ to 50 days in increments of 10 days, relative to S1.
Negative levels are marked by dashed contours.
The gray-scale background image shows the HFF F140W epoch2 version 1.0 mosaic.}
\label{fig:compareTDzoomrefsdal}
\end{center}
\end{figure*}

Figure~\ref{fig:compareMASS} shows the convergence (i.e., surface mass
density in units of the lensing critical density) maps. There are
striking qualitative differences.  The Zit-g map is significantly
rounder than the others. The Die-a map has significantly more
structure, notably two overdensities near SY/1.3 and at the bottom
right of the map. These features were to be expected based on the
assumptions used by their methods. The Grillo, Oguri, and Sharon
convergence maps are the most qualitatively similar. This is perhaps
unsurprising since the three codes are based on fairly similar assumptions.

Magnification maps are shown in Fig.~\ref{fig:compareMAG}. The regions
of extreme magnification are qualitatively similar, even though,
similarly to the convergence maps, the Zit-g model is overall rounder,
while the Die-a model has more structure.

The time-delay surfaces are illustrated at three zoom levels to highlight
different features. Fig.~\ref{fig:compareTD} shows the global topology
of the time-delay surfaces, which is very similar for all models, with
minima near 1.1 and SY/1.3 and a saddle point near SX/1.2. As was the
case with convergence and magnification, the Zit-g and Die-a time-delay 
surfaces are rounder and have more structure, respectively,
than those produced by the other models.

Zooming in on the region of SX/1.2 and 1.1 in
Fig.~\ref{fig:compareTDzoom} reveals more differences. The locations
of the minimum near SN Refsdal and of the saddle point near 1.2 are
significantly different for the Zit-g model, seemingly as a result of
the different contribution of the bright galaxy to the NW of 1.2.

A further zoom-in on the region of the known images is shown in
Figure~\ref{fig:compareTDzoomrefsdal}. The time-delay surface contour
levels are shown in step of 10 days to highlight the behavior
relevant for the cross configuration. Whereas the ``simply-parametrized''
models are topologically very similar to each other, the Die-a and
Zit-g models are qualitatively different. The time-delay surface is
shifted upward, probably as a result of the nearby perturber
highlighted in the previous paragraph. We stress that all of the models
here are global models, developed to reproduce the cluster potential
on larger scales. Hence, local differences should be expected, even
though of course they are particularly important in this case.
  
\subsection{Comparing Model Predictions with Measured Time Delays and Magnification Ratios}
\label{ssec:cross}

Before proceeding with a quantitative comparison, we emphasize once
again that the uncertainties discussed in this section include only
statistical uncertainties. Furthermore, in the comparison we neglect
for computational reasons the covariance between the predictions for
each of the images, both in time delays and in
magnification. Systematic uncertainties will be discussed in
Section~\ref{sec:discussion}.

\begin{figure}[]
\centerline{
\includegraphics[width=0.49\textwidth]{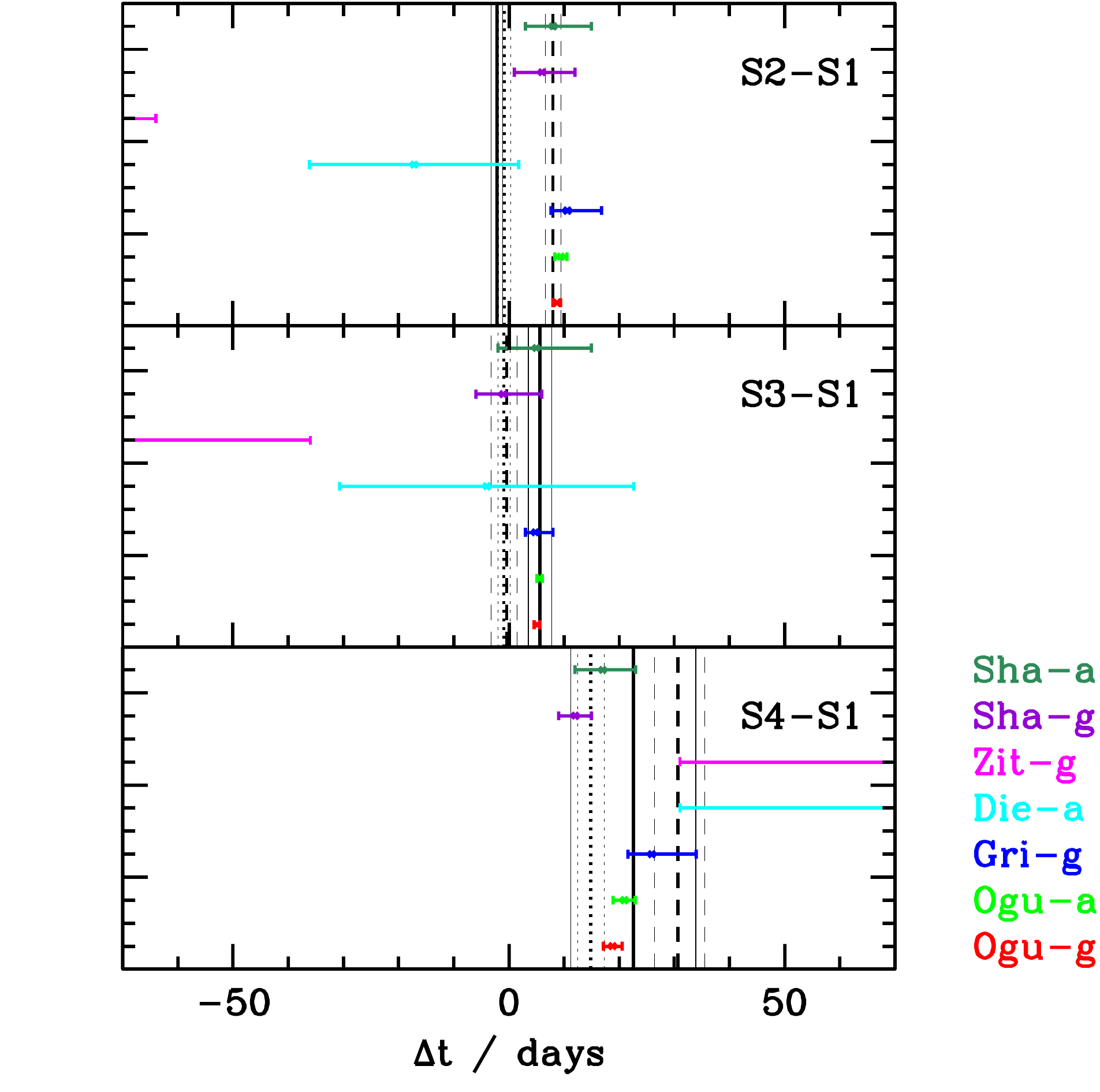}}
\caption{Observed (solid vertical line represents the preliminary measurements; dashed vertical line represents the updated template-based measurements; dotted vertical line represents the updated polynomial-based measurements; thin lines represent the 68\% confidence range for each measurement) and predicted (points with error bars) time delays for the images in the cross configuration, relative to S1.
Uncertainties represent the 68\% confidence interval. \label{fig:tdcross}}
\end{figure}

Figure~\ref{fig:tdcross} compares the measured time delays with those
predicted by the models for the cross configuration. We stress that
the measurements were not used in the construction of the models (or
known to the modelers), and therefore they can be considered an
independent test of the models. The time delay between S2 and S1 (and
to some extent that between S3 and S1) is very short, and in fact not
all the models agree on the ordering of the two images. The time delay
between S4 and S1 is longer and better behaved, with all the models
agreeing on the order of the images and with the measured value within
the uncertainties. Overall, the models are in reasonable agreement with
the measurements, even though formally some of them are in statistical
tension. This tension indicates that the uncertainties for some of the
parametric models are underestimated. 

\begin{figure}[]
\centerline{
\includegraphics[width=0.49\textwidth]{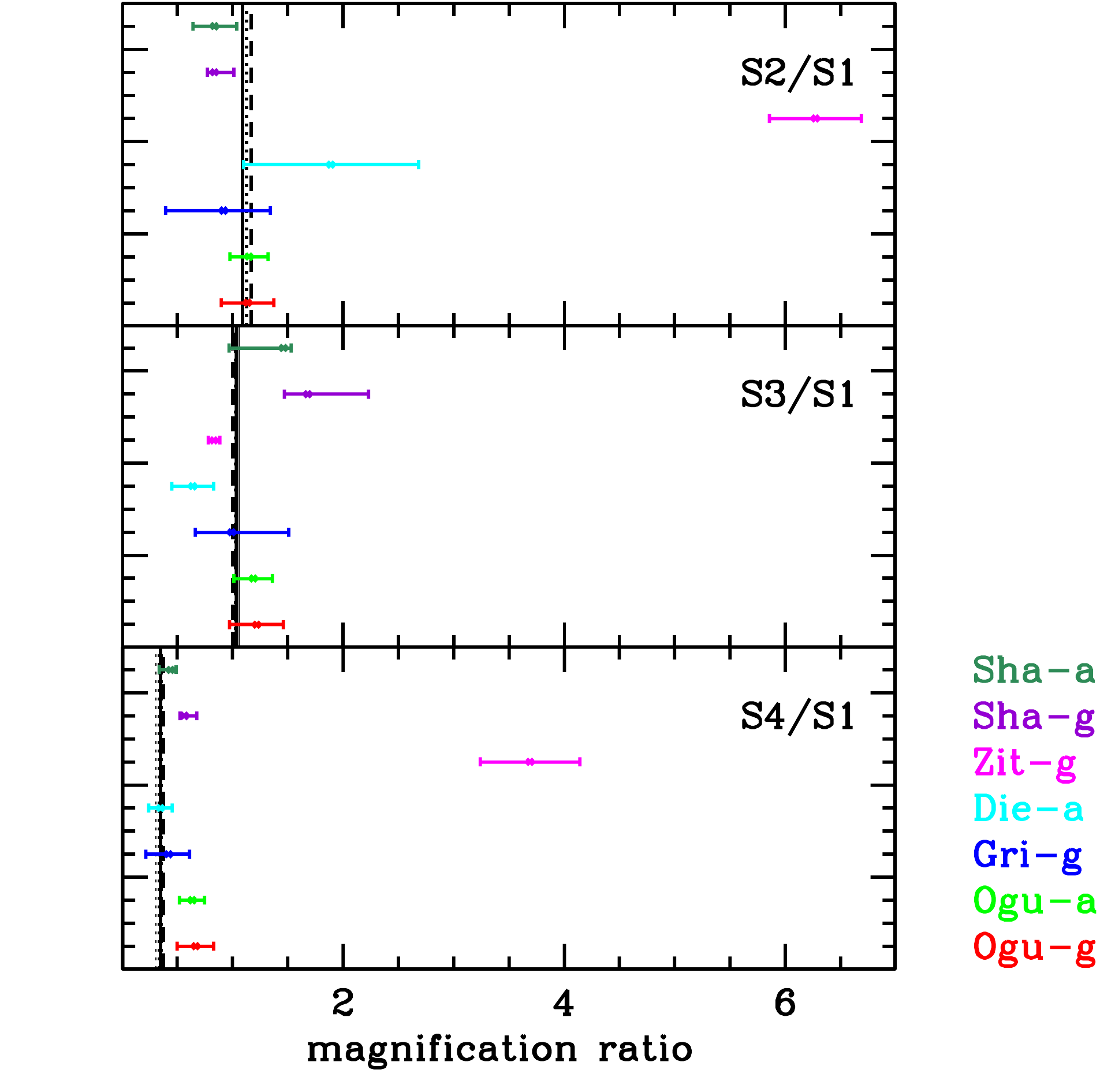}}
\caption{Observed (lines as in Figure~\ref{fig:tdcross}) and predicted (points with error bars) magnification ratios (absolute values) for the images in the cross configuration, relative to S1.
\label{fig:mucross}}
\end{figure}

Interestingly, the models appear to predict rather accurately the
observed magnification ratios (Fig.~\ref{fig:mucross}), even though
these quantities should be more sensitive to systematic uncertainties
arising from millilensing and microlensing effects than time delays.

Overall, the Zit-g model stands apart from the rest, predicting
significantly different time delays and magnification ratios, and
larger uncertainties. This qualitative difference is consistent with
the different topology of the time-delay surface highlighted in the
previous section. Quantitatively, however, the Zit-g model predictions
are in broad agreement with the measurements if one considers the 95\%
credible interval. Collectively, the ``simply-parametrized'' models seem
to predict smaller uncertainties than the others, especially the Ogu-g
and Ogu-a ones. This is expected, considering that they have less
flexibility than the free-form model. What is surprising, however, is
that they also obtain the smaller RMS residual scatter in the
predicted vs. observed image positions (Table~\ref{tab:models}). The
Zit-g light-traces-mass model is perhaps the least flexible, in the
sense that it cannot account for systematic variations in the
projected $M/L$ ratio. This appears to be reflected in its
overall largest RMS residual scatter. When comparing the Die-a to
the Zit-g model, we note that the former uses significantly more
constraints than the latter. This may explain why, even though Die-a
is in principle more flexible, it ends up estimating generally smaller
uncertainties than Zit-g.

\subsection{Forecasts for SN Refsdal: Peak Appearance and Brightness}
\label{ssec:long}

\begin{figure}[]
\centerline{
\includegraphics[width=0.49\textwidth]{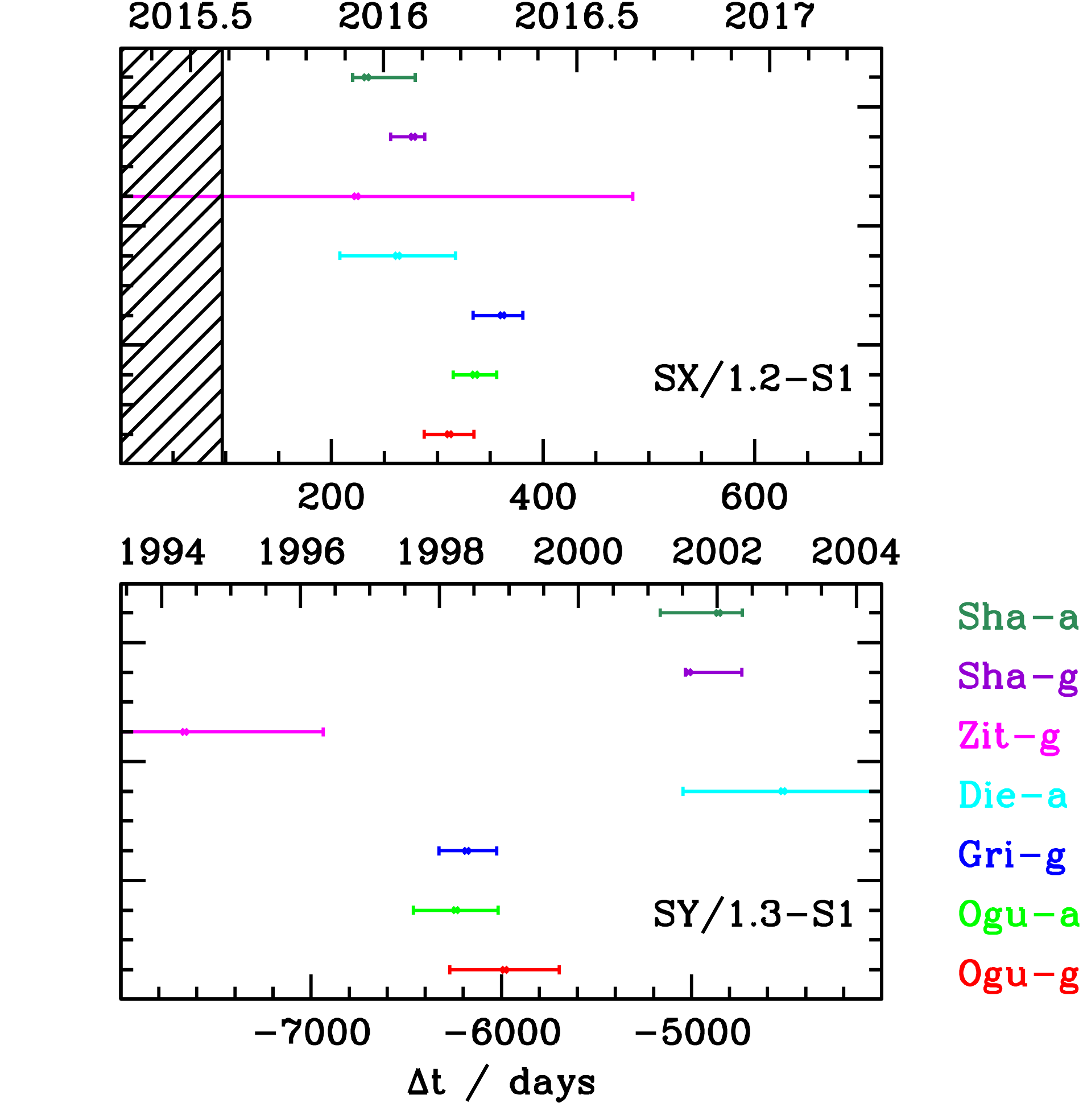}}
\caption{Predicted time delays for the more distant images, relative to S1. The top scale gives the expected date of the peak brightness of the image, with an uncertainty of $\pm20$ days given by the uncertainty in the date of the peak of the observed images (to preserve full blindness we adopt here the preliminary peak measurement April 26 2015, and not the revised measurement April 13 2015; they are consistent within the uncertainties of $\pm20$ days). The hatched region is ruled out by past {\it HST} observations.
\label{fig:tdlong}}
\end{figure}

\begin{figure}[]
\centerline{
\includegraphics[width=0.49\textwidth]{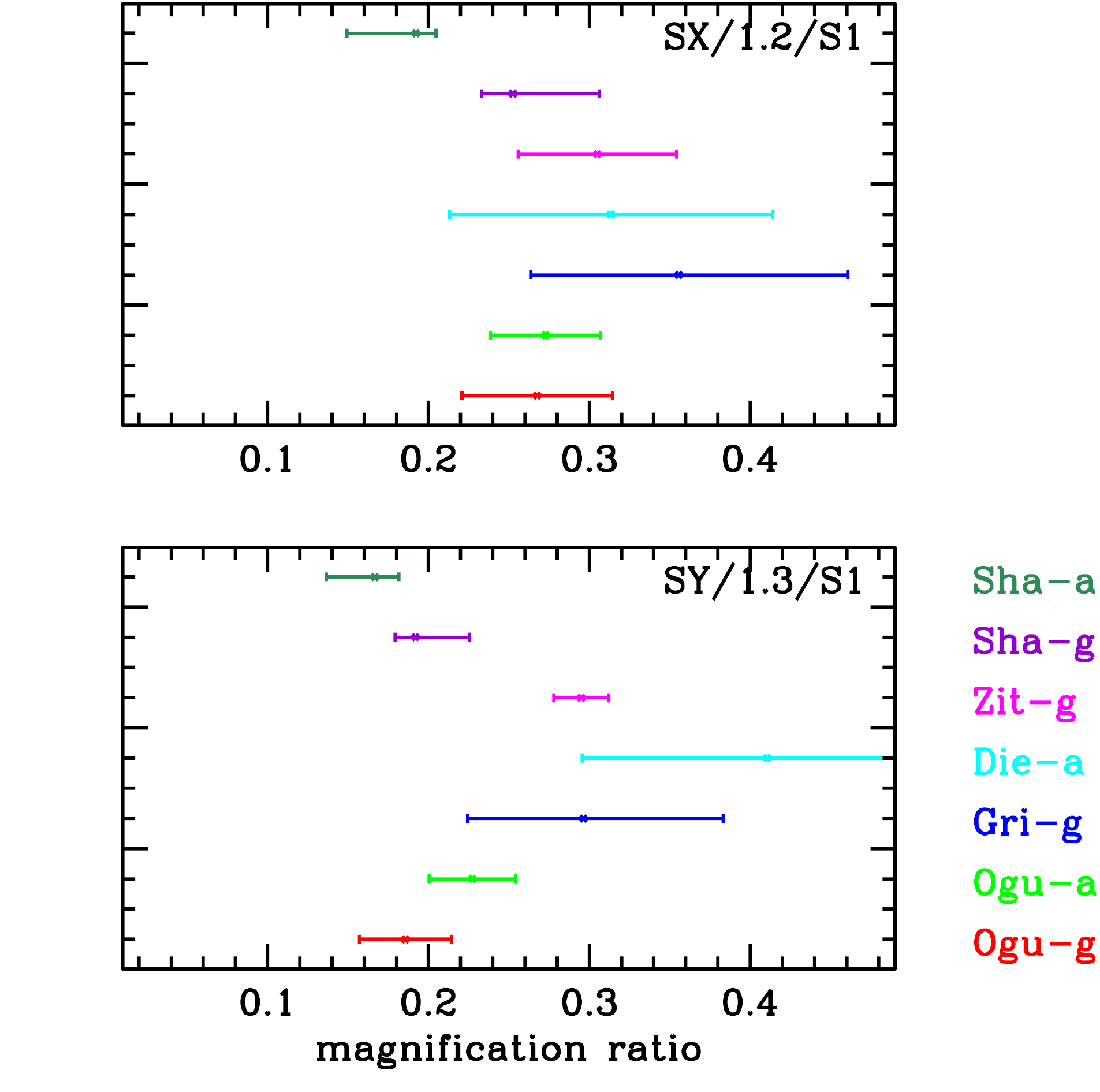}}
\caption{Predicted magnification ratios (absolute values) for the more distant images, relative to S1.
\label{fig:mulong}}
\end{figure}

Figure~\ref{fig:tdlong} compares the prediction for the next
appearance of SN Refsdal, near image 1.2 of the spiral galaxy
(hereafter SX/1.2). All the models considered here predict the image
to peak between the end of 2015 and the first half of 2016. We note
that S1 was first discovered six months before its peak with F160W AB
magnitude $\sim$25.5 \citep{Kel++15}, and it peaked at
F160W $\approx$ 24.5 AB (Kelly et al. 2015, in prep.; Rodney et
al. 2015, in prep.). Image SX/1.2 is predicted to be
approximately 1/3 as bright as image S1 (Figure~\ref{fig:mulong}), so
it should be $\sim$26.7~mag six months before peak and
$\sim 25.7$~mag at peak. No image is detected in the vicinity of SY/1.3 in data
taken with \HST\ up until \M\ became unobservable at the end of July,
allowing us to rule out predicted peak times until January 2016.

Remarkably, the models are in excellent mutual agreement regarding the
next appearance of SN Refsdal. All of the predictions agree on the first
trimester of 2016 as the most likely date of the peak. Sha-a is the
only one that predicts a slightly fainter flux with a magnification
ratio ($0.19^{+0.01}_{-0.04}$) as opposed to the $\sim$1/3 value
predicted by the other models. Interestingly, Zit-g has the largest
uncertainty on time delay, but not on magnification ratio. As in the
case of the cross configuration, the ``simply-parametrized'' models yield
the smallest uncertainties.

Unfortunately, the model-based estimates of the past appearance of SN
Refsdal cannot be tested by observations. The image near 1.3 (hereafter
SY/1.3) is estimated to have been significantly fainter than S1, and
thus undetectable from the ground, at a time when WFC3-IR was not
available. The images of \M\ taken in the optical with ACS in April
2004 (GO-9722, PI Ebeling; 3$\sigma$ limit F814W AB = 27.0 mag) are not
sufficiently deep to set any significant constraints, considering the
peak brightness of S1 in F814W was $\sim$27 mag, and we
expect SY/1.3 to be 0.75--2 mag fainter.  As a purely
theoretical exercise it is interesting to notice that the time delay
varies dramatically between models, differing by almost 10 years
between the Zit-g and the Sha-a, Sha-g, and Die-a models. Remarkably,
and similarly to what was seen for the cross configuration, the
magnifications are in significantly better agreement.

\section{Discussion}
\label{sec:discussion}

In this section we briefly discuss our results, first by
recapitulating the limitations of our analysis
(Section~\ref{ssec:limits}), and then by comparing them with previous
work (Section~\ref{ssec:comp}).

\subsection{Limitation of the Blind Test and of the Models}
\label{ssec:limits}

SN Refsdal gives us a unique opportunity to test our models
blindly. However, in order to draw the appropriate conclusions from
this test, we need to be aware of the limitations of both the test and
the models.

The first limitation to keep in mind is that this test is very
specific. We are effectively testing point-like predictions of the
lensing potential and its derivatives. Similarly to the case of SN
``Tomas'' \citep{Rod++15}, it is very hard to generalize the results of
this test even to the strong lensing area shown in our maps. More
global metrics should be used to infer a more global assessment of the
quality of the models. An example of such a metric is the RMS scatter
between the image positions given in Table~\ref{tab:models}, even
though of course even this metric does not capture all of the features 
of a model. For example, the RMS does not capture how well the model
reproduces time delays and magnifications, in addition to positions,
and one could imagine trading one for the other. 

It is also important to remind ourselves that whereas the
magnification and time delays at specific points may vary
significantly between models, other quantities that are more relevant
for statistical use of clusters as cosmic telescopes, such as the area
in the source plane, are much more stable
\citep[e.g.,][]{Wan+15}. And, of course, other quantities such as
colors and line ratios are not affected at all by gravitational
lensing. It would be interesting to find ways to carry out true
observational tests of more global predictions of lens models. One way
to achieve this would be to carry out tests similar to those afforded
by SN Tomas and SN Refsdal on a large sample of clusters. Another
possibility could be to reach sufficiently deep that the statistical
properties of the background sources (e.g., the luminosity function)
are measured with sufficient precision and small enough cosmic
variance to allow for meaningful tests of model uncertainties.
Alternatively, tests against simulated data are certainly informative
(e.g., Meneghetti et al. 2015, in prep.), although their results
should also be interpreted with great care, as they depend crucially
on the fidelity of the simulated data and the cross-talk between
methods used to simulate the data and those used to carry out the
inference.

The second limitation to keep in mind is that the uncertainties listed
in this paper are purely statistical in nature. As for the case of
image positions --- where the RMS scatter is typically larger than
the astrometric precision of the image positions themselves
\citep[consistent with the fact that there are residual systematics in
cluster lens modeling owing to known effects such as substructure,
e.g.,][]{Bra++09} --- we should not expect the time delays and
magnifications to be perfectly reproduced by the models either.  The
spread between the different model predictions gives us an idea of the
so-called model uncertainties, even though unfortunately they cannot
be considered an exact measurement. The spread could be exaggerated by
inappropriate assumptions in some of the models, or underestimated if
common assumptions are unjustified.

\begin{figure}[]
\centerline{
\includegraphics[width=0.49\textwidth]{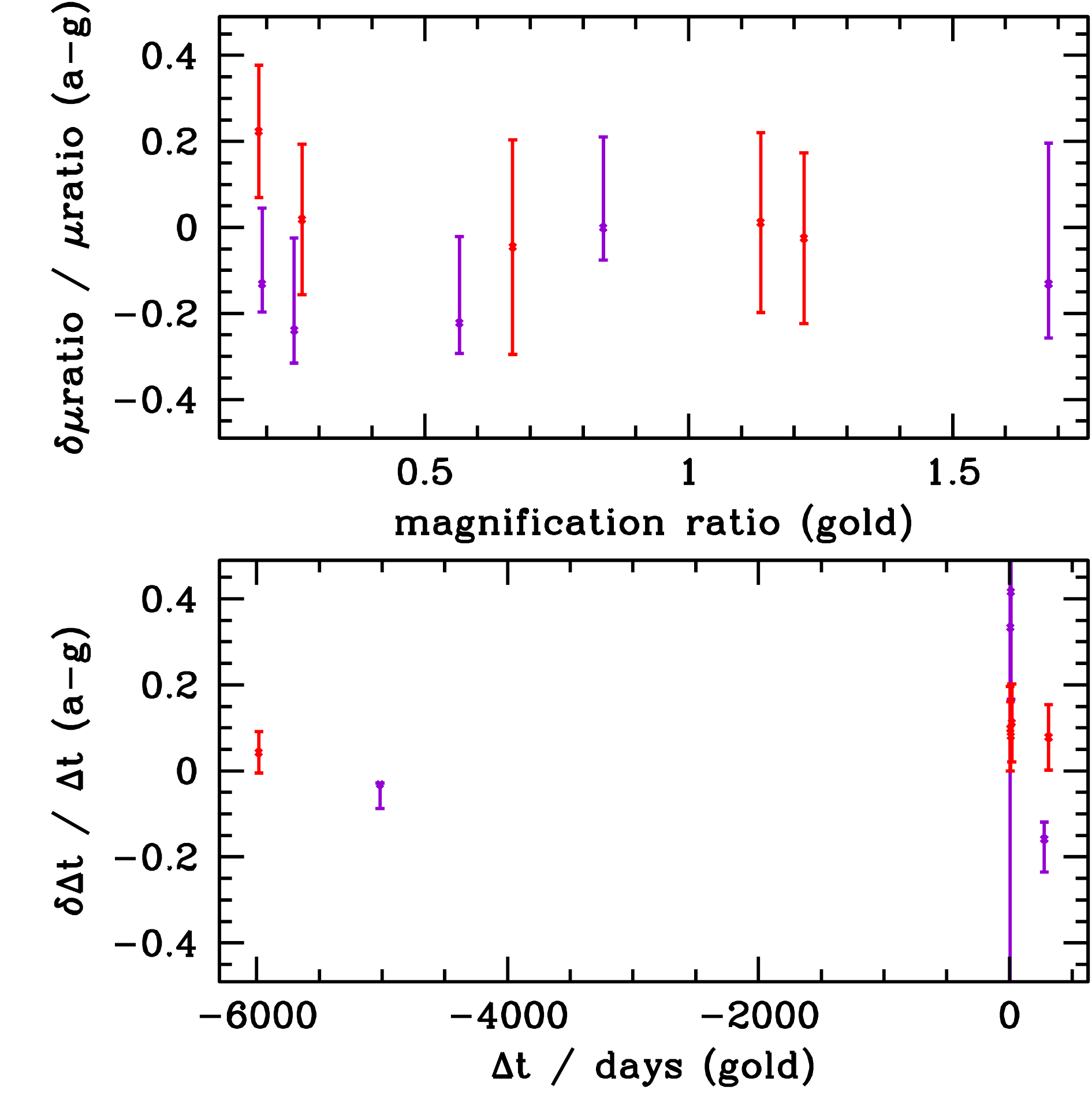}}
\caption{Relative change in predicted time delays (bottom) and magnification ratio (top) between ``a'' and ``g'' models. Dark purple points represent the models by Sharon et al., while red points represent the models by Oguri et al. Vertical bars represent the statistical uncertainties on the ``g'' models.
\label{fig:avsg}}
\end{figure}

We can use the fact that Oguri et al. and Sharon et al. each
submitted two models to estimate the uncertainty relative to the
choice of multiple images. By comparing the predictions of the Sha-a
and Sha-g and of the Ogu-a and Ogu-g models, we can measure how much
the predictions of the models change by adding nonspectroscopically
confirmed images to the gold sample, keeping everything else fixed. As
shown in Figure~\ref{fig:avsg}, the predicted magnification ratios
change by less than the statistical uncertainties.  The changes in
predicted time delays are slightly larger, comparable to the estimated
statistical uncertainties, although the relative change in time delays
is perhaps not the best metric for the short time delays in the
Einstein cross configuration. As can be seen in
Figures~\ref{fig:tdcross} and~\ref{fig:tdlong}, the absolute change in
predicted time delays is typically within the estimated statistical
uncertainties. From this test we conclude that in this case deciding
whether to consider a secure but not spectroscopically confirmed set
of multiple images introduces an uncertainty that is subdominant with
respect to the statistical uncertainties. This is consistent with our
expectation that the enlarged set of multiple images does not contain
false candidates. In interpreting this result, however, we have to
keep in mind the locality of this test and the fact that the nearest
images to the observed and predicted images of SN Refsdal are the
knots of its host galaxy, all at the same known spectroscopic
redshift. Thus, it would have been surprising to find a large
difference at these locations.

As already mentioned in the introduction, other potential sources of
uncertainty are related to the mass-sheet degeneracy and its
generalizations \citep{FGS85,SS13,SS14}, the effects of structure
along the line of sight \citep{DHB05}, and multiplane lensing
\citep{S14,Mcc++15}. All of the models considered here are single-plane
lens models. They break the mass-sheet degeneracy by assuming that the
surface mass density profile goes to zero at infinity with a specific
radial dependency.

On the scale of the known images of SN Refsdal, the measured time
delays and magnification ratios give us a way to estimate these
residual uncertainties. The reasonably good agreement between the
model prediction and measurements shows that these (systematic)
``unknown unknowns'' are not dominant with the respect to the
(statistical) ``known unknowns.'' However, since the agreement is not
perfect, we conclude that the ``unknown unknowns'' are not negligible
either. We can perhaps use the experience gathered in the study of
time delays of lensed quasars to estimate the amplitude of the 
line-of-sight uncertainties. On scales similar to that of the known 
images of SN Refsdal, they are believed to be up to $\sim10$\% before 
corrections for galaxies not in clusters
\citep{Suy++10,Gre++13,Collett:2013p34320,Suy++14}.  
In numerical simulations, the line-of-sight effects appear to increase with 
the measured overdensity of galaxies \citep{Gre++13}, so it is possible
that they are larger for an overdense region like that of \M. 

On galaxy scales, breaking the mass-sheet degeneracy using stellar
kinematics and physically motivated galaxy models appears to produce
results consistent with residual uncertainties on the order of a few
percent \citep{Suy++14}. On cluster scales, the degeneracy is partly
broken by the use of multiple images at different redshifts
\citep[e.g.,][]{BLS04}. However, in the absence of nonlensing
data, we cannot rule out that the residual mass-sheet degeneracy is
the dominant source of systematic uncertainty. Assessing the
uncertainties related to multiplane effects would require knowledge of
the mass distribution in three dimensions and is beyond the scope of
the present work. Thus, multiplane lensing cannot be ruled out as a
significant source of systematic uncertainty for the prediction of the
time delay and magnification ratios of the known images of
SN Refsdal. As far as the future image of SN Refsdal is concerned,
future observations will tell us how much our uncertainties are
underestimated owing to unknown systematics.

Finally, we remind the reader that although for this analysis we kept
fixed the cosmological parameters, they are a (subdominant) source of
uncertainty. To first order, the time-delay distance is proportional
to the Hubble constant, so there is at least a 3\% systematic
uncertainty \citep{Rie++11,Fre++12} on our predicted time delays
\citep[and typically 5--10\% when considering all of the other parameters, 
depending on assumptions and priors;][]{Suy++13,Suy++14}.

\subsection{Comparison with Previous Models}
\label{ssec:comp}

We can get a quantitative sense of the improvement of the mass models
as a result of the new data by comparing how the prediction of the
time delay and magnification ratios have changed for the teams who had
previously published predictions.

\subsubsection{Previous Models by Members of our Team}

The Zit-g model updates the models developed by A.Z. for the SN 
Refsdal discovery paper \citep{Kel++15}. The Zit-g model supersedes
the estimates of time delays and magnifications given in the original
paper by providing predictions as well as quantitative uncertainties.

The update of the \cite{Ogu2015} model presented here changes the time
delays for S2, S3, S4, SX, SY from 9.2, 5.2, 22.5, 357, $-6193$ days to
$8.7\pm0.7$, $5.1\pm0.5$, $18.8\pm1.7$, $311\pm23.6$, $-5982\pm287$
days, respectively (for the Ogu-g model, see plot for Ogu-a). Thus,
the predicted time delays have changed by less than 1--2$\sigma$,
with the inclusion of additional data. The magnification ratios have
been similarly stable. The main effect of the additional data has been
to reduce the uncertainties.

\begin{deluxetable*}{lcccccccccc} \tablecolumns{11}
\tablewidth{0pt} 
\tablecaption{Summary of Predicted Time Delays and Magnification Ratios \label{tab:predictions}}
\tablehead{
\colhead{Model}    & \colhead{$\Delta t_{21}$} & \colhead{$\Delta t_{31}$}     & \colhead{$\Delta t_{41}$}& \colhead{$\Delta t_{\rm X1}$} & \colhead{$\Delta t_{\rm Y1}$} & \colhead{$\mu$(2)/$\mu$(1)} & \colhead{$\mu$(3)/$\mu$(1)} & \colhead{$\mu$(4)/$\mu$(1)} & \colhead{$\mu$(X)/$\mu$(1)} & \colhead{$\mu$(Y)/$\mu$(1)}}
Die-a & -17$\pm$19 & -4.0$\pm$27 & 74$\pm$43 & 262$\pm$55 & -4521$\pm$524 	& 1.89$\pm$0.79 & 0.64$\pm$0.19 & 0.35$\pm$0.11 & 0.31$\pm$0.10 & 0.41$\pm$0.11 \\
Gri-g & 10.6$^{+6.2}_{-3.0}$ & 4.8$^{+ 3.2}_{-1.8}$ & 25.9$^{+8.1}_{-4.3}$ & 361$^{+ 19}_{-27}$ & -6183$^{+160}_{-145}$ 	& 0.92$^{+0.43}_{-0.52}$ & 0.99$^{+0.52}_{-0.33}$ & 0.42$^{+0.19}_{-0.20}$ & 0.36$^{+ 0.11}_{-0.09}$ & 0.30$^{+0.09}_{-0.07}$ \\
Ogu-g & 8.7$\pm$0.7 & 5.1$\pm$0.5 & 18.8$\pm$1.7 & 311$\pm$24 & -5982$\pm$287 	& 1.14$\pm$0.24 & 1.22$\pm$0.24 & 0.67$\pm$0.17 & 0.27$\pm$0.05 & 0.19$\pm$0.03 \\
Ogu-a & 9.4$\pm$1.1 & 5.6$\pm$0.5 & 20.9$\pm$2.0 & 336$\pm$21 & -6239$\pm$224 	& 1.15$\pm$0.17 & 1.19$\pm$0.17 & 0.64$\pm$0.11 & 0.27$\pm$0.03 & 0.23$\pm$0.03 \\
Sha-g & 6$^{+6}_{-5}$ & -1$^{+7}_{-5}$ & 12$^{+3}_{-3}$ & 277$^{+11}_{-21}$ & -5016$^{+281}_{-15}$ 	& 0.84$^{+0.18}_{-0.06}$ & 1.68$^{+0.55}_{-0.21}$ & 0.57$^{+0.11}_{-0.04}$ & 0.25$^{+0.05}_{-0.02}$ & 0.19$^{+0.03}_{-0.01}$ \\ 
Sha-a & 8$^{+7}_{-5}$ & 5$^{+10}_{-7}$ & 17$^{+6}_{-5}$ & 233$^{+46}_{-13}$ & -4860$^{+126}_{-305}$ 	& 0.84$^{+0.20}_{-0.19}$ & 1.46$^{+0.07}_{-0.49}$ & 0.44$^{+0.05}_{-0.10}$ & 0.19$^{+0.01}_{-0.04}$ & 0.17$^{+0.02}_{-0.03}$  \\
Zit-g & -161$\pm$97 & -149$\pm$113 & 82$\pm$51 & 224$\pm$262 & -7665$\pm$730 	& 6.27$\pm$0.41 & 0.83$\pm$0.05 & 3.69$\pm$0.45 & 0.31$\pm$0.05 & 0.30$\pm$0.02 \\
\enddata
\tablecomments{For each model we list the predicted observables: time delays (in days) and absolute values of the magnification ratios, relative to image S1. The listed uncertainties only include random errors. Systematic errors are nonnegligible and described in detail in Section~\ref{ssec:limits}.
}
\end{deluxetable*}
\clearpage

The update of the \citet{S+J15} model presented here changes the time
delays for S2, S3, S4, SX, SY from $2.0^{+10}_{-6}$,
$-5.0^{+13}_{-7}$, $7.0^{+16}_{-3}$, $237^{+37}_{-50}$
$-4251^{+369}_{-373}$ days to $6\pm6$, $-1^{+7}_{-3}$, $12\pm3$,
$277^{+11}_{-21}$, $-5016^{+281}_{-15}$ days, respectively (for the
Sha-g model, see plot for Sha-a). Thus, the predicted time delays have
changed by less than 1--2$\sigma$, with the inclusion of additional
data, especially the new spectroscopic redshifts (the list of multiple
images is very similar). The magnification ratios have been similarly
stable.  The main effect of the additional data has been to reduce the
uncertainties.

\citet{Die++15} do not give time delays for the cross configuration, owing 
to the limitations inherent to keeping the $M/L$ of the galaxy in the
middle of the cross fixed to the global value.  Their predictions for
the long delays SX and SY have changed with the inclusion of new data
from $375\pm25$ to $262\pm54$ days, and from $-3325\pm762$ to
$-4521\pm524$. Interestingly, the uncertainties in the future delay
have increased with the new data, which may be caused by the correction
of previously erroneous inputs, like the redshift of system 3, and
also to the increased range of models and grid parameters considered
here.  The fact that the predictions changed by more than the
estimated uncertainties is consistent with our previous conclusion
that the statistical uncertainties underestimate the total
uncertainty.

\subsubsection{Jauzac et al.}

During the final stages of the preparation of this manuscript,
\citet{Jau++15} posted on the arxiv another independent model of
\M. Their model is based on a subset of the data presented here, 
and different sets of multiple images and knots in the spiral host
galaxy. Comparing only the systems with spectroscopic redshifts, our
analyses agree on multiply imaged systems 1, 2, 3, 4, 5, 110 (22 in their
nomenclature), and in rejecting the identification of system 12 as
multiply imaged. We do not use system 9, for which they obtain a
spectroscopic redshift of 0.981. We obtain spectroscopic redshifts for
systems 13 and 14 (1.24 and 3.70), which contradict their model
redshifts of 1.34 and 2.88.  Their catalog comprises 57
spectroscopically confirmed cluster members, while ours consists of
\Nzmem. Thus, the \citet{Jau++15} model is not directly comparable to the
models presented here. However, it provides a useful additional
comparison for this forecast. We note that \citet{Jau++15} include the
main developers of Lenstool, the ``simply-parametrized'' lens modeling
software used by Sharon et al. for the analysis presented in this
paper. The difference in the predictions between the two teams
highlights how systematic differences can arise from input data and
modelers' choices, as well as from assumptions of each modeling method.

The \citet{Jau++15} model\footnote{We refer here to version 3 of the
\citet{Jau++15} paper, which appeared on the arxiv on 2015 October 13. 
The predictions have changed significantly between
versions 2 (2015 October 1) and 3, owing to the improved treatment of
the cluster galaxy nearest to the cross configuration.} predicts time
delays and magnification ratios that are significantly different from
the ones actually observed from the cross configuration. When
comparing their prediction with observations, all of the limitations
discussed in Section~\ref{ssec:limits} should be kept in mind, as they
apply to the Jauzac et al.\ models as well. Furthermore, as
\citet{Jau++15} point out, their predictions for the cross configuration are
very sensitive to the mass density profile assumed for the cluster
galaxy closest to it. Therefore, the random uncertainties
underestimate the total uncertainties.

Their predicted time delays for S2--S1, S3--S1, and S4--S1 are
$90\pm17$, $30\pm35$, and $-60\pm41$ days (respectively), to be
compared with the measured values given in
Table~\ref{tab:measurements}.  Considering their statistical
uncertainties, which are much larger than the measurement
uncertainties, the time delays are within $\sim$5$\sigma$, 1$\sigma$,
and 2$\sigma$ of the measurements.  The disagreement with the S2--S1
time delay is especially remarkable considering that all the other
models predict the two images to be almost simultaneous.  The flux
ratios ($0.86\pm0.13$, $0.89\pm0.11$, and $0.42\pm0.05$ for S2/S1,
S3/S1, S4/S1, respectively) are also somewhat in tension with the
measured values, although the disagreement is in line with that of the
models presented in this paper.  It would be interesting to update the
\citet{Jau++15} model, correcting the redshifts of systems 13 and 14
to see if those misidentifications could be at the root of the
discrepancy.  Overall, it is interesting to note that
\cite{Jau++15} predict the magnifications with higher accuracy
than the time delays, even though magnifications are potentially more
sensitive to local substructure (millilensing) and microlensing
effects.

The time delay predicted by \citet{Jau++15} for image SX/1.2 is
significantly longer than for the models presented here, pushing the
next appearance of the peak to the middle of 2016. The time delay of
image SY/1.3 is shorter than for most models presented here, but
unfortunately not short enough to be testable with archival
observations. Incidentally, \citet{Jau++15} also predict SY/1.3 to be
fainter than in the models presented here ($0.16\pm0.02$ of the
brightness of S1), which makes this prediction even more difficult to
test with archival data.

\section{Summary} 
\label{sec:summary} 
SN Refsdal gives us a unique opportunity to carry out a truly blind test
of cluster-scale gravitational lens models. In order to make the most
of this opportunity, we have used an unprecedented combination of
imaging and spectroscopic data as input for 7 lens models, based on 5
independent techniques. The models have been tested against
independent measurements of time delays and magnification ratios for
the known images of SN Refsdal and used to predict its future (and past)
appearance. Our main results can be summarized as follows.

\begin{enumerate}
\item We have collected \Nztot\ spectroscopic redshifts in the field of \M\ from VLT-MUSE \citep{Gri++15}, Keck DEIMOS, and {\it HST}-WFC3 \citep[Brammer et al. 2015, in prep.][]{Schmidt:2014p33661,Tre++15} observations. These include \Nzmem\ spectroscopic cluster members and 23 multiple images of 10 different galaxies.
\item We have collected measurements of time delays and magnification ratios for the known images of SN Refsdal (Rodney et al. 2015, in prep.).
\item We have compiled and expanded a list of candidate multiply imaged galaxies and multiply imaged knots in the host galaxy of SN Refsdal. All images have been vetted by a group of expert classifiers, resulting in a list of gold and silver-quality images. 
\item The seven lens models have remarkably good fidelity, with residual RMS scatter between observed and predicted image positions ranging between 0\farcs16 and 1\farcs3.
\item The model predictions agree reasonably well with the observed delays and magnifications of SN Refsdal (within 68--95\% uncertainty, or 10 days in the case of S2--S1), showing that unknown systematics are comparable to or smaller than the calculated statistical uncertainties.
\item All models predict that an image of SN Refsdal will appear near the SX/1.2 location between the submission of this paper and the beginning of 2016. The most likely time for the peak is the first trimester of 2016. 
Given the slow rise of the light curve of SN Refsdal and the predicted brightness of SX/1.2, the image could appear as soon as \M\ is visible again by {\it HST}-WFC3 at the end of October 2015. 
\item The past appearance of SN Refsdal near position SY/1.3 would have been too faint to be detectable in existing archival images, and thus cannot be tested.
\end{enumerate}

There are two possible outcomes to the work presented in this
paper. First, our predictions could be proven correct. This outcome
would be an encouraging sign that all the efforts by the community to
gather data and improve lens modeling tools are paying off. If,
alternatively, our predictions turn out to be wrong, we will have to
go back to the drawing board, having learned an important lesson about
systematic uncertainties.

\acknowledgements

The authors thank Raphael Gavazzi, Mathilde Jauzac, and Jens Hjorth
for insightful comments on the manuscript. This work is supported by
NASA through grants HST-GO-13459 and HST-GO-14041 from the Space
Telescope Science Institute (STScI), which is operated by the
Association of Universities for Research in Astronomy, Inc., under
NASA contract NAS 5-26555. T.T. gratefully acknowledges the
hospitality of the American Academy in Rome and of the Observatorio di
Monteporzio Catone, where parts of this manuscript were written.
T.T. is supported by the Packard Foundation in the form of a
Packard Research Fellowship.  J.M.D. is grateful for support from the
consolider project CSD2010-00064 and AYA2012-39475-C02-01 funded by
the Ministerio de Economia y Competitividad. A.V.F.'s research is
supported by the Christopher R. Redlich Fund, the TABASGO Foundation,
and NSF grant AST-1211916.  C.G. acknowledges support by the VILLUM
FONDEN Young Investigator Programme through grant no. 10123.  Support
for A.Z. was provided by NASA through Hubble Fellowship grant
\#HST-HF2-51334.001-A awarded by STScI. The work of M.O. was supported
in part by World Premier International Research Center Initiative (WPI
Initiative), MEXT, Japan, and Grant-in-Aid for Scientific Research
from the JSPS (26800093). Financial support for this work was provided
to S.A.R. by NASA through grants HST-HF-51312 and HST-GO-13386 from
STScI. A.H. is supported by NASA Headquarters under the NASA Earth and
Space Science Fellowship Program, grant ASTRO14F-0007. R.J.F. 
gratefully acknowledges support from NSF grant
AST-1518052, NASA grants HST-GO-14041 and HST-GO-13386, and the Alfred
P. Sloan Foundation.  This work was supported in part by a NASA Keck
PI Data Award (PID 47/2014B\_N125D, PI Jha).

We are very grateful to the staff of the {\it Hubble Space Telescope}
for their assistance in planning, scheduling, and executing the 
{\it HST} observations used in this work. We thank the STScI and ESO
directors for granting Directory Discretionary time to allow for
timely follow-up observations of SN Refsdal. Some of the data
presented herein were obtained at the W. M. Keck Observatory from
telescope time allocated to NASA through the agency's scientific
partnership with the California Institute of Technology and the
University of California; the Observatory was made possible by the
generous financial support of the W. M. Keck Foundation. The authors
wish to recognize and acknowledge the very significant cultural role
and reverence that the summit of Mauna Kea has always had within the
indigenous Hawaiian community. We are most fortunate to have the
opportunity to conduct observations from this mountain.

\section*{Note added in proof}

After the acceptance of this manuscript, it was discovered that the
predictions of the Zit-g lens model are inaccurate, due to numerical
resolution insufficient to correctly resolve the vicinity of the
Einstein Cross configuration S1-S4.  The quantities affected are the
panels labeled Zit-g in Figures 4,5,6,7,8; the points labeled Zit-g in
figures 9,10,11,12; and the last line of Table 6.  While calculations
at higher resolution yield qualitatively the same results,
nevertheless they have smaller uncertainties and are in better
agreement with the observed S1-S4 time delays and magnification ratio.

The numerical inaccuracy of the Zit-g model does not affect in any way
the other results in the paper, including redshifts, arcs and knots
identifications, measured time delays, and the predictions of the six other
lens models. Furthermore, since the qualitative description of the Zit-g
model and its predictions are correct, the discussion and
conclusions of the paper are not affected in any way.

The error was discovered after the blind test deadline (30 October
2015, date of the first HST observations of the field). Thus, in order
to preserve the blindness of the models described in this paper, the
corrected results will be presented in a future publication.


\end{document}